\documentclass[linenumbers]{aastex62}
\usepackage{times}
\usepackage{epsfig,graphics}
\usepackage{amsmath, amssymb, graphicx, multirow}
\usepackage{textcomp}
\usepackage{natbib}
\usepackage{float}
\usepackage{color}
\usepackage{epstopdf}
\usepackage[countmax]{subfloat}
\usepackage{booktabs}
\usepackage[caption=false]{subfig}
\usepackage{mwe}
\usepackage{longtable}
\newsubfloat{figure}
\shortauthors{Kalita et al.}

\accepted{ApJS}
\shorttitle{Optical variability of BZCAT blazars}
\shortauthors{Kalita et al. 2021}
\begin{document}

\title{\Large {\bf Optical variability of a newly discovered blazar sample from the BZCAT Catalog}} 

\correspondingauthor{Nibedita Kalita}
\email{nibeditaklt1@gmail.com}

\author[0000-0002-9323-4150]{Nibedita Kalita}
\affiliation{Key Laboratory for Research in Galaxies and Cosmology, Shanghai Astronomical Observatory, Chinese~Academy of Sciences, 80 Nandan Road, Shanghai 200030, China}

\author{Alok C. Gupta}
\affiliation{Aryabhatta Research Institute of Observational Sciences (ARIES), Manora Peak, Nainital, 263002, India}

\author{Minfeng Gu}
\affiliation{Key Laboratory for Research in Galaxies and Cosmology, Shanghai Astronomical Observatory, Chinese~Academy of Sciences, 80 Nandan Road, Shanghai 200030, China}


\begin{abstract}
 
In an optical monitoring program to characterize the variability properties of blazar, we observed 10 sources from the Roma-BZCAT catalogue for 26 nights in V and R bands during October 2014 to June 2015 with two telescopes located in India. The sample includes mainly newly discovered BL Lacs where the redshift of some sources are not known yet. We present the results of flux and color variations of the sample on intraday and short time scales obtained by using the power-enhanced F-test and the nested-ANOVA tests, along with their spectral behavior. We find significant intraday variability in the single FSRQ in our sample, having an amplitude of variation $\sim 12\%$. Although a few of BL Lacs showed probable variation in some nights, none of them passes the variability tests at 99.9\% significance level. We find that 78\% of the sample showed significant negative colour--magnitude correlations i.e., a redder-when-brighter spectral evolution. Those which do not show strong or clear chromatism, predominantly exhibit a redder-when-brighter trends. Unlike on hourly timescales, the high synchrotron peaked (HSP) blazars in the sample (BZGJ0656+4237, BZGJ0152+0147 and BZBJ1728+5013) show strong flux variation on days to months timescales, where again we detect a decreasing trend of the spectral slope with brightness. We observe a global steepening of the optical spectrum with increasing flux on intranight timescale for the entire blazar sample. Non-variability in the BL Lacs in our sample could be resulted by distinct contribution from the disk as well as from other components in the studied energy range.  

\end{abstract}

\section{{\bf Introduction}}

\noindent
Blazars represent a small subset of radio-loud (RL) active galactic nuclei (AGNs) which exhibit strong flux and spectral variability in the complete electromagnetic (EM) spectrum, large ($>$ 3\%) and variable polarization from radio to UV wavelengths, and usually display core-dominated radio structures. Classically BL Lacertae objects (BL Lacs) and flat-spectrum radio quasars (FSRQs) are collectively known as blazars. In optical spectrum, BL Lacs either have very weak (equivalent width $<$ 5$\AA$) or no emission lines, while FSRQs show strong emission lines \citep[e.g.,][]{1991ApJS...76..813S, 1996MNRAS.281..425M, 1978PhyS...17..265B, 1997A&A...327...61G}. The observed radiative output of blazars from radio to high energy $\gamma$-rays, is dominated by nonthermal emission from a relativistic jet pointing towards us making an angle $<$10$^{\circ}$ with the observer's line of sight \citep{1995PASP..107..803U}. The Doppler boosted broadband spectral energy distribution (SED) of blazars is characterize by two distinct components in the $\nu$--$\nu$$F_{\nu}$ representation \citep{1998MNRAS.299..433F,1998MNRAS.301..451G}. The first component which usually peaks at near-infrared (NIR)/optical or UV/X-rays wavelengths ascribed to synchrotron emission from relativistic particles accelerating within the inner part of the jet, and the second component which peaks in GeV to TeV energies, commonly believed to be arises from inverse-Compton (IC) scattering of low energy seed photons off the synchrotron emitting charged particles \citep{1998A&A...333..452K}. Besides the classical classification, blazars are further divided based on the peak frequency of synchrotron hump $\nu_{sync}^{peak}$ in the SEDs of a large sample of Fermi-LAT blazars, into low-synchrotron-peaked (LSP) blazars with $\nu_{sync}^{peak} \leq$ 10$^{14}$ Hz; intermediate-synchrotron-peaked (ISP) blazars with 10$^{14}$ $< \nu_{sync}^{peak} <$ 10$^{15}$ Hz; and high-synchrotron-peaked (HSP) blazars with $\nu_{sync}^{peak} \geq$ 10$^{15}$ Hz \citep{2010ApJ...716...30A}. In a recent study with a much larger sample as compared to the previous study, the synchrotron peak frequency  for upper and lower limit for ISP and HSP blazars have been slightly modified from $10^{15}$ Hz to $10^{15.3}$ Hz, respectively \citep{2016ApJS..226...20F}. The relation between the different class of blazars and so the classification, is somewhat fuzzy as the observed source's properties can change depending on intensity of emission from the jet and thus, still a matter of debate.\\
   
Blazars show flux variation on diverse timescales ranging from as short as a few tens of seconds to as long as several decades. Blazars variability can be broadly classified into three classes, namely, micro or intra-day or intra-night variability, short-term variability, and long-term variability. Variation in flux of up to a few tenths of a magnitude over the course of a day or less is known as intra-day variability (IDV; \citep{1995ARA&A..33..163W}), micro-variability \citep[e.g.,][]{1989Natur.337..627M}, or intra-night variability \citep[e.g.,][]{1995MNRAS.274..701G}. In the manuscript, hereafter we will call it IDV. Short term variability (STV) and long term variability (LTV) can show up to five magnitude change in flux, and usually have timescales from weeks to several months and several months to years, respectively \citep{2004A&A...422..505G}. \\

Optical IDV in blazars is usually intrinsic by nature and may be due to the interaction of shocks with small-scale particles or magnetic field irregularities present in the jet \citep[e.g.,][]{2014ApJ...780...87M,2015JApA...36..255C}. There are only a few blazars (e.g. AO 0235+164) that has revealed foreground-absorbing systems \citep{1987ApJ...318..577C,1996A&A...314..754N} which makes it a gravitationally lensed system, and in such sources detected IDV will be the mixture of intrinsic as well as extrinsic by nature. STV and LTV are usually a mixture of intrinsic and extrinsic mechanisms. On these variability timescales, intrinsic mechanisms involve shocks propagating down twisted jets or plasma blobs moving through some helical structure in jets \citep{2008Natur.452..966M}. Extrinsic mechanisms involve the geometrical effects which result in bending of the jets \citep[e.g.,][]{2016ApJ...820...12P,2012MNRAS.421.1861V}. \\

A new catalog of 2728 blazars called Roma-BZCAT was presented based on multi-frequency surveys and on an extensive review of the literature \citep{2009A&A...495..691M}. In the catalog Blazars were classified as BL Lacs (BZB), FSRQs (BZQ), and blazars of uncertain/transitional type (BZU). The catalog is recently updated which contains 3561 sources \citep{2015Ap&SS.357...75M}. Majority of these blazars were never studied for optical variability properties on diverse timescales. As a pilot long term project to search for optical photometric variability properties of selected blazars, we carried out photometric monitoring using our 1.04 meter and 1.3 meter optical telescopes in India for a blazar sample during October 2014 to October 2015 and are presenting our first result here. \\
  
The paper is structured as follows. In Section 2, we provide information about our new optical photometric 
observational data and its analysis. In Section 3, we present various analysis technique we use to find variability in the blazars light curves (LCs). In Section 4, we present the results, and the discussions are given in Section 5. We summarize our results in Section 6.

\begin{table*}
\centering
\caption{\bf Details of the optical observing facilities.}\vspace{0.2cm} 
\label{tab1}

\begin{tabular}{llllllllll} \hline \hline

Telescope&                 CCD model&     Chip size&        Pixel size&  Scale&                 Field&         Gain&   Readout Noise & Typical seeing\\
      &                              & (pixels$^{2}$)     & ($\mu$m)&  (arcsec pixel$^{-1}$) & (arcmin$^2$)   & ($e^-$ ADU$^{-1}$) & ($e^-$ rms)    & (arcsec)      \\\hline
ST        & Wright 2K     & 2048$\times$2048 & 24        &  0.37                & $13\times13$ & 10           & 5.3          & 1 -- 2.8      \\
ST                   & Tektronics 1k & 1024$\times$1024 & 24        &  0.37                & $6\times6$   & 11.98        & 7            & 1.4 -- 2.6    \\
DFOT  & Andor 2K      & 2048$\times$2048 & 13.5      &  0.535               & $18\times18$ & 1.4          & 4.1          & 1.2 -- 2      \\\hline\\
\end{tabular}
{\bf Notes:} Both telescopes are located at ARIES, Nainital, India. ST and DFOT stand for 1.04 m Sampurnanad Telescope and Devasthal Fast Optical Telescope, respectively. \\
\end{table*}

\begin{table*}
\centering
\caption{\bf Basic parameters of the Roma BZCAT catalog blazars studied in our sample.}\vspace{0.2cm}
\label{tab2}

\begin{tabular}{cccclcccccl} \hline \hline
Source name & Popular name &R.A. ($\alpha_{2000.0}$) &Dec. ($\delta_{2000.0}$) &$z$ &R$_{mag}$ & Source&SED type & X-ray flux \\
            &              &(h m s)                  &( $^{o}$ $\arcmin$ $\arcsec$ )&     &    &class & &0.1--2.4 keV        \\
   (1)      &   (2)        &    (3)           &        (4)        & (5) & (6)       & & (7)      &  (8)       \\\hline
BZGJ0152+0147 &PMN J0152+0146&01 52 39.5 &+01 47 17.1 &0.08  &11.8 & BL Lac  &HSP    & 3.13  \\  
BZBJ0509+0541 &TXS 0506+056  &05 09 25.9 &+05 41 35.3 &...   &14.3 & BL Lac  &ISP    & 0.56  \\   
BZGJ0656+4237 &4C +42.22     &06 56 10.6 &+42 37 02.7 &0.059 &11.1 & BL Lac  &HSP    & 1.35  \\   
BZGJ0737+5941 &...           &07 37 30.0 &+59 41 03.1 &0.041 &11.8 & BL Lac  &...    & 0.8   \\
BZGJ0754+3910 &...           &07 54 37.0 &+39 10 47.6 &0.096 &15.7 & BL Lac  &...    & 0.44  \\
BZGJ0831+5400 &...           &08 31 00.4 &+54 00 23.5 &0.062 &15.3 & BL Lac  &...    & ...   \\ 
BZGJ1154+1225 &...           &11 54 10.4 &+12 25 09.6 &0.081 &13.8 & BL Lac  &...    & ...   \\
BZQJ1229+0203 &3C 273        &12 29 06.7 &+02 03 08.6 &0.158 &12.9 & FSRQ    &LSP    & 63.1  \\
BZBJ1725+1152 &1H 1720+117   &17 25 04.4 &+11 52 15.2 &...   &14.9 & BL Lac  &HSP    & 11.5  \\
BZBJ1728+5013 &IZw 187       &17 28 18.6 &+50 13 10.4 &0.055 &13.9 & BL Lac  &HSP    & 20.4  \\\hline\\
\end{tabular} \\   
{\bf Columns:} (1) Roma BZCAT catalog name of the blazars; (2) name taken from Fermi-AGN catalog; (3) right ascension; (4) declination;\\ (5) redshift (6) apparent R band magnitude taken from BZCAT; (7) classification of the blazars based on spectral energy distribution (SED): high-energy synchrotron peaked (HSP), low-energy synchrotron peaked (LSP) and intermediate synchrotron peaked (ISP) blazar, (8) in the unit of $\times$ $10^{-12}$ erg $cm^{-2} s^{-1}$. {\bf Note:} ... = not available. 
\vspace{0.5cm}  
\end{table*}


\begin{table*}

\centering
\caption{\bf Observing log of the studied Roma BZCAT catalogue blazars.}
\label{tab3}
\small
\begin{tabular}{cccccc} \hline \hline
Source name && Date of Obs.& Telescope & Exposure time    & Data Points\\ 

(BZCAT)     && (dd.mm.yyyy)&           & (hour)           & (V, R)\\\hline 

BZGJ0152+0147  & &21.12.2014    &DFOT  &3 hr 40 m  &26, 26 \\
               & &22.12.2014    &DFOT  &2 hr 15 m  &18, 18 \\ 
               & &23.12.2014    &DFOT  &4 hr 15 m  &34, 34 \\
               & &25.12.2014    &ST    &2 hr 30 m  &14, 14 \\
               & &26.12.2014    &ST    &--         &1, 1   \\
               & &27.12.2014    &ST    &3 hr 20 m  &16, 16 \\ 
               & &28.12.2014    &ST    &--         &3, 3   \\ 
               & &31.12.2014    &ST    &--         &1, 1   \\  

BZBJ0509+0541  
               & &28.12.2014    &ST    &5 hr 30 m  &31, 31 \\
               & &31.12.2014    &ST    &1 hr 30 m  &10, 10   \\ 
                                                                                                 
BZGJ0656+4237  & &15.10.2014    &ST    &1 hr 40 m  &10, 10 \\   
               & &16.10.2014    &ST    &2 hr 20 m  &11, 11 \\   
               & &28.10.2014    &ST    &2 hr       &9, 9   \\   
               & &29.10.2014    &ST    &2 hr       &9, 9   \\   
               & &15.11.2014    &DFOT  &--         &1, 1   \\  
               & &16.11.2014    &DFOT  &6 hr 10 m  &40, 40 \\  
               & &21.11.2014    &DFOT  &1 set      &1, 1   \\  
               & &22.11.2014    &DFOT  &5 hr 20 m  &38, 38 \\  
               & &28.11.2014    &ST    &--         &1, 1   \\  
               & &21.12.2014    &DFOT  &--         &1, 1   \\  
               & &22.12.2014    &DFOT  &--         &1, 1   \\
               & &23.12.2014    &DFOT  &--         &1, 1   \\ 
               & &25.12.2014    &ST    &5 hr 10 m  &24, 24 \\  
               & &31.12.2014    &ST    &--         &1, 1   \\               

BZGJ0737+5941  & &19.04.2015    &ST    &2 hr 30 m  &11, 11 \\
                
BZGJ0754+3910  & &26.12.2014    &ST    &4 hrs 10 m &19, 19 \\
                
BZGJ0831+5400  & &27.12.2014    &ST    &5 hr 20 m  &26, 26 \\ 
               & &28.12.2014    &ST    &3 hr       &18, 18 \\
                
BZGJ1154+1225  & &20.03.2015    &ST    &3 hr       &15, 15 \\

BZQJ1229+0203  & &20.04.2015    &ST    &4 hr 20 m  &26, 26 \\
                
BZBJ1725+1152  & &14.06.2015    &ST    &3 hr 40 m  &23, 23 \\
                
BZBJ1728+5013  & &19.03.2015    &ST    &--         &1, 1  \\
               & &20.03.2015    &ST    &1 hr       &9, 9   \\
               & &19.04.2015    &ST    &3 hr 15 m  &20, 20 \\ 
               & &20.04.2015    &ST    &4 hr       &22, 22 \\
               & &28.05.2015    &ST    &4 hr 40 m  &21, 21 \\
               & &11.06.2015    &ST    &6 hr       &1, 65  \\
               & &12.06.2015    &ST    &5 hr 30 m  &1, 32  \\
               & &14.06.2015    &ST    &--         &1, 1   \\
               & &15.06.2015    &ST    &--         &1, 1   \\\hline
                
\end{tabular} \\ 
\end{table*}

\begin{table*}
\centering
\caption{\bf Positions and magnitudes of the comparison stars used to generate the DLCs of the blazars.}
\label{tab4}
\small
\begin{tabular}{llllllcclc} \hline \hline

Source name& Ref. star, \&&&R.A. ($\alpha_{2000.0}$) &&Dec. ($\delta_{2000.0}$)  &&V$_{mag}$ &&R$_{mag}$ \\
           & Comp. stars  &&(h m s)              &&( $^{o}$ $\arcmin$ $\arcsec$ )&&          &&          \\
           &  (1)         &&   (2)               &&    (3)                       && (4)      && (5) \\\hline

BZGJ0152+0147$^{*}$  &Ref. Star && 01 52 59.94  && +01 46 24.54  && 15.97  &&   15.07       \\  
                     &CS1       && 01 52 52.35  && +01 47 16.38  && 15.07  &&   14.41       \\  
                     &CS2       && 01 52 15.63  && +01 48 34.67  && 16.09  &&   15.51       \\  

BZBJ0509+0541$^{***}$&Ref. Star && 05 09 23.86  && +05 39 46.28  && 15.24  &&   --          \\ 
                     &CS1       && 05 09 30.35  && +05 42 57.25  && 16.05  &&   --          \\  
                     &CS2       && 05 09 33.17  && +05 43 52.12  && 16.23  &&   --          \\  
                                                                                
BZGJ0656+4237$^{**}$ &Ref. Star && 06 56 17.19  && +42 35 52.91  && 15.46  &&   14.53       \\ 
                     &CS1       && 06 56 8.34   && +42 37 38.92  && 16.40  &&   15.58       \\  
                     &CS2       && 06 56 1.87   && +42 38 35.06  && --     &&   --         \\ 
                                                                                
BZGJ0737+5941$^{**}$ &Ref. Star && 07 37 41.9   && +59 42 26.8   && 15.41  &&   14.48       \\ 
                     &CS1       && 07 37 41.5   && +59 41 9.26   && 14.69  &&   14.12       \\ 
                     &CS2       && 07 37 40.69  && +59 40 31.4   && 15.66  &&   15.26       \\  
                                                                                                                                                                
BZGJ0754+3910$^{*}$  &Ref. Star && 07 54 38.99  && +39 12 43.62  && 16.38  &&   16.02       \\ 
                     &CS1       && 07 54 45.88  && +39 12 28.56  && 16.49  &&   16.09       \\ 
                     &CS2       && 07 54 55.43  && +39 10 40.67  && 16.94  &&   16.60       \\ 
                                                                                
BZGJ0831+5400$^{*}$  &Ref. Star && 08 30 42.11  && +54 00 36.51  && 16.25  &&   15.74       \\  
                     &CS1       && 08 30 57.85  && +53 59 47.50  && 17.21  &&   16.83       \\  
                     &CS2       && 08 30 32.48  && +54 02 04.46  && 16.89  &&   16.43       \\  
                                                                                
BZGJ1154+1225$^{*}$  &Ref. Star && 11 54 29.72  && +12 23 38.46  && 17.15  &&   16.63       \\ 
                     &CS1       && 11 54 05.34  && +12 22 22.09  && 16.89  &&   16.33       \\ 
                     &CS2       && 11 54 32.00  && +12 26 39.78  && 16.54  &&   15.73       \\  
                                                                                
BZQJ1229+0203$^{*}$  &Ref. Star &&  12 29 8.39  && +02 00 18.95  && 13.71  &&   13.97       \\ 
                     &CS1       &&  12 29 3.19  && +02 03 18.15  && 13.62  &&   13.29       \\  
                     &CS2       &&  12 29 2.85  && +02 02 15.99  && 14.89  &&   14.29       \\  
                                                                                
BZBJ1725+1152$^{**}$ &Ref. Star && 17 25 5.75   && +11 49 50.44  && 15.87  &&   15.32       \\  
                     &CS1       && 17 24 56.05  && +11 52 30.84  && 15.82  &&   15.34       \\  
                     &CS2       && 17 25 16.1   && +11 52 30.83  && --     &&   --          \\

BZBJ1728+5013$^{**}$ &Ref. Star && 17 28 14.24  && +50 12 41.17  && 15.34  &&   15.01       \\ 
                     &CS1       && 17 28 14.96  && +50 13 52.60  && 16.16  &&   15.72       \\ 
                     &CS2       && 17 28 16.26  && +50 11 50.20  && 16.31  &&   15.98       \\\hline\\ 
                                                                                
\end{tabular} \\   
{\bf Columns:} (1) Reference and comparison stars; (2) right ascension; (3) declination; (4) \& (5) apparent V and R band magnitudes, respectively extracted from $^{*}$Sloan Digital Sky Survey (SDSS), $^{**}$Pan-STARRS1 and taken from $^{***}$Frankfurt Quasar Monitoring\footnote{http://quasar.square7.ch/fqm/0506+056.html}. For transformation of SDSS and Pan-STARRS1 to Johnson-Cousins UBVRI system magnitude, we used formula given by \citet{2006A&A...460..339J} and \citet{2018BlgAJ..28....3K}, respectively. {\bf Note:} -- = not available 
\vspace{0.5cm}  
\end{table*}

\section{{\bf Observations and Data Reductions}}
\noindent
We observed a sample of ten blazars selected from the Roma--BZCAT catalogue based on their visibility from the observing site and apparent brightness. Our observing sites have a visibility window of declination $- 10^{\circ}$ to $+ 70^{\circ}$ and we choose R-band magnitude to be $<$ 16 mag for our sample to obtain better accuracy on measurements. The observations were carried out with two telescopes; the 1.04 m Sampurnanad Telescope (ST) and the 1.3 m Devasthal Fast Optical Telescope (DFOT), located at Aryabhatta Research Institute of observational sciencES (ARIES), Nainital, India between October 2014 and October 2015 for a total of 26 nights. Both of these telescopes are equipped with CCD detectors and standard Johnson UBV and Cousins RI filters system. The details of the telescopes parameters are given in Table \ref{tab1}.\\

The sample of blazars used in this study along with their basic information are presented in 
Table \ref{tab2}. Photometric images of the blazar fields were collected in the V and R pass bands in an alternative sequence, with exposure times ranging from 80 to 300 seconds, depending on the target's brightness and visibility of the sky. Throughout the observing run, the typical seeing was $\sim$ 2.0 arcsec (varied between 1.5 to 2.5 arcsec), and the air mass values were between 1.00 and 2.2. Several bias frames were taken at regular intervals covering the whole night and twilight sky flats were taken in the pass bands in each observing night. The observation log is given in Table \ref{tab3}. As mentioned in the previous section, most of these sources have not been studied at any frequency yet, thus they do not have pre-defined standard field stars. This leads us to opt for the differential photometry method for data reduction.

\subsection{{\bf Differential photometry}}
The raw CCD images were reduced following the standard routines in IRAF\footnote{IRAF is distributed by the National Optical Astronomy Observatories, which are operated by the Association of Universities for Research in Astronomy, Inc., under cooperative agreement with the National Science Foundation.} (Image Reduction and Analysis Facility) software. The Pre-processing of the CCD images were done by bias subtraction, flat-fielding and cosmic rays removal. In order to do so, first, we generated a master bias frame by taking median of all the good quality bias frames taken in that particular observing night. Later this master bias was subtracted from all twilight flat frames and the image frames. In the next step, a master flat in each filter was generated by median combine of all the flat frames in a single pass band and then normalized in each band separately. These normalized master flats for each passband was used to remove pixel to pixel inhomogeneities in the source frames. All the image frames containing source were further cleaned by removing cosmic ray contamination.\\

Aperture photometry was carried out to obtain the instrumental magnitude of the blazar and comparison stars in the source fields using DAOPHOT II (Dominian Astronomical Observatory Photometry) software \citep{1987PASP...99..191S}. In the process, we used four different concentric apertures radii as multiples of the Full Width at Half Maximum (FWHM) of the stars in the image frames, i.e., $\approx$ 1$\times$ FWHM, 2$\times$ FWHM, 3$\times$ FWHM and 4$\times$ FWHM. For sky background subtraction, we used a source free annulus region of radii approximately 5 times of the source FWHM. The data reduced with different aperture radii were found to be in good agreement. However, it was noticed that the best signal to noise ratio (S/N) was obtained with an aperture radius of $\approx$ 2 $\times$ of the typical FWHM, which was thus adopted in our photometric measurements. In the field of each source, first we selected 5 comparison stars and among them choose only 3 which showed least variation during the observing run. Finally, we used the one which has brightness closest to the blazar as the reference star to construct differential instrumental magnitude light curves (DLCs: Blazar - comparison star). Positions and magnitudes of the reference and comparison stars used to generate the DLCs of the blazars are given in Table \ref{tab4}.\\

\begin{figure*}[hbt!]  
\centering
\hspace*{0.5cm}
\mbox{\subfloat{\includegraphics[scale=0.4]{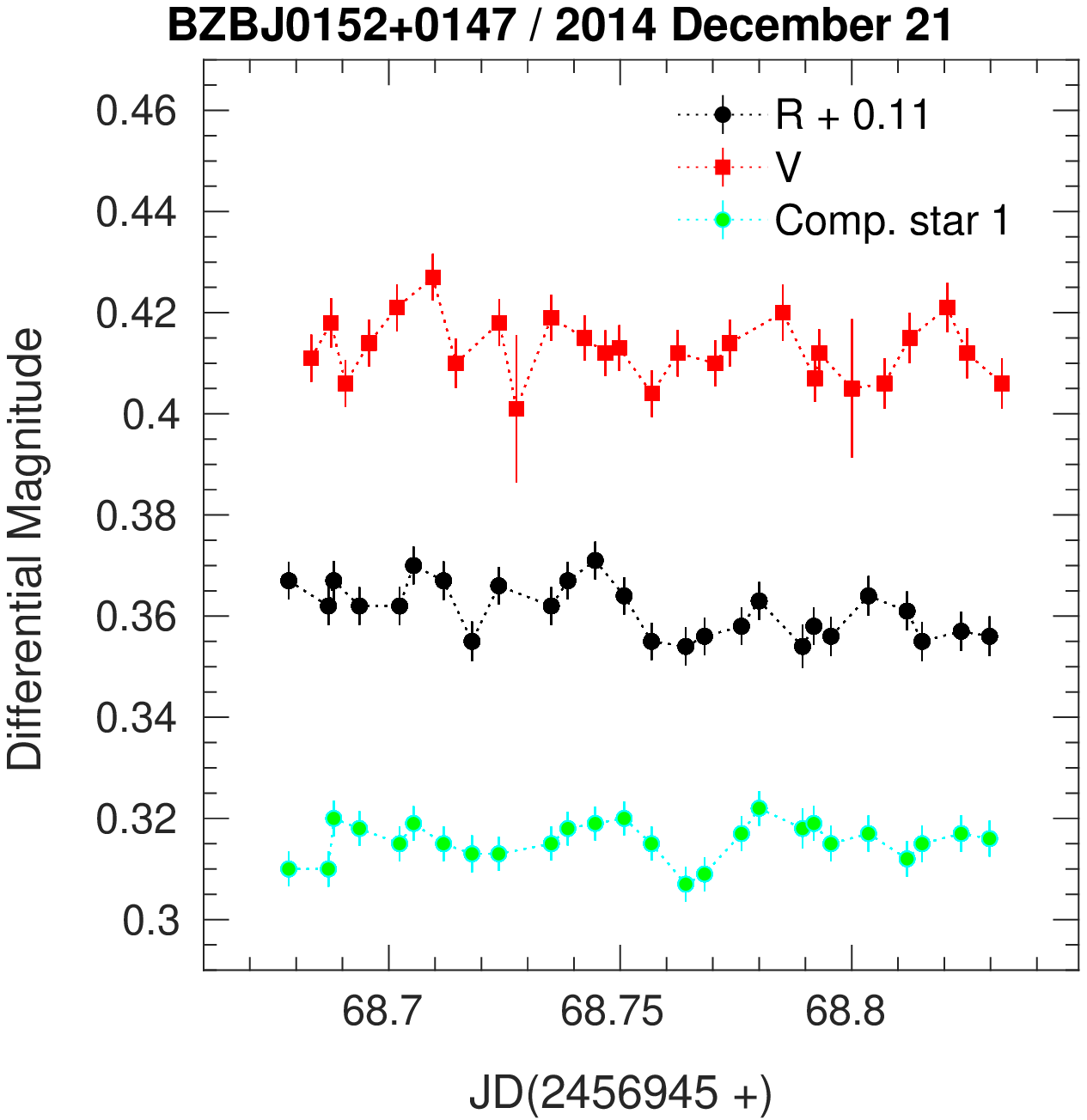} }\quad
\subfloat{\includegraphics[scale=0.4]{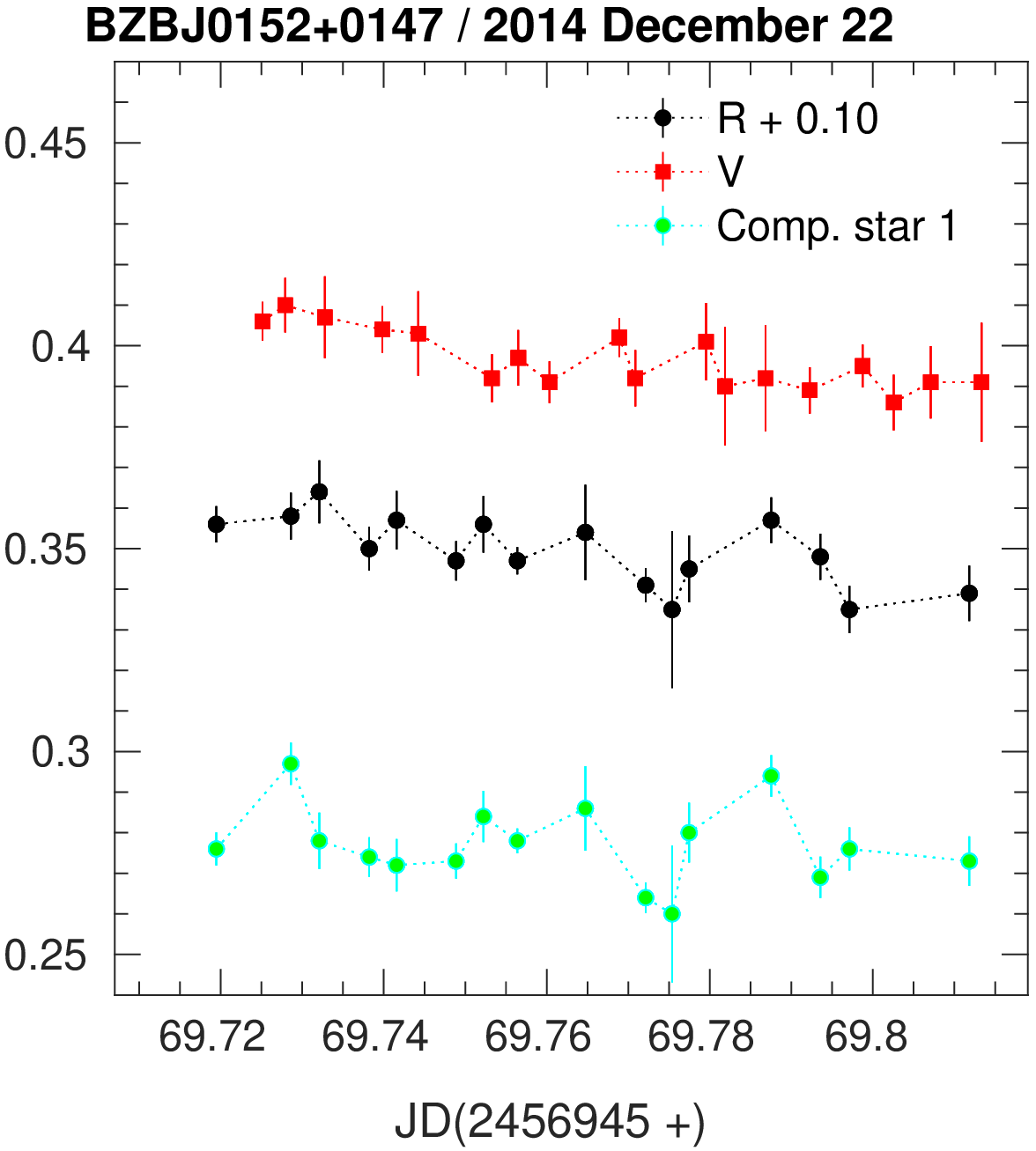} }\quad
\subfloat{\includegraphics[scale=0.4]{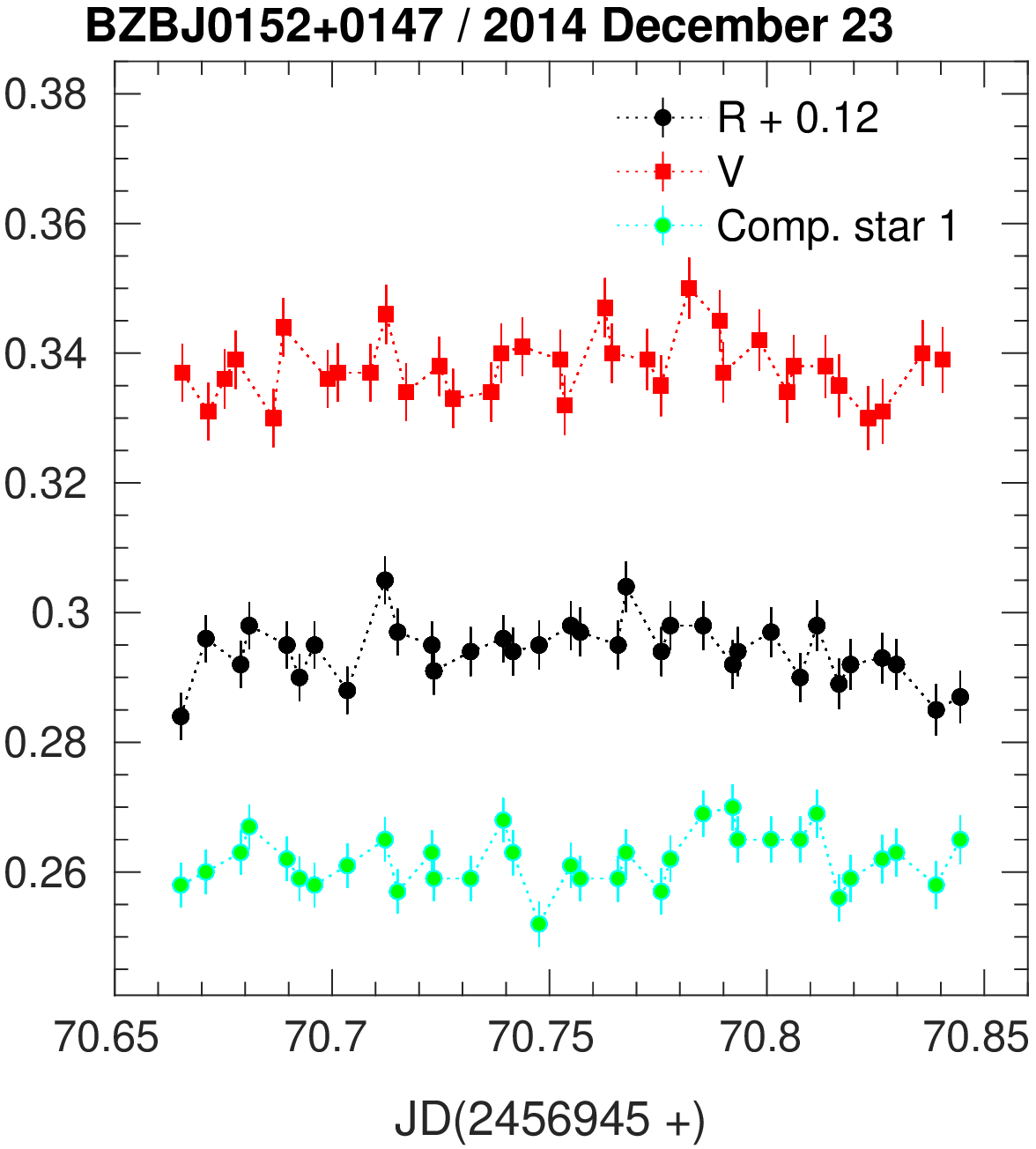} }}
\newline
\hspace*{0.5cm}
\mbox{\subfloat{\includegraphics[scale=0.4]{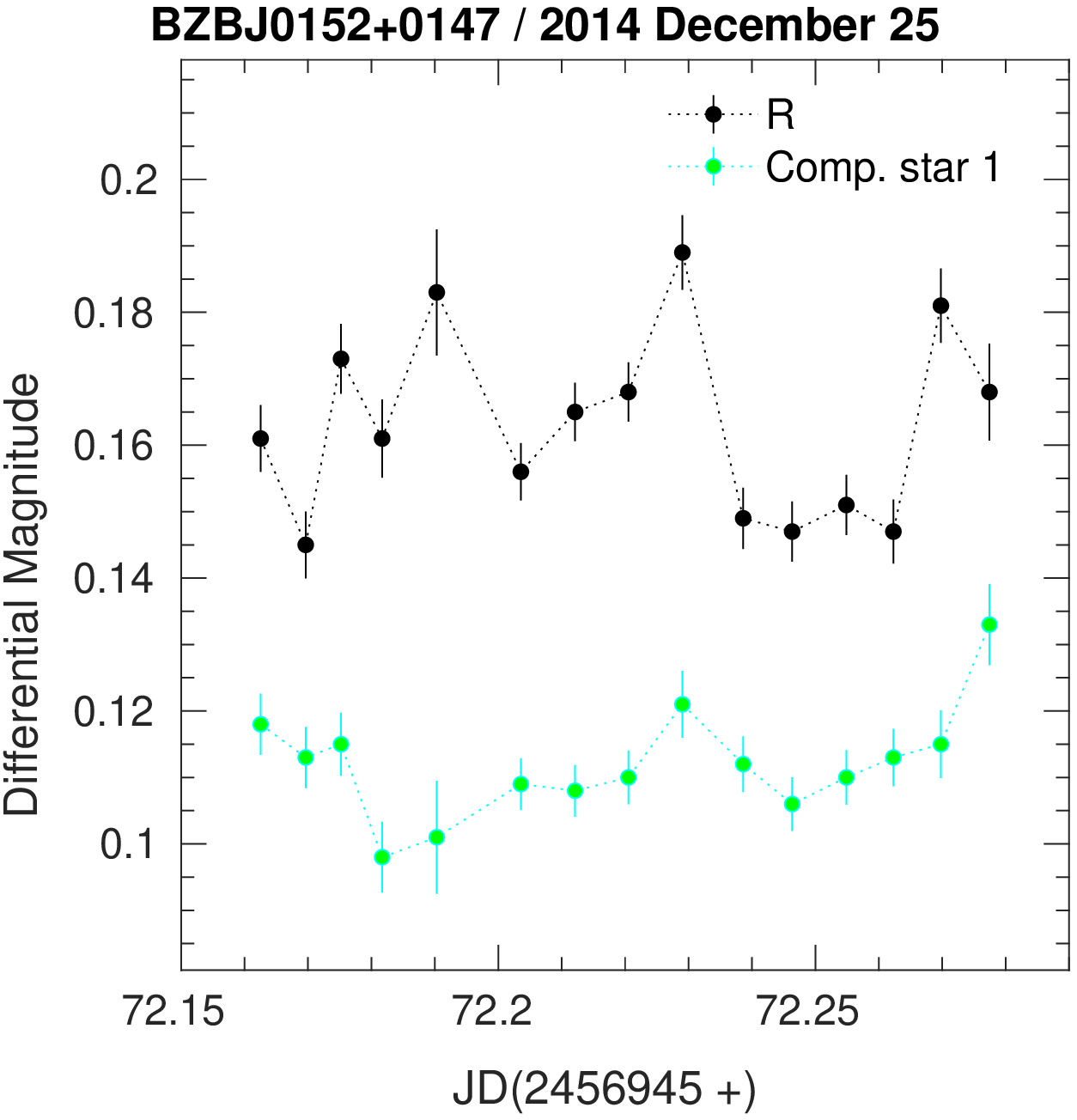} }\quad
\subfloat{\includegraphics[scale=0.4]{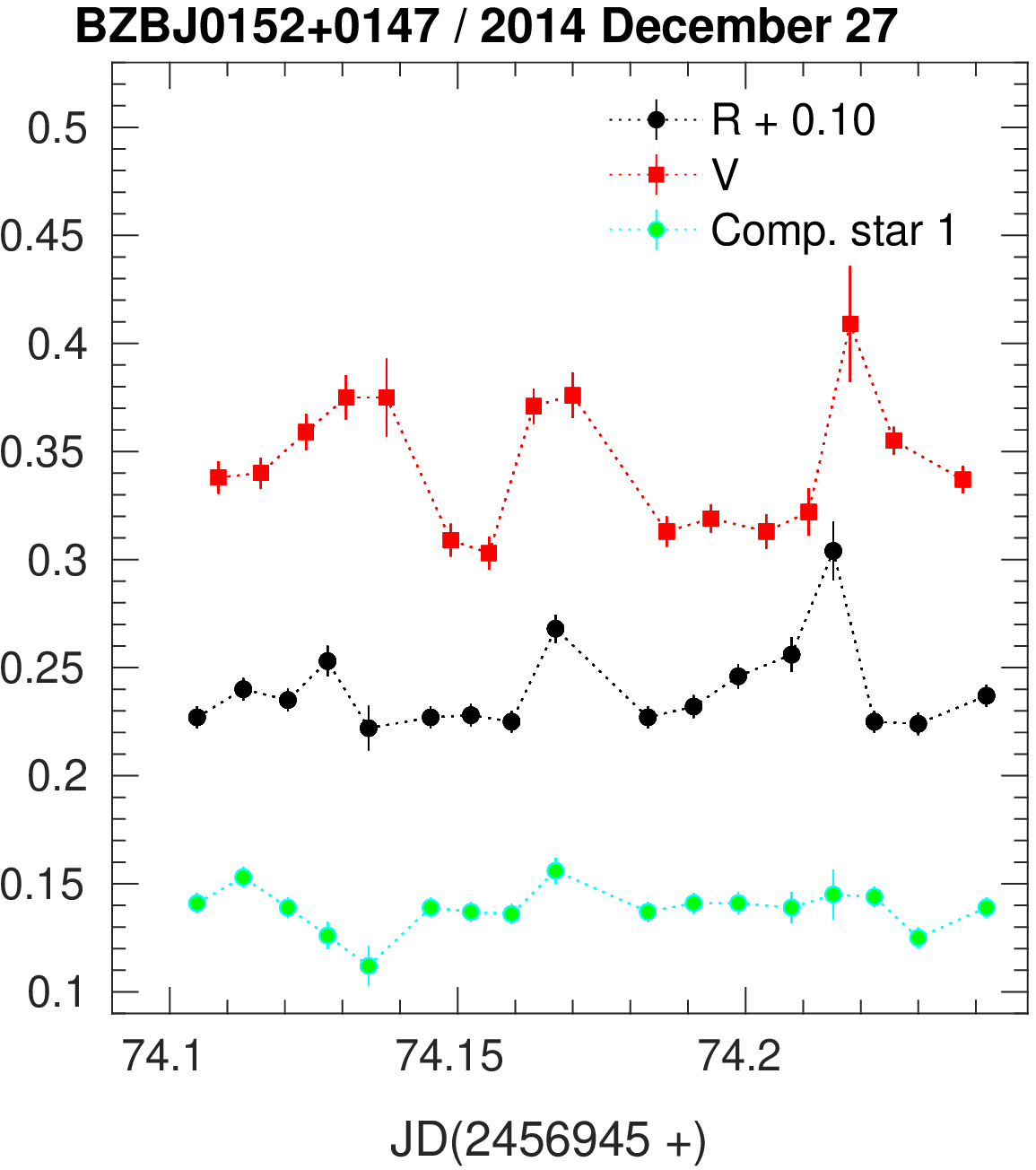} }\quad
\subfloat{\includegraphics[scale=0.4]{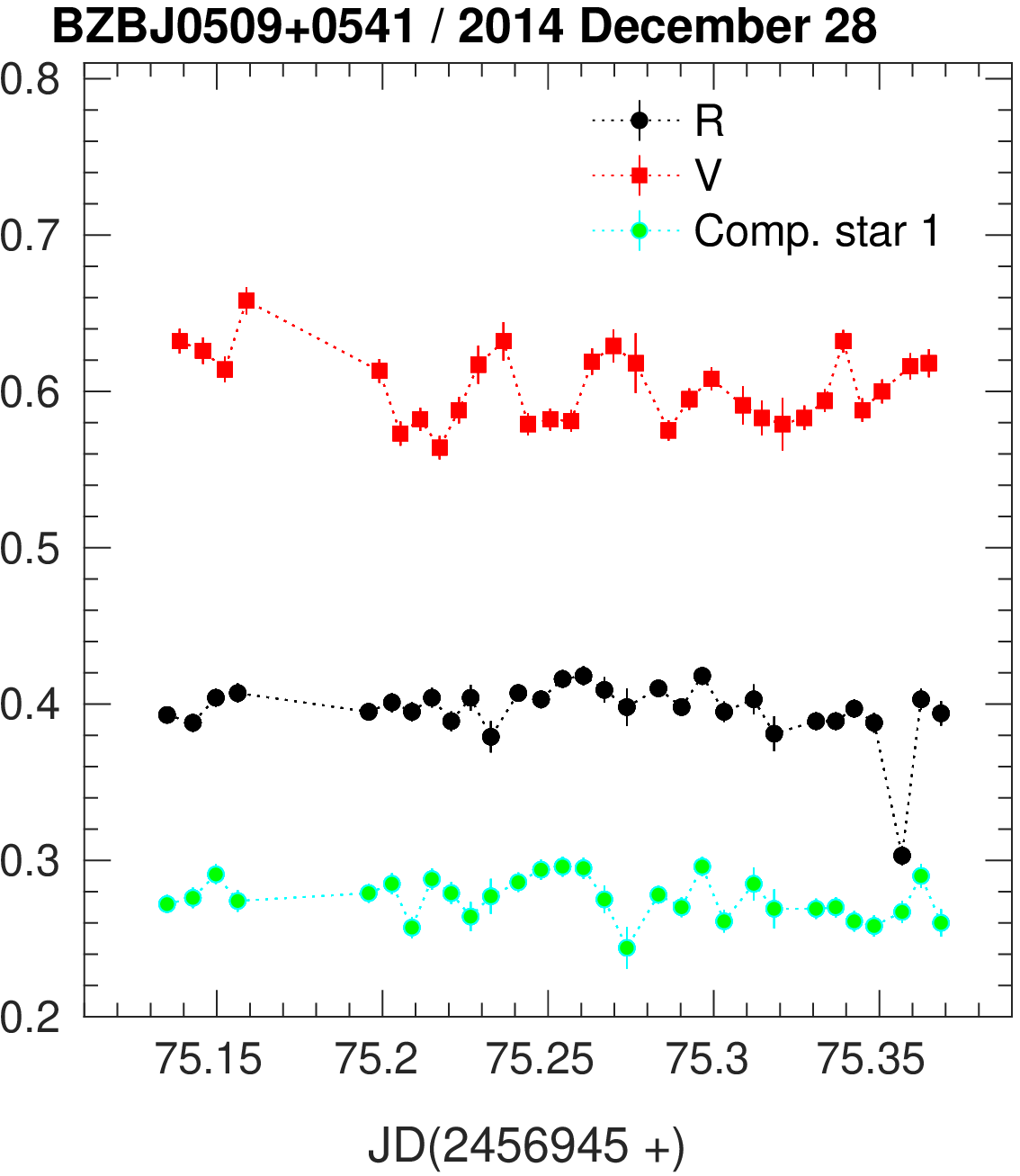} }}
\newline
\hspace*{0.5cm}
\mbox{\subfloat{\includegraphics[scale=0.4]{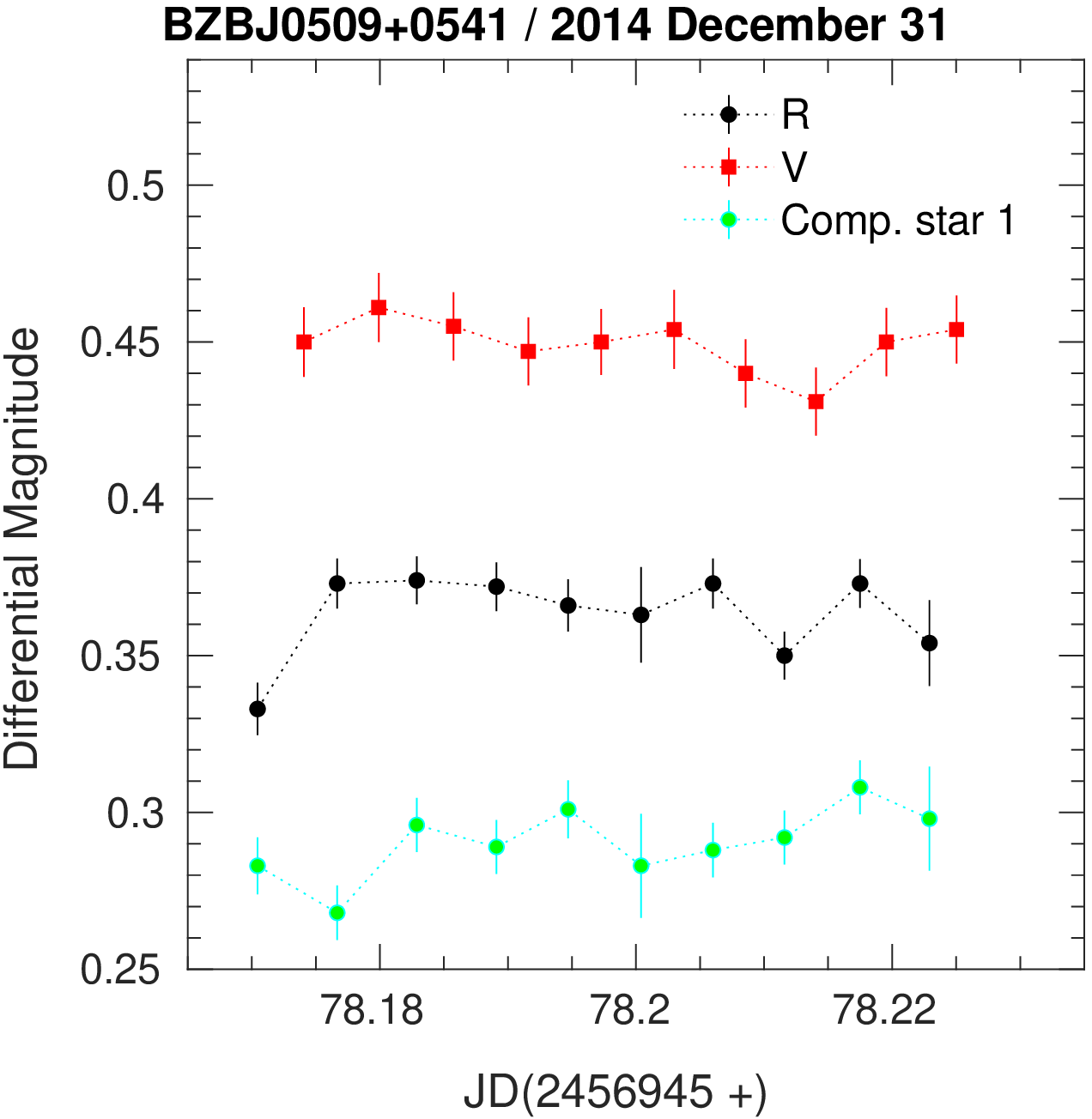} }\quad
\subfloat{\includegraphics[scale=0.4]{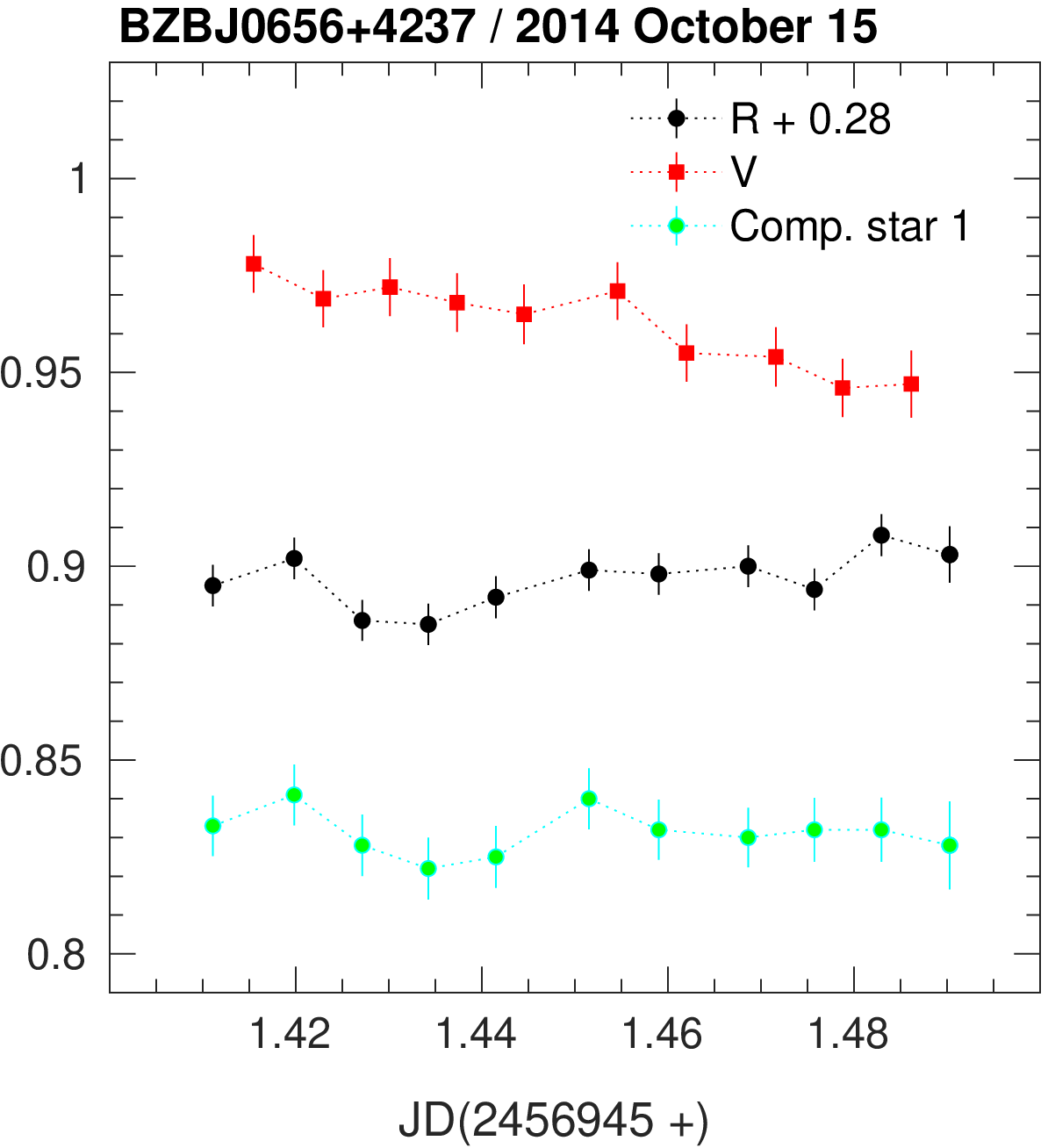} }\quad
\subfloat{\includegraphics[scale=0.4]{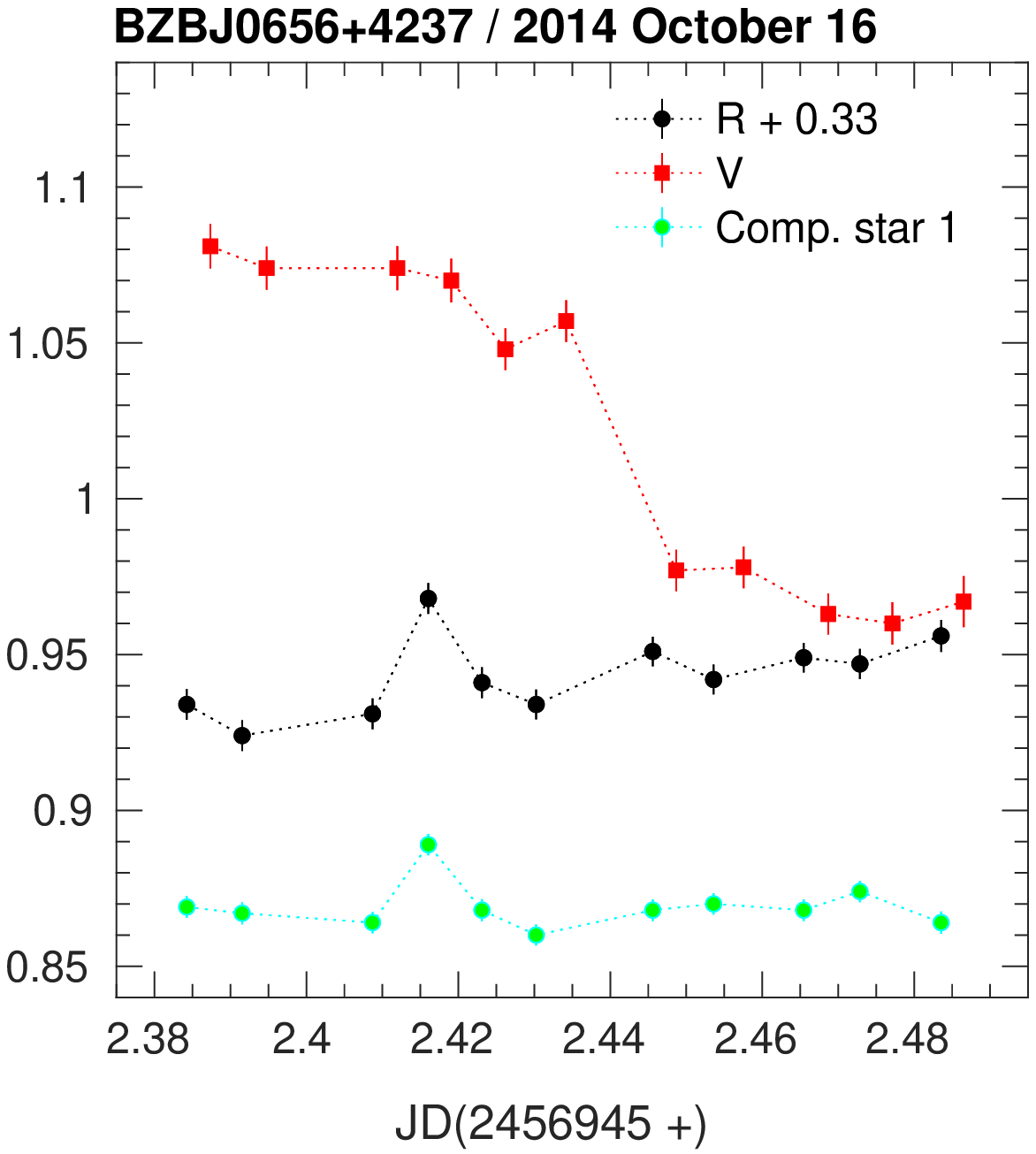} }}
\newline
\hspace*{0.0cm}
\mbox{\subfloat{\includegraphics[scale=0.4]{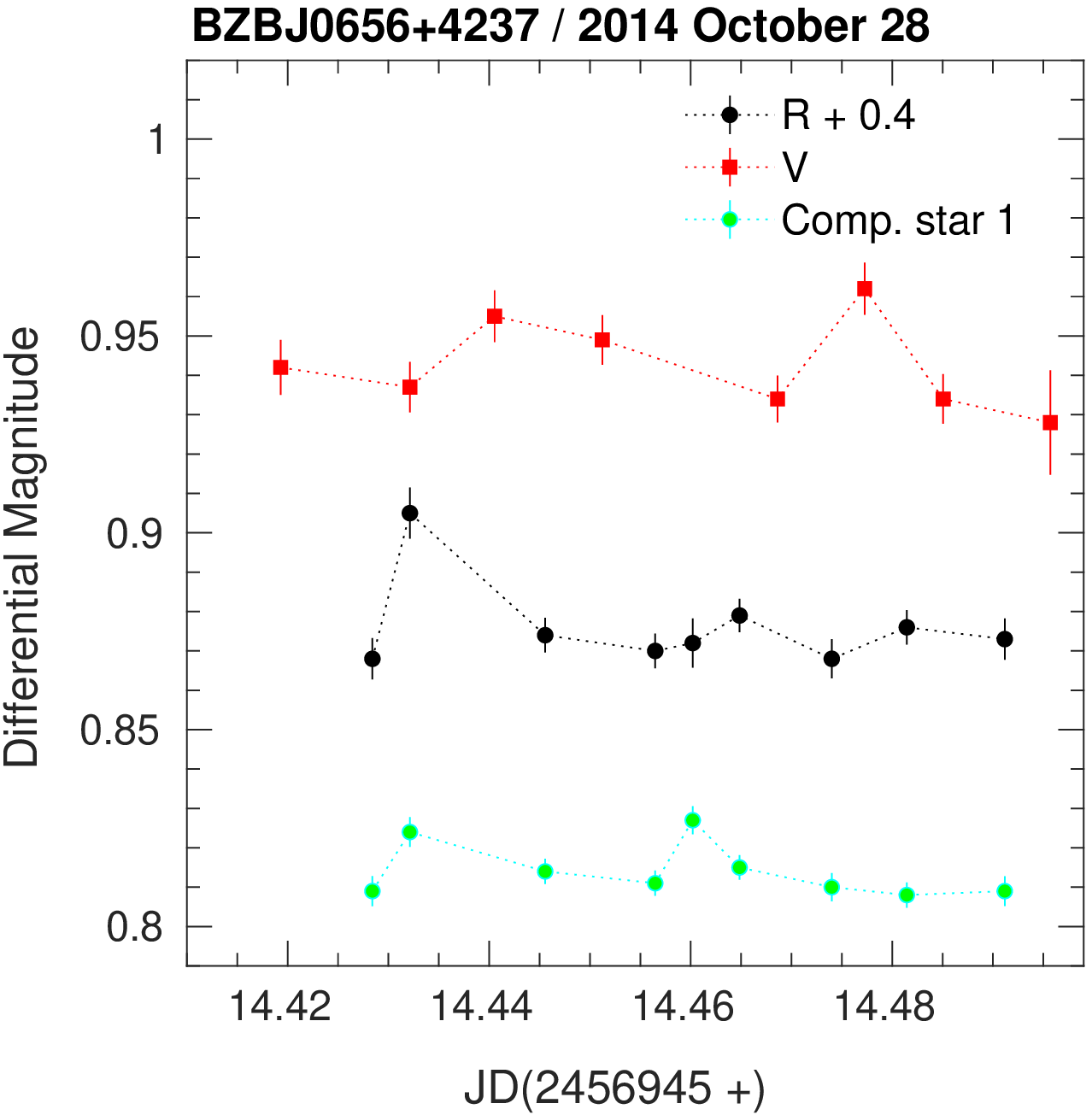} }\quad
\subfloat{\includegraphics[scale=0.4]{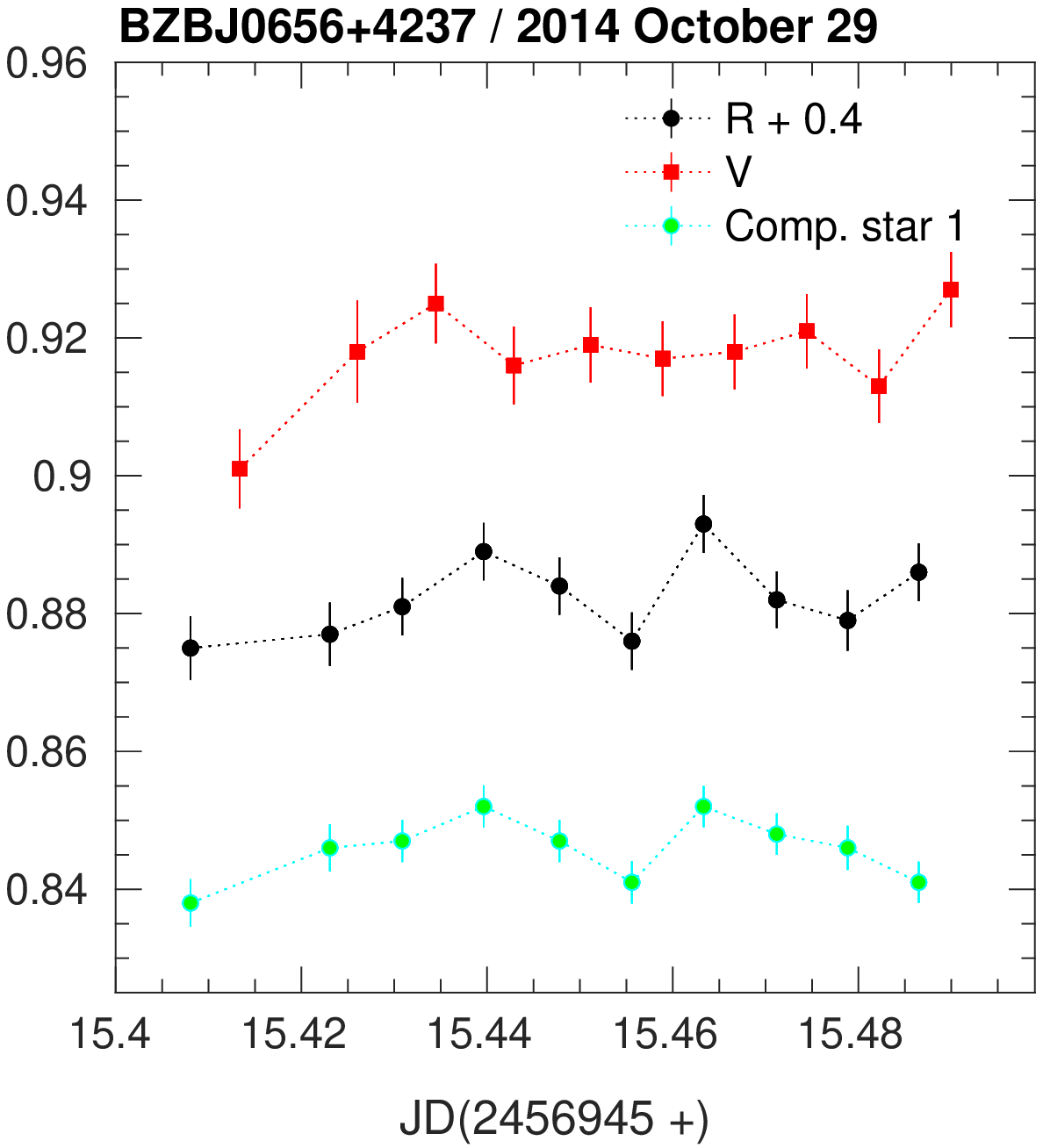}}\quad
\subfloat{\includegraphics[scale=0.4]{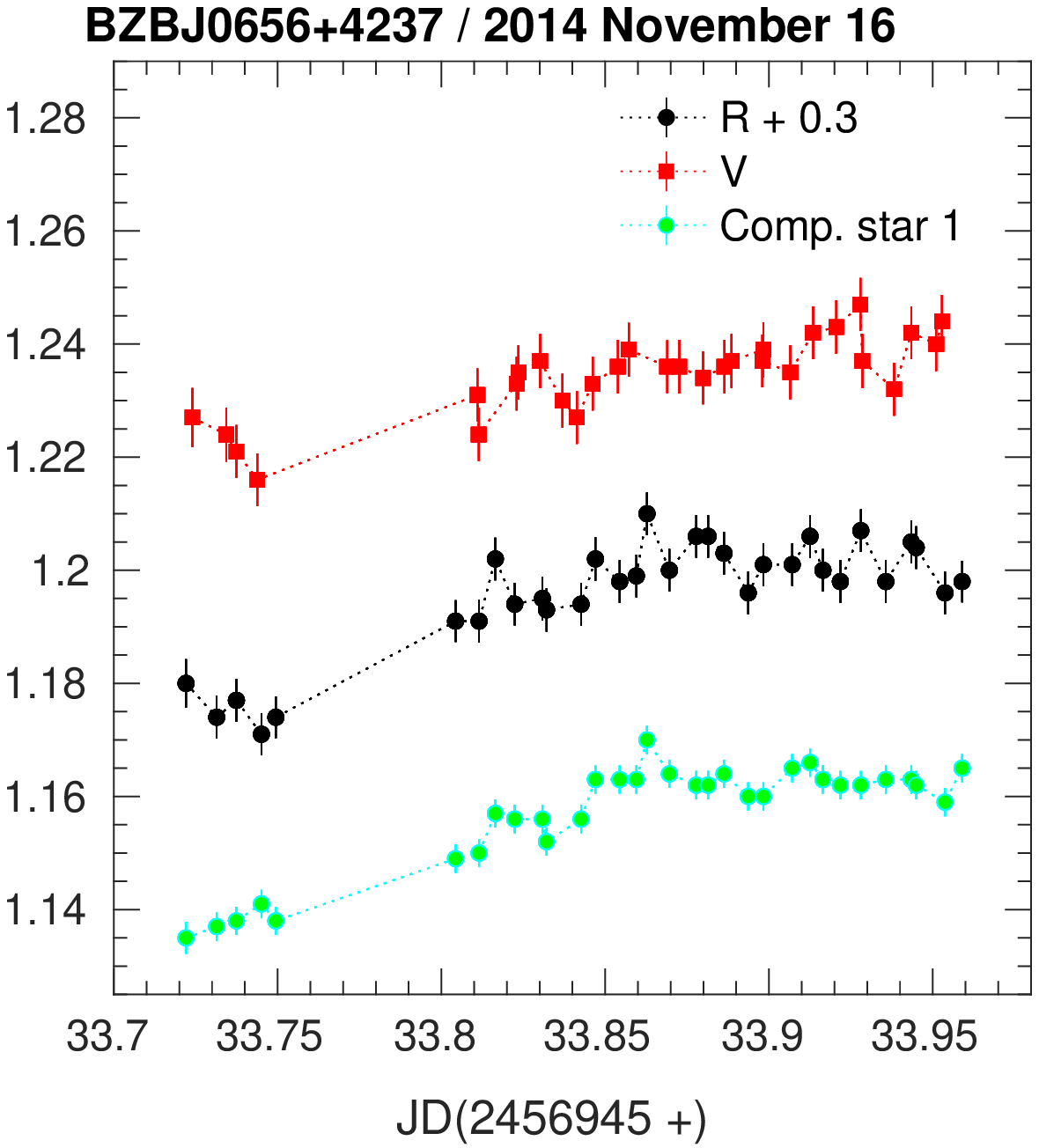} }}
\caption{Differential optical magnitude light curves in V and R filters of the blazar sample during the period October 2014 -- June 2015. DLCs of the comparison star 1 (CS1-reference star) is shown at the bottom of individual plots. The DLCs are evenly sampled using a bin size of 80--300 s. In the plots, either V or R filters are shifted  whenever required, by a constant value for better visibility, which are displayed in the upper right corner of the respective plots along with source  names and their observing dates. }
\end{figure*}

\begin{figure*}[hbt!]  
\centering
\ContinuedFloat
\hspace*{0.7cm}
\mbox{\subfloat{\includegraphics[scale=0.4]{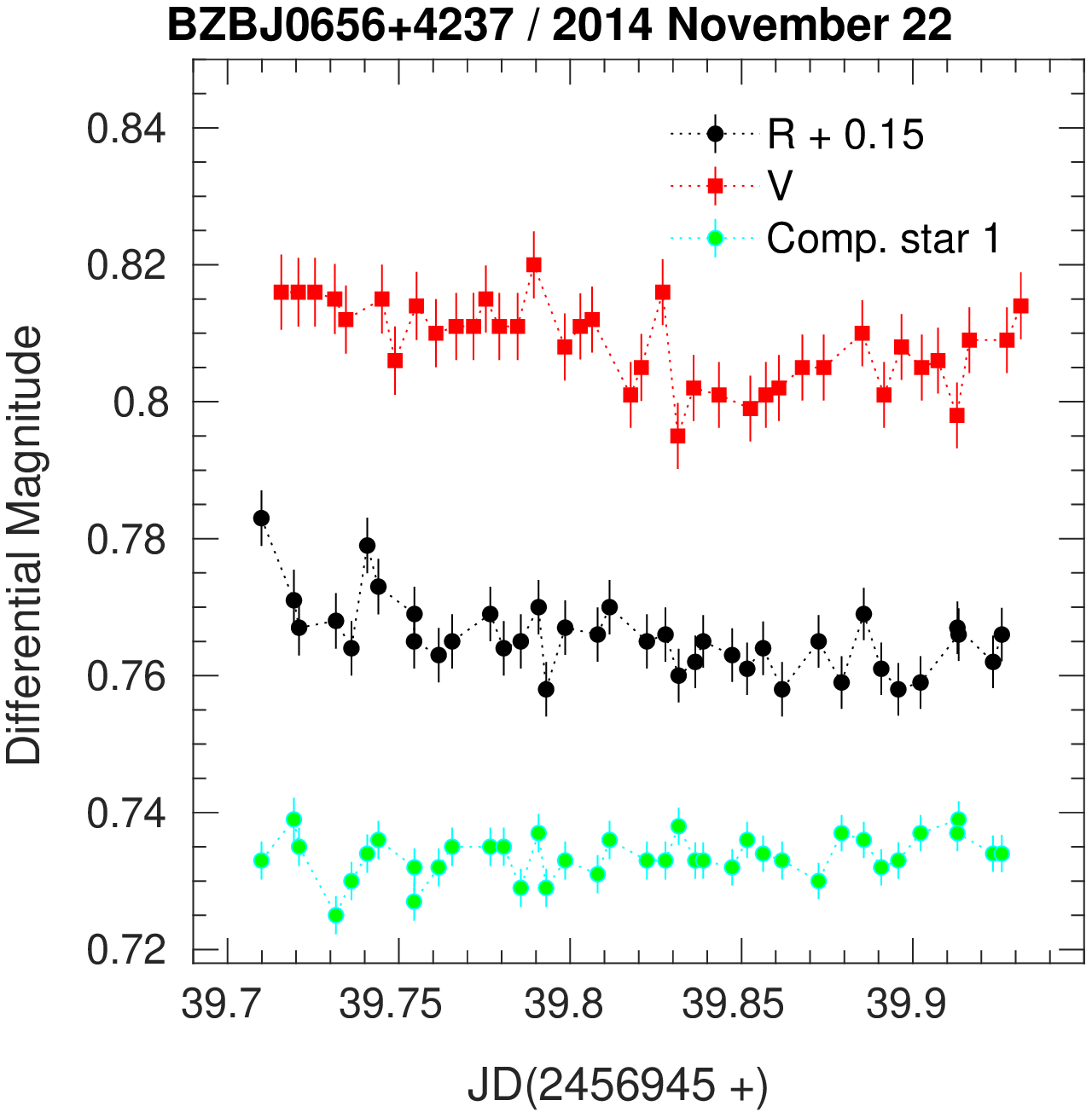}}\quad
\subfloat{\includegraphics[scale=0.4]{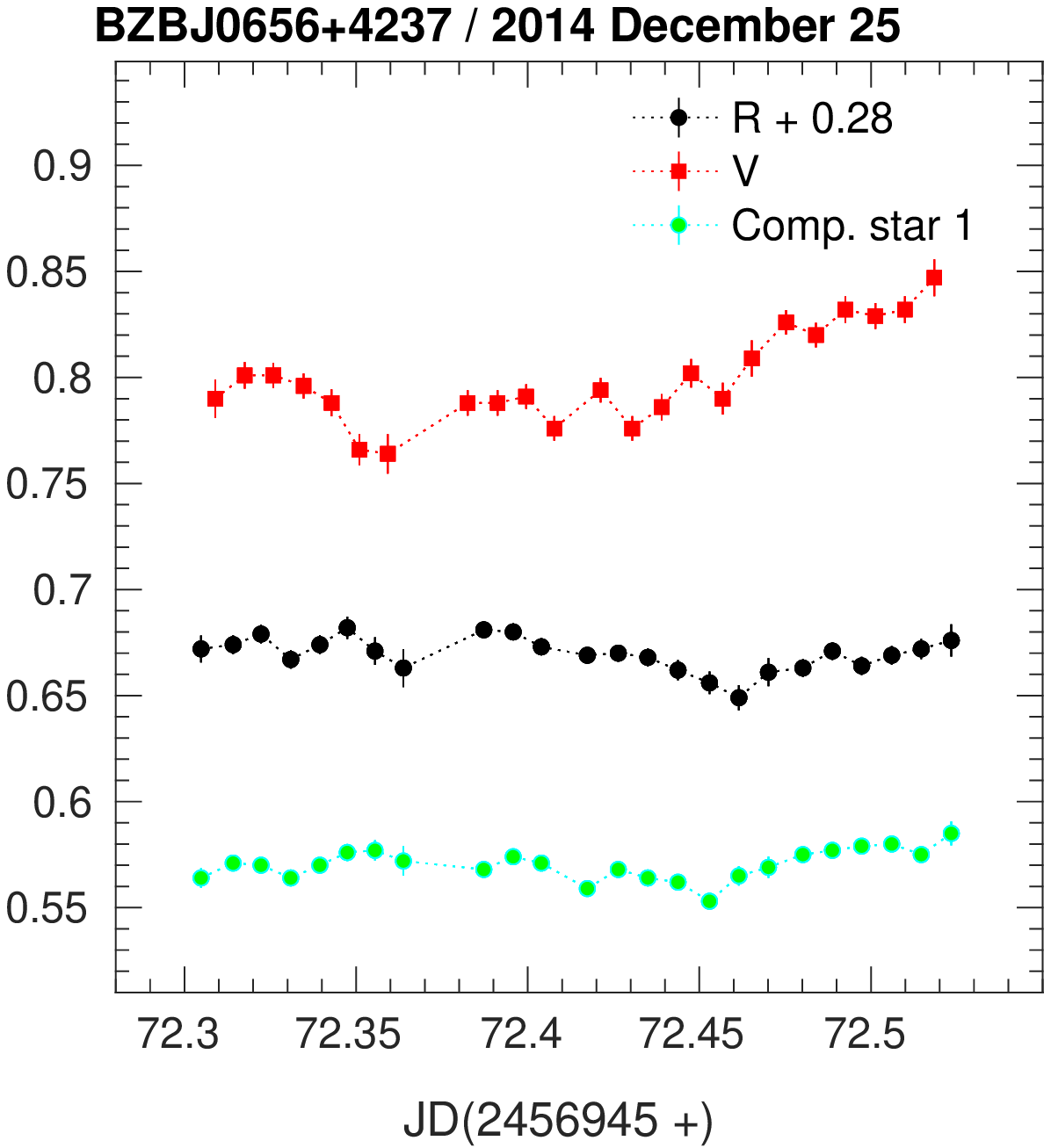}}\quad
\subfloat{\includegraphics[scale=0.4]{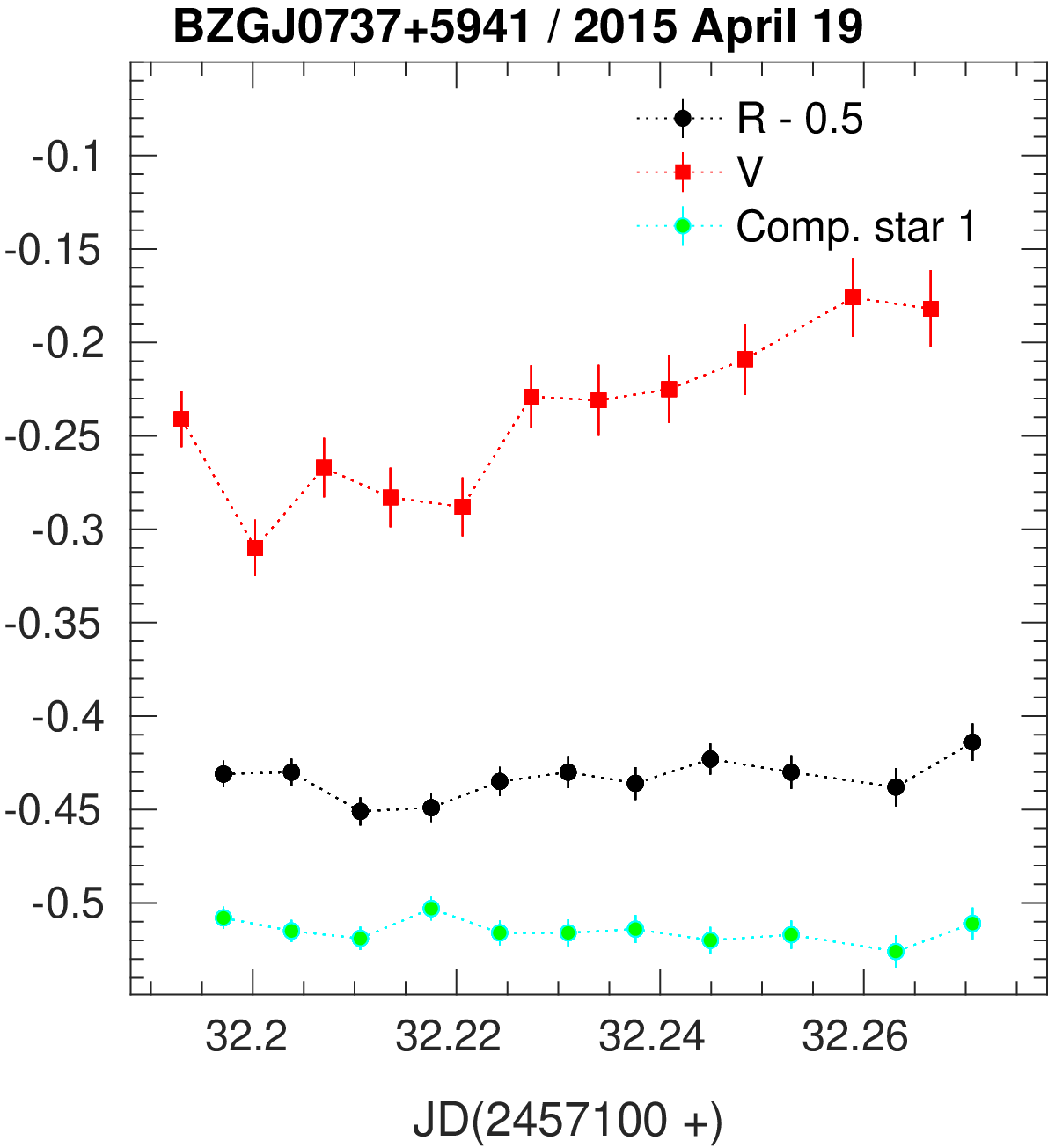} }}
\newline
\hspace*{0.8cm}
\mbox{\subfloat{\includegraphics[scale=0.4]{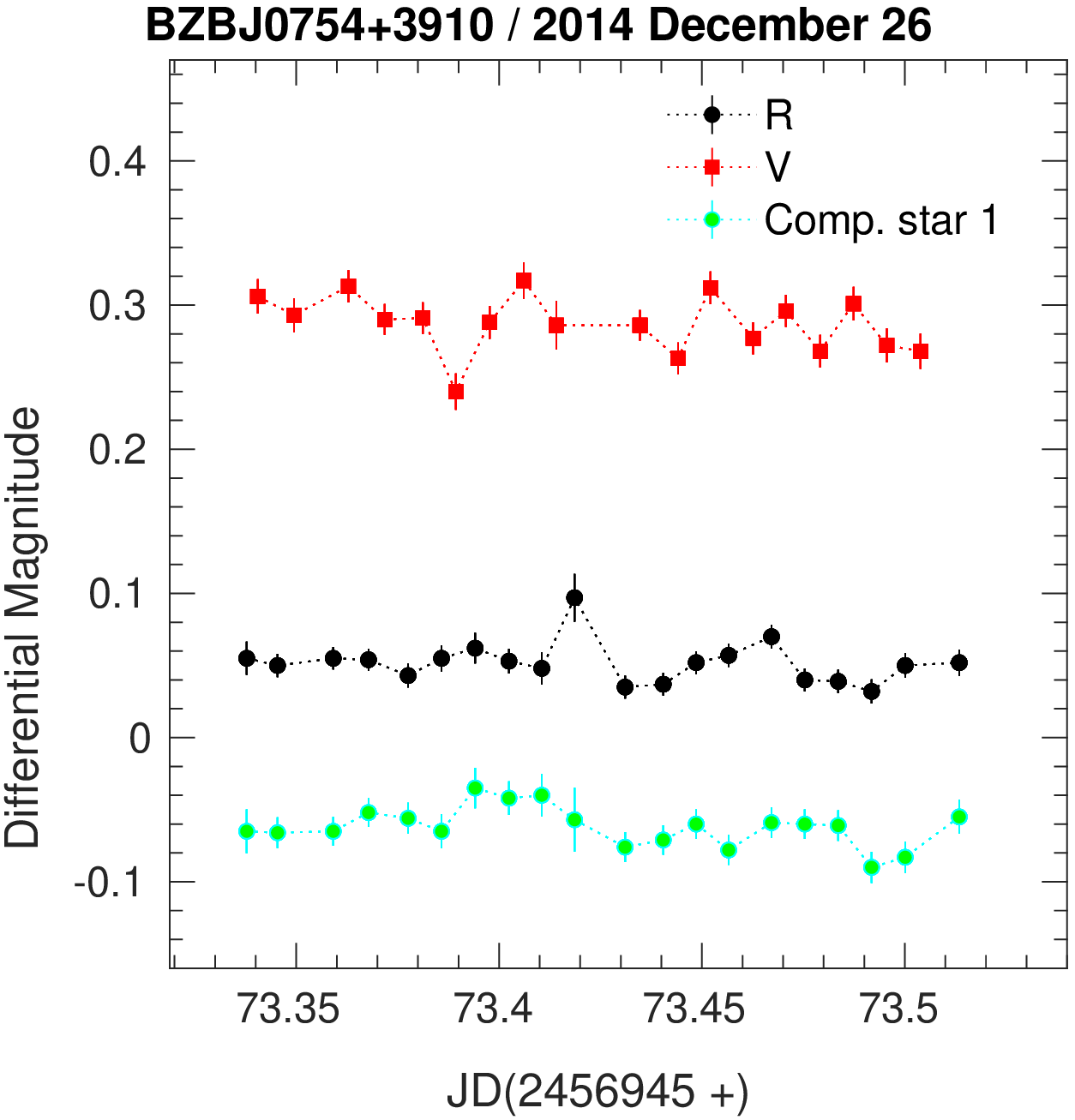} }\quad
\subfloat{\includegraphics[scale=0.4]{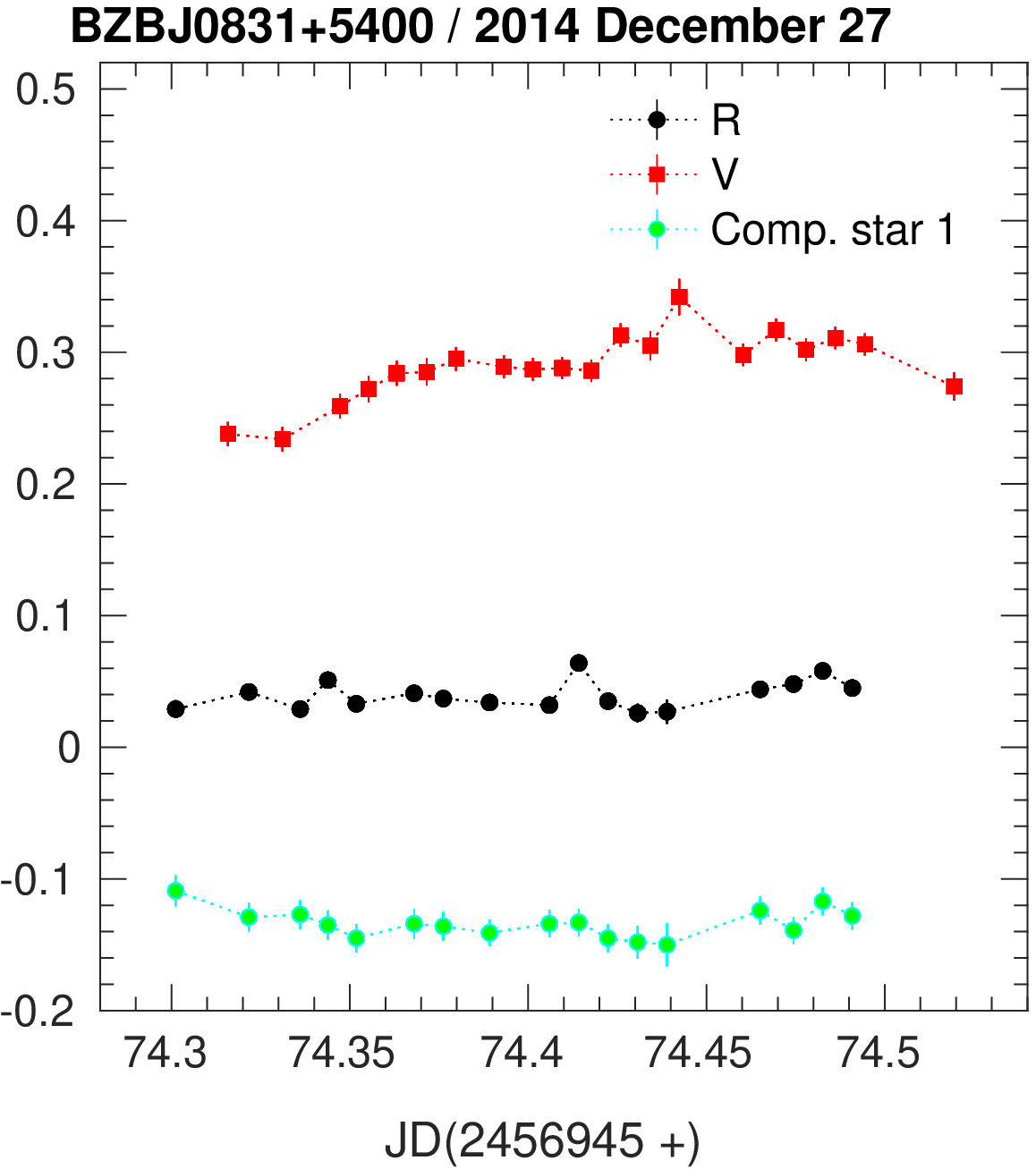} }\quad
\subfloat{\includegraphics[scale=0.4]{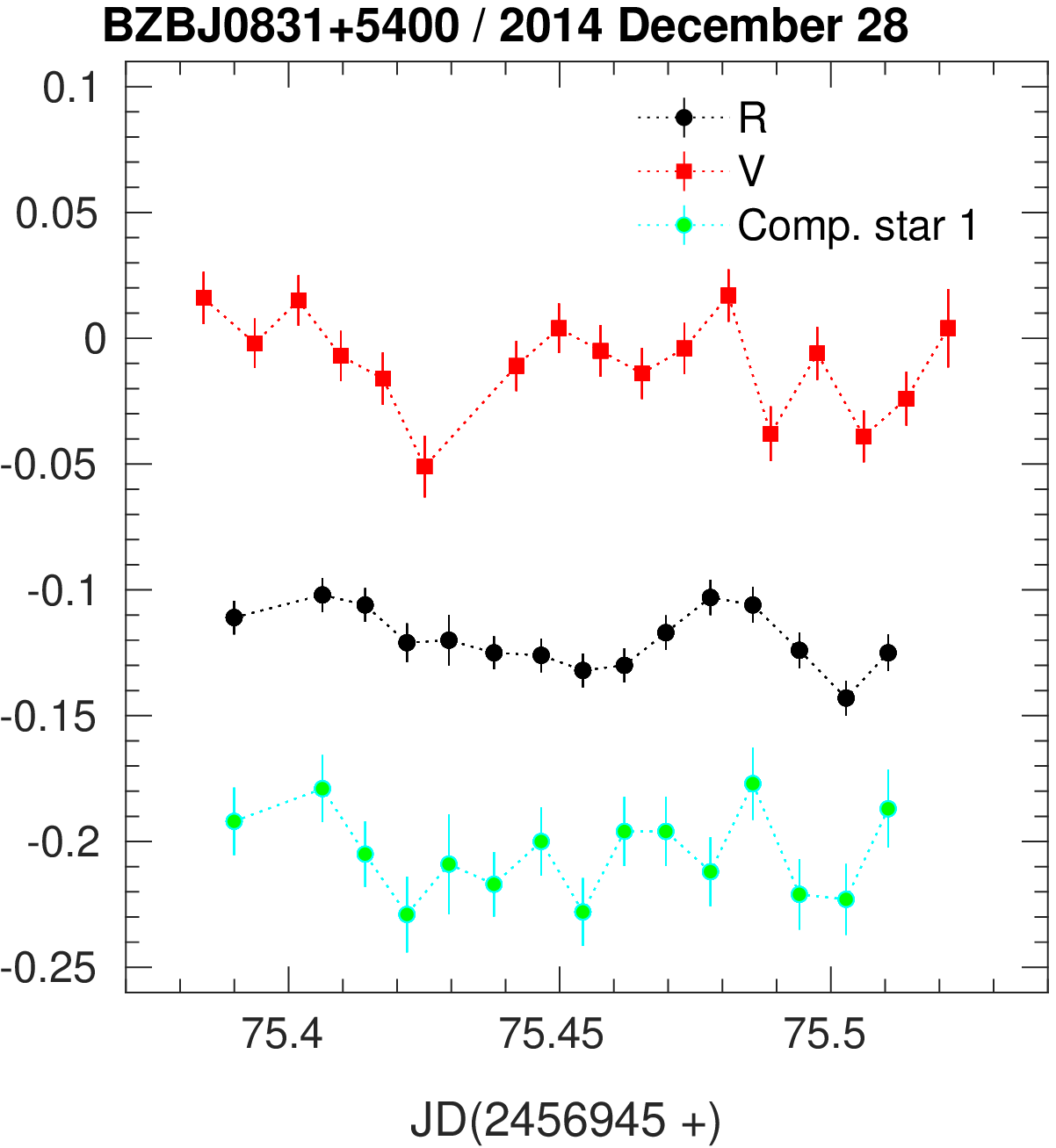} }}
\newline
\hspace*{0.6cm}
\mbox{\subfloat{\includegraphics[scale=0.4]{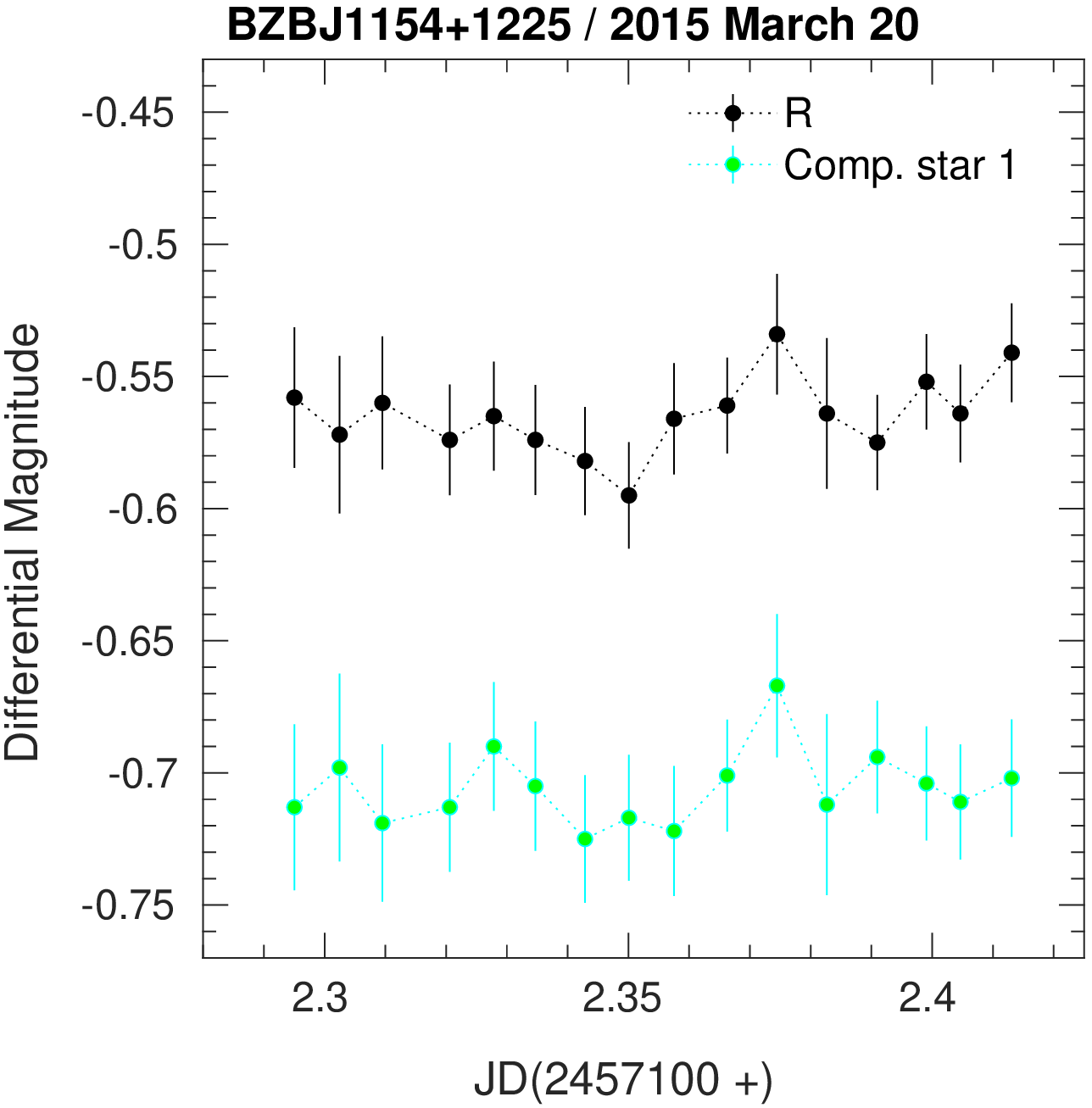} }\quad
\hspace*{0.1cm}
\subfloat{\includegraphics[scale=0.4]{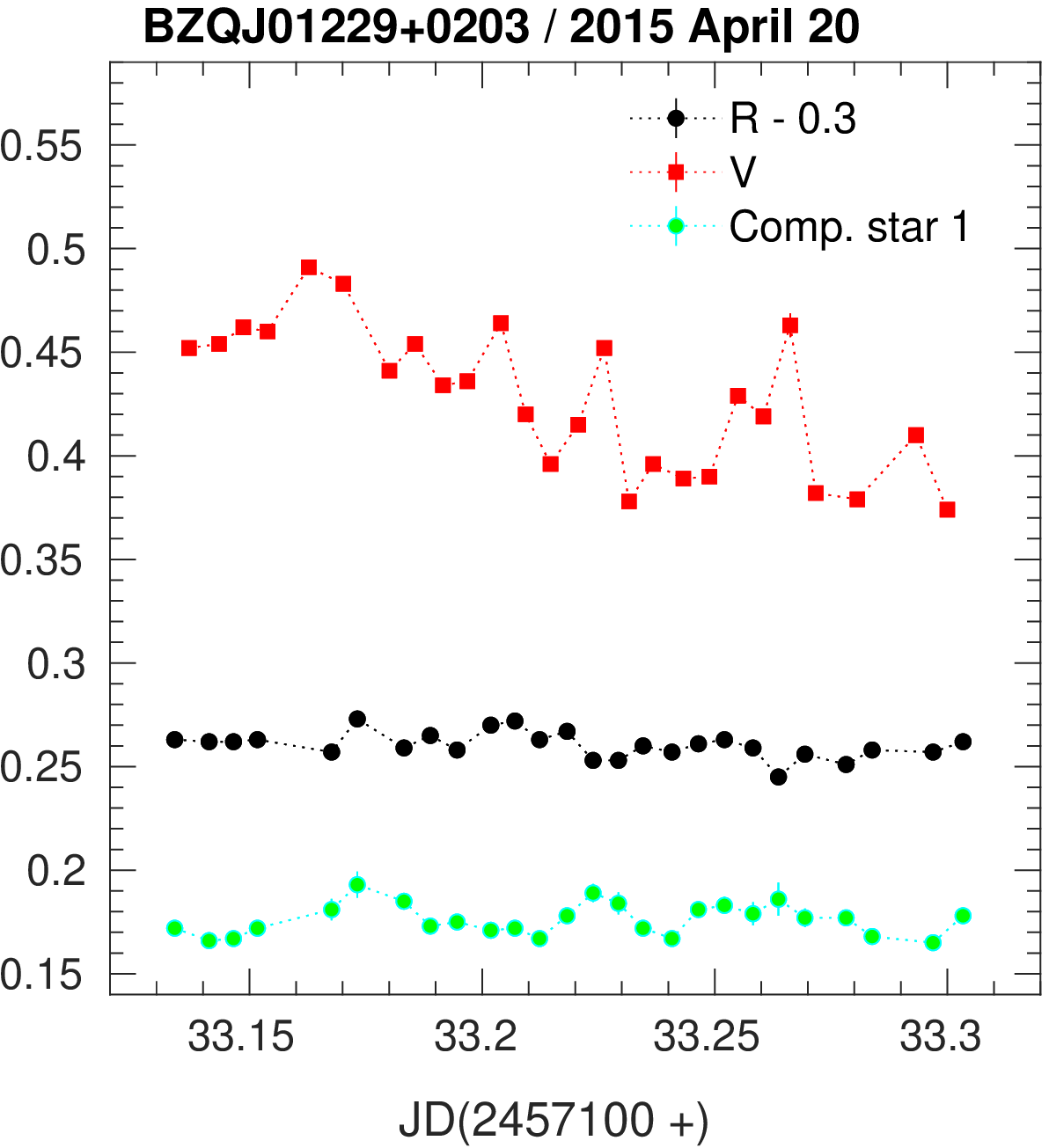}}\quad
\hspace*{0.1cm}
\subfloat{\includegraphics[scale=0.4]{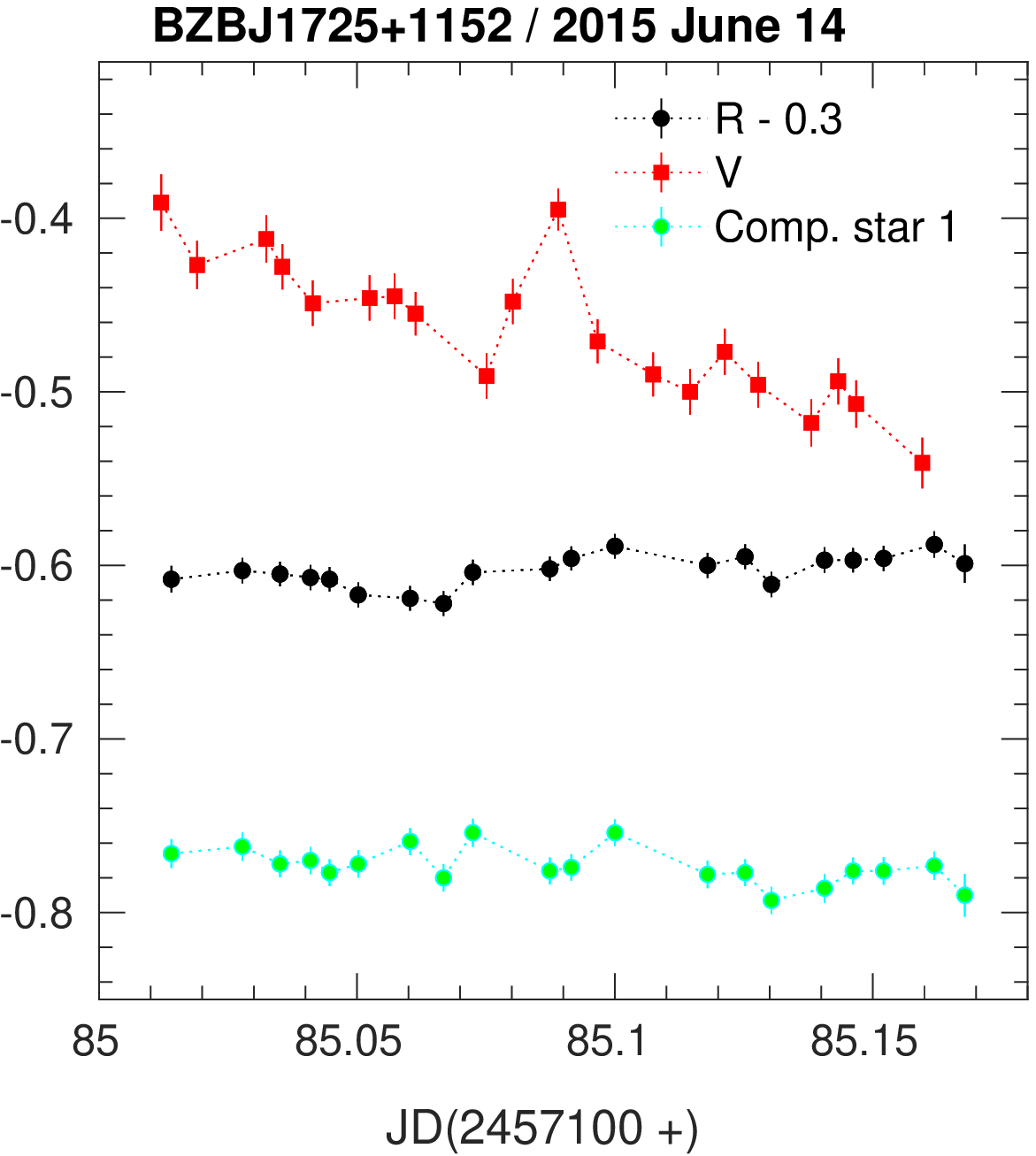} }}
\newline
\hspace*{0.2cm}
\mbox{\subfloat{\includegraphics[scale=0.4]{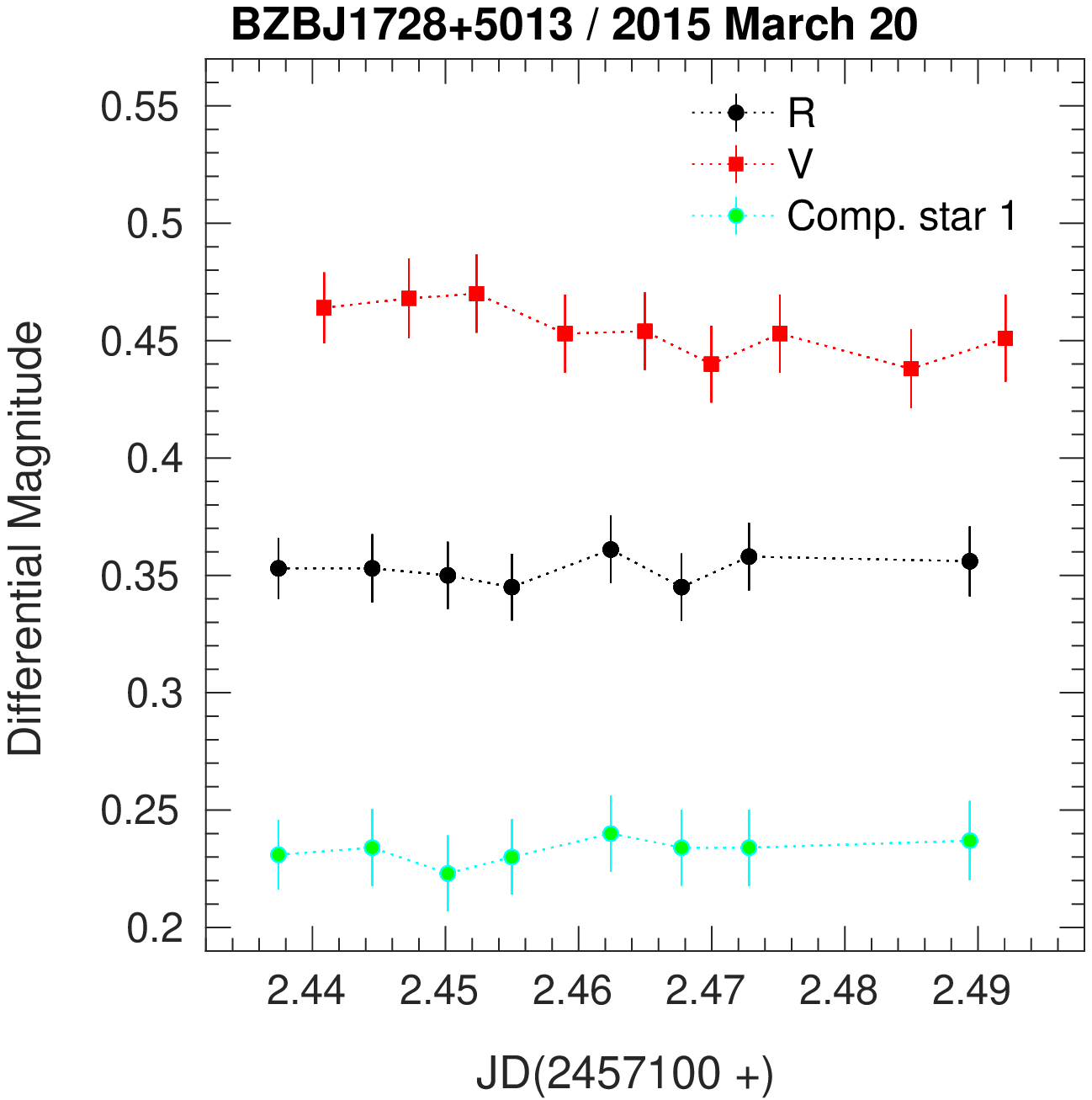}}\quad
\hspace*{0.3cm}
\subfloat{\includegraphics[scale=0.4]{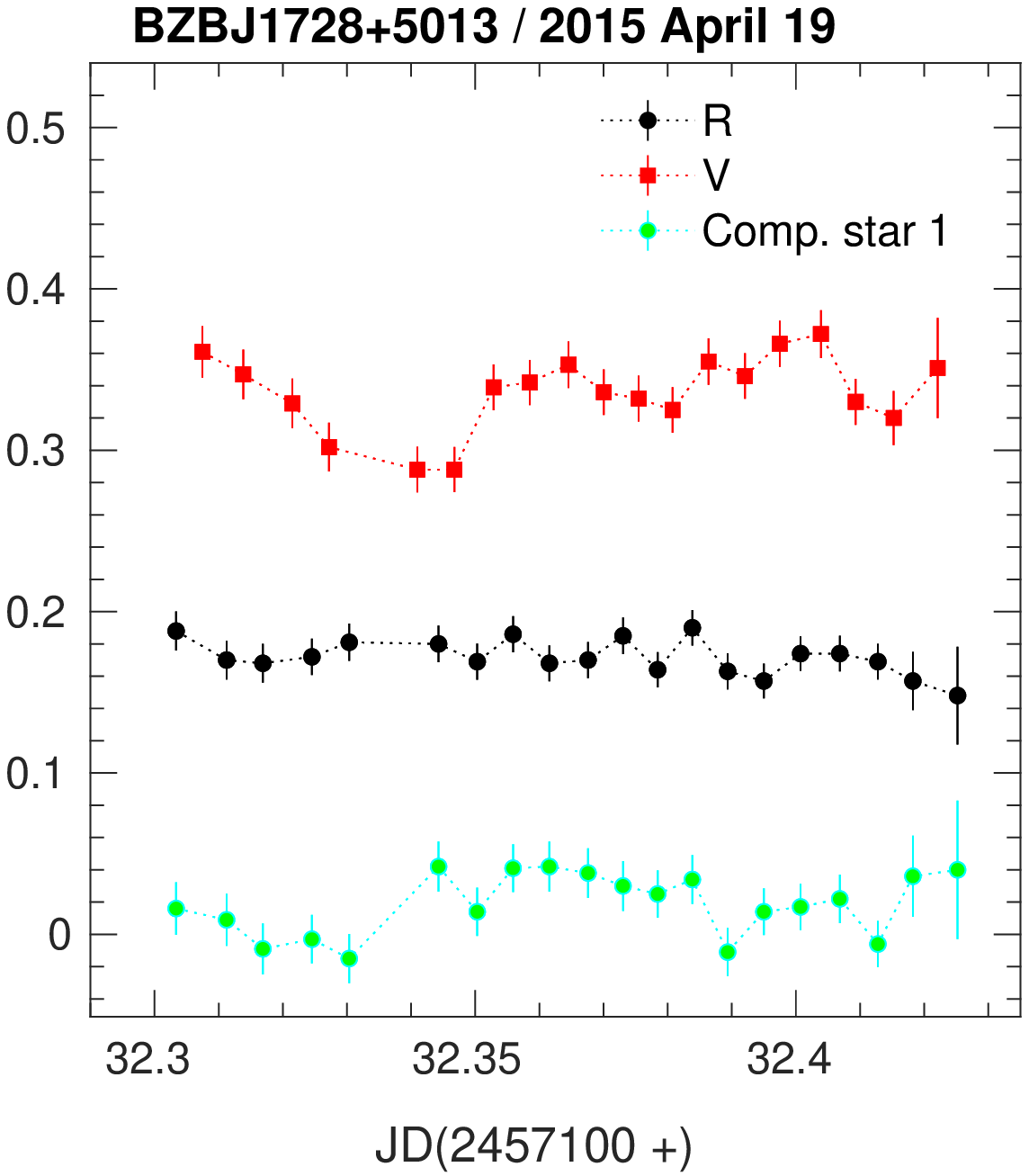} }\quad
\hspace*{0.2cm}
\subfloat{\includegraphics[scale=0.4]{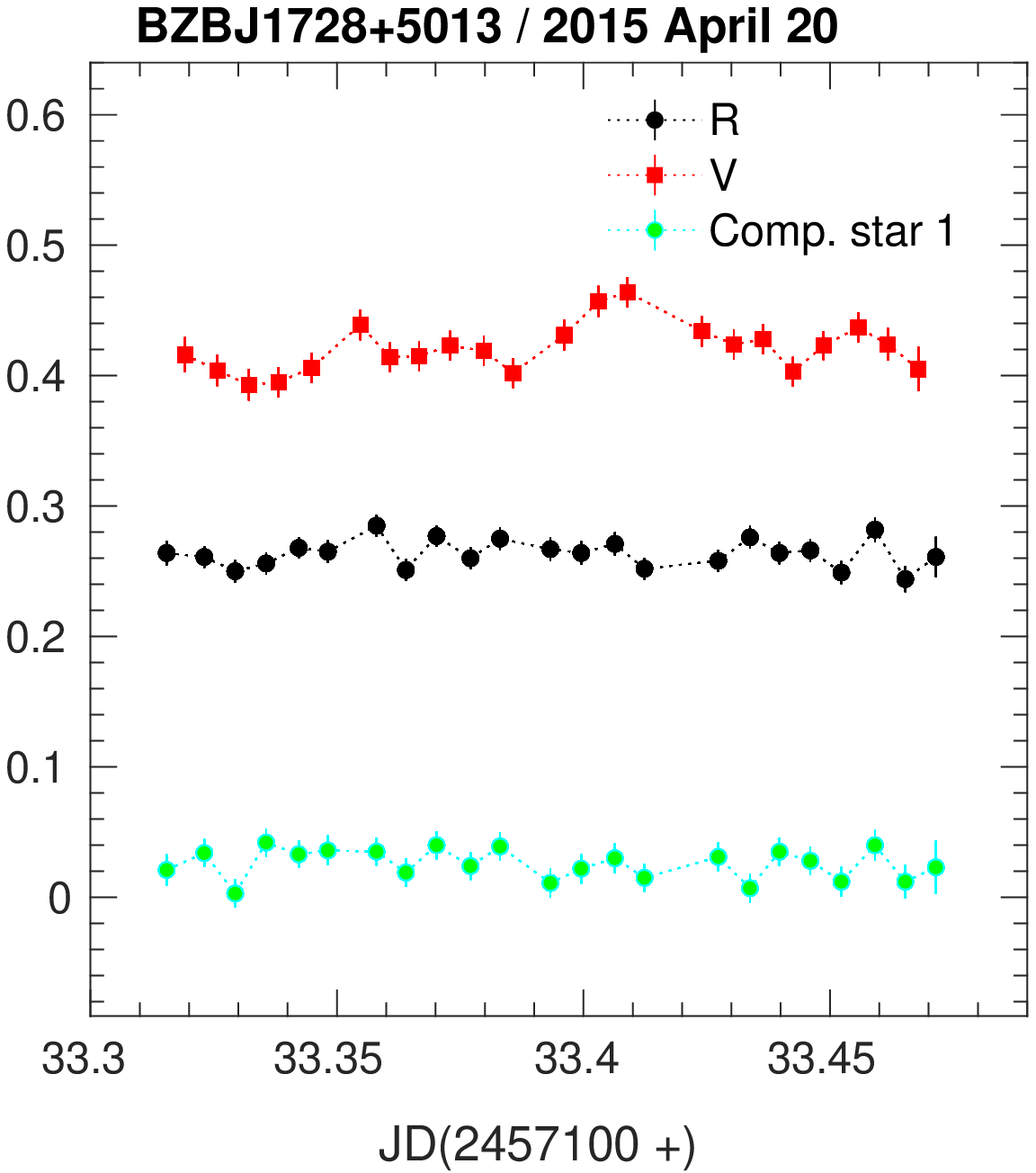} }}
\caption{Continued}
\end{figure*}

\begin{figure*}[hbt!]  
\centering
\ContinuedFloat
\mbox{\subfloat{\includegraphics[scale=0.4]{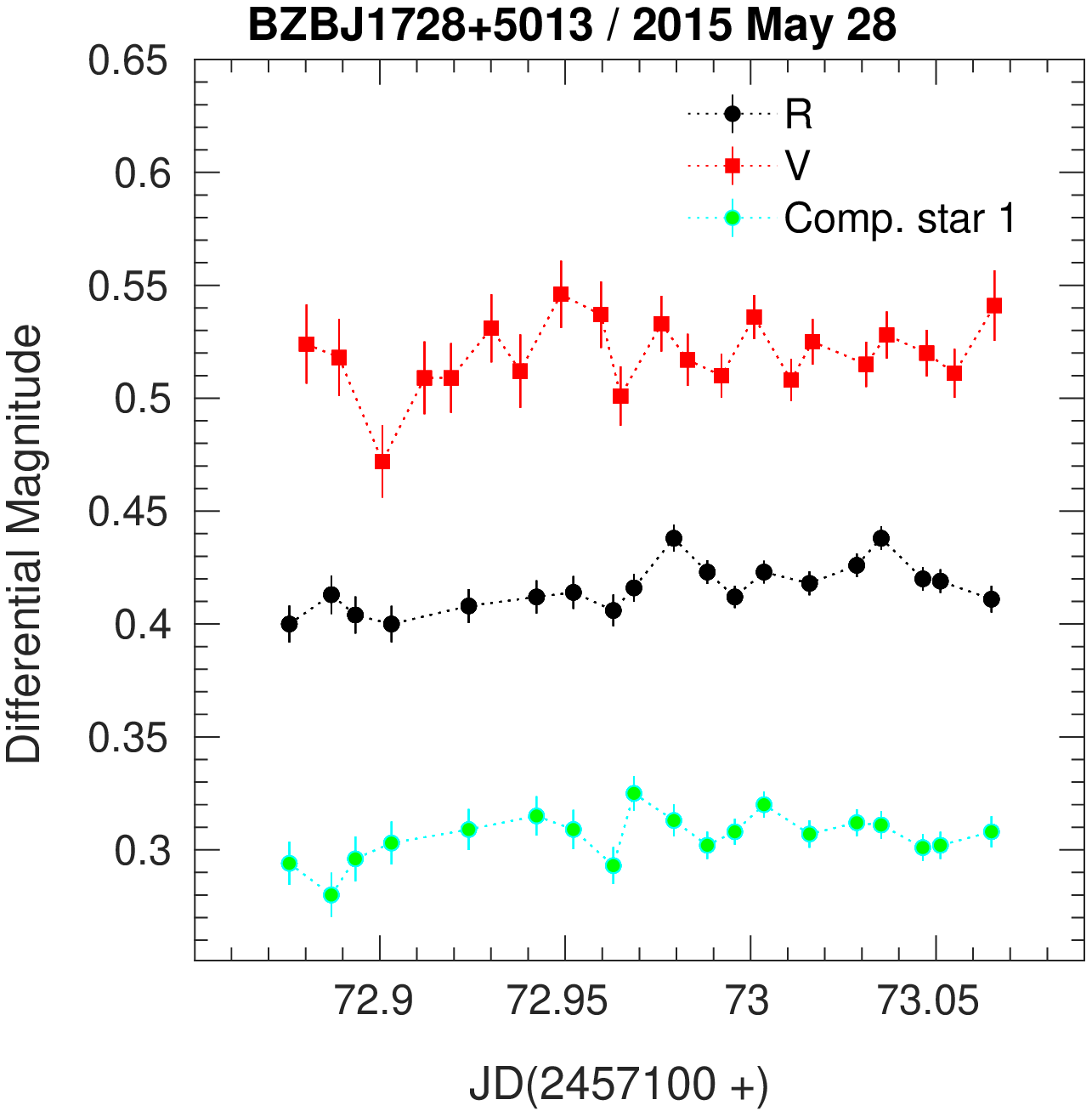} }\quad
\subfloat{\includegraphics[scale=0.4]{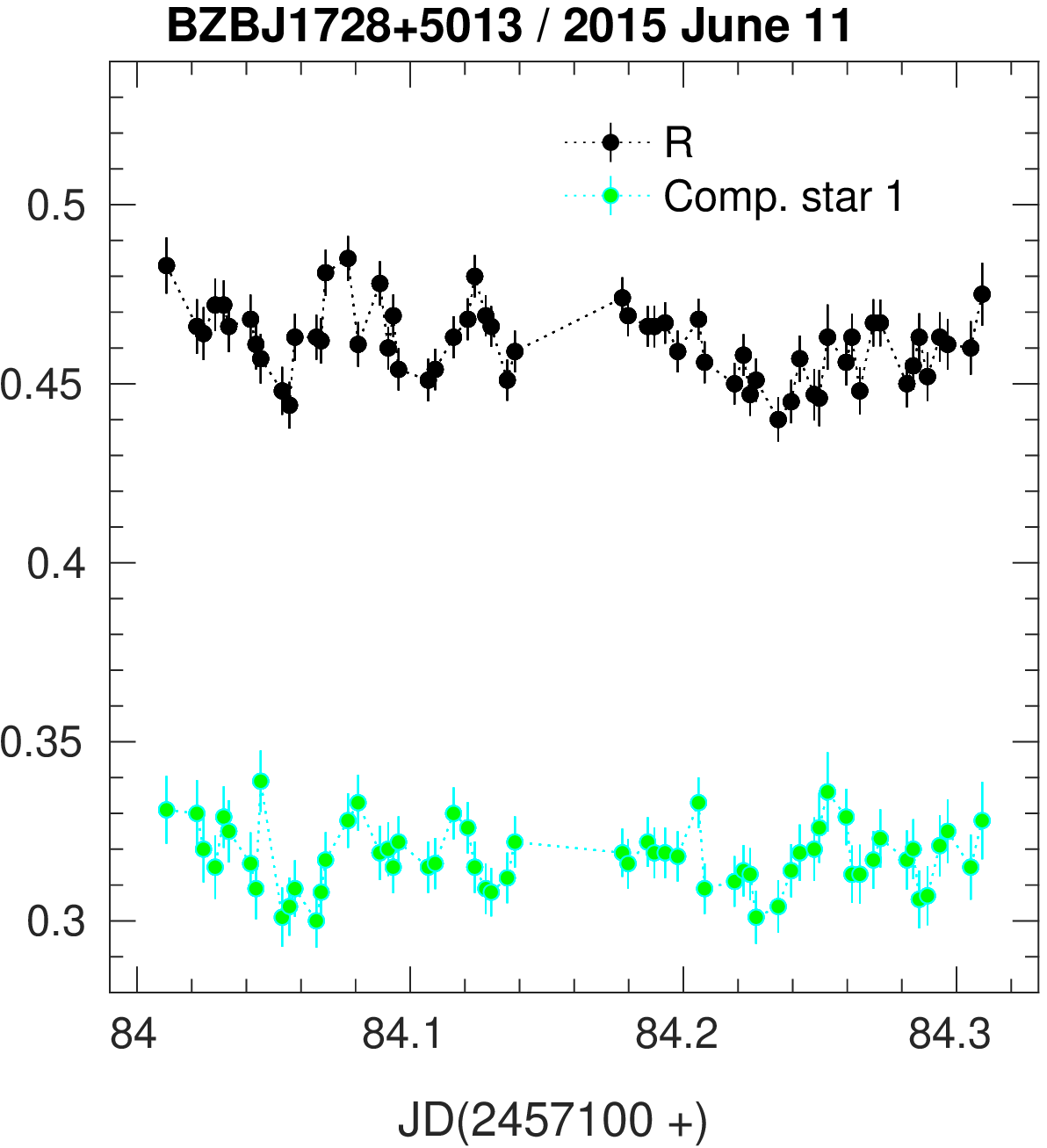} }\quad
\subfloat{\includegraphics[scale=0.4]{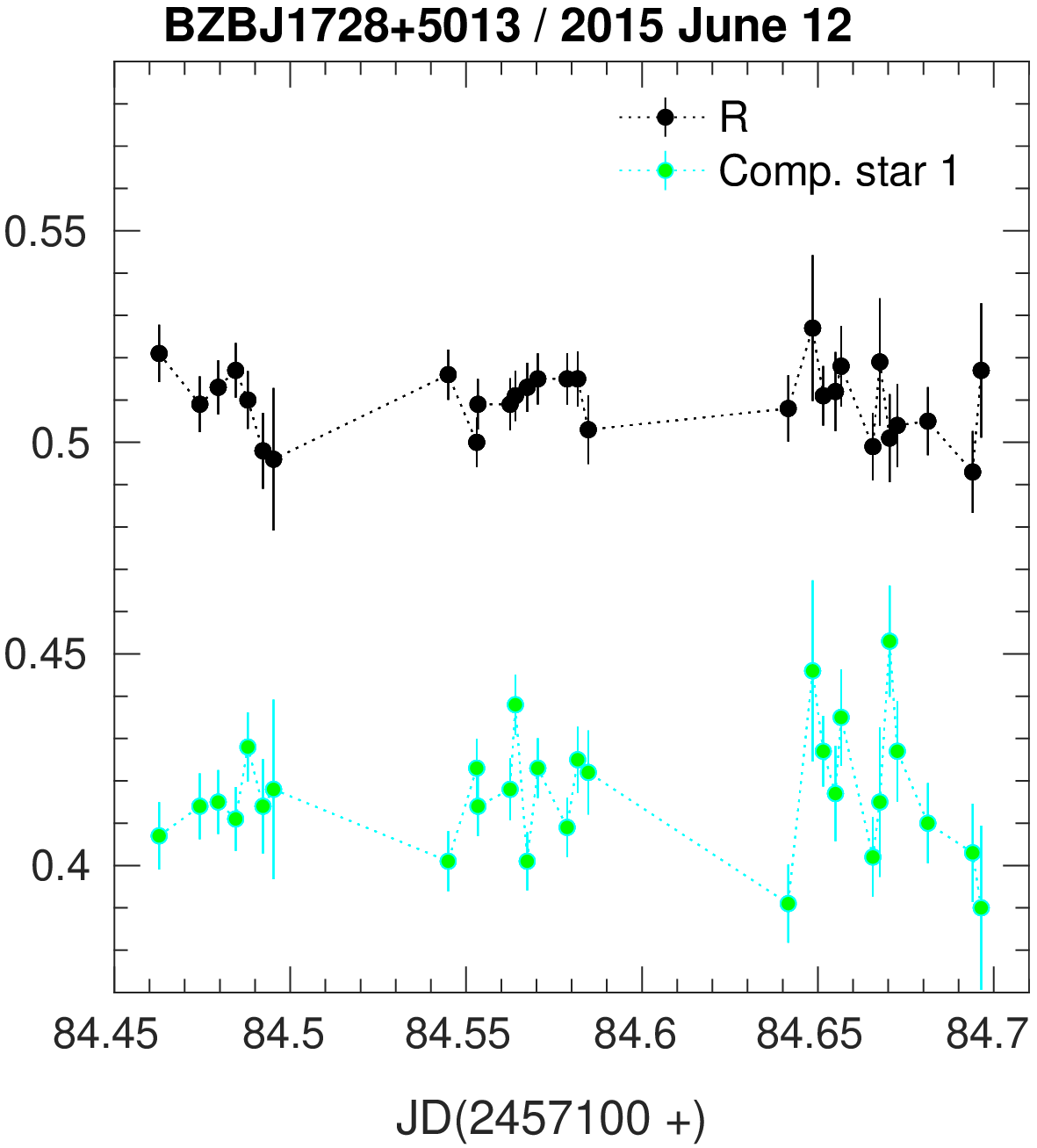} }}
\caption{Continued}
\end{figure*}

While selecting the comparison stars, we keep in mind that large difference in brightness between the blazar and comparison stars might lead to false detection of IDV in the DLCs \citep{2007MNRAS.374..357C}. Thus, all our comparison stars lies within $\pm$ 1 magnitude of the blazar. To construct DLCs on intraday and short timescales for a particular blazar, we used the same comparison and/or reference stars throughout the entire data set and the same was done for variability tests. All DLCs on intraday timescale for the whole blazar sample are displayed in Figure 1. The magnitude errors were estimated via simple error propagation. The colour coding in the plots correspond to different energy bands. \\

In our sample, 3 sources; BZBJ0656+4237, BZBJ0152+0147 and BZBJ1728+5013 have more than 10 nightly observations that expanses over time periods of week to months. This makes it possible to check for short term variability (STV) for these blazars. We constructed the short timescale DLCs. In the plots, each data points belong to different nightly observations and represent daily averaged differential magnitude, estimated using either at least 3 CCD images, or multiple images whenever the source was observed for longer period. The V--R DCIs are also estimated in the same way described before. These light curves are shown in Figure 2. The V-R CIs variation with time and R-band magnitude for these 3 blazars are also shown in the middle and bottom panels of the plots, respectively. We used a total of 82 DLCs on intraday timescales; 50 in V and R bands and 23 in V--R colour bands and 9 DLCs on short timescale in V, R and V--R bands in this study.

\section{{\bf Statistical tests to study Variability}}
Detection of micro-variability in AGN with precision and accuracy requires not only careful handling of the data, but also using robust and powerful statistical methods for the analysis. In the past, several statistical tests have been introduced and used for this purpose, among which the C-, F- and $\chi^{2}$-tests have gained massive popularity for searching optical variability in blazars. However, these methods have their own limitations and many caveats, which leads to low power and/or less reliable results, which are pointed out and discussed carefully in \citet{2014AJ....148...93D, 2015AJ....150...44D}. The authors suggested that by increasing the number of comparison stars in the F-test and analysis of variance test can significantly enhance both the power and the reliability of the test results, which they showed via simulations of several variable quasars. Considering the robustness over other conventional tests, we choose these two methods in the current work, which are briefly discussed in the next two sections.

\subsection{{\bf Power-enhanced F-test}} 
The conventional F-test compares two sample variances, one for the variable source, i.e., blazar in our case, and the other for the non-variable comparison star (CS) in the source field. In the modified F-test, i.e., the Power-enhanced F-test \citep[here after PEF-test][]{2014AJ....148...93D}, instead of a single CS, we compare the source differential  variance (i.e., estimated from the differential light curve) to the combined differential variance of multiple CSs. 

If the source is observed $N_j$ times along with $k$ number of CSs in a single night observation, then, for each observation $i$ of the $j$th CS, the scaled square deviation is given by 

\begin{equation}
s_{j,i}^2 = \omega_j(m_{j,i}-\bar{m_j})^2,
\end{equation}

where $\omega_j$ is a scaling term used to scale the variance of $j$th star to the level of the source \citep{2011MNRAS.412.2717J}, $m_{j,i}$ and $\bar{m_j}$ are the differential and mean magnitude of the star. Now, the combined variance of the CSs can be calculated by stacking all the scaled square deviation for all observations and CSs, which can be written as 

\begin{equation}
s_c^2 = \frac{1}{(\sum_{j=1}^k N_j)-k}\sum_{j=1}^{k} \sum_{i=1}^{N_i} s_{j,i}^2,
\end{equation}

And comparing the above parameter to the differential variance of the source, $s_{src}^2$, gives the PEF-values as

\begin{equation} 
F_{enh} = \frac{s_{src}^2}{s_c^2},
\end{equation}

This F-statistic values with $\nu_{src} = N_{src}-1$ degrees of freedom (dof) in the numerator and $\nu_c =(\sum_{j=1}^k N_j)-k$ dof in the denominator are compared to critical F-value ($F_{critical}$) at 99\% (or 95\%, depending on the situation and requirement) confidence level. If $F_{enh}$ $\geq$ $F_{critical}$ value for a particular light curve, then we consider that light curve as variable, otherwise non-variable.

In our case, for a particular blazar, the source and all comparison stars have the same number of observations in their time series measurements irrespective of hourly or nightly observations hence, the dofs reduces to $\nu_{src} = N-1$ and $\nu_c = k(N-1)$, respectively. For calibration, we selected several stars from the source field of view, which are either slightly brighter or dimmer than or have similar brightness as the blazar. From those, we selected 3 least variable stars for comparison and reference.  

\subsection{{\bf Nested ANOVA test}}
The ANOVA test requires groups of replicated observations and it compares the dispersion of the individual differential magnitudes of the source within the groups and dispersion between the groups without using any comparison stars. In nested ANOVA test \citep{2015AJ....150...44D}, several field stars are used as reference to estimate the blazar differential photometry. Thus, we will have one more star in this analysis than that in the {\it PEF}-test. Here, we used three reference stars, the same stars used as comparison + reference in {\it PEF}-test. We divided the time series observations into different temporal groups, $a$, where each group contain $b = 5$ observations. Larger time lapse might affect the time resolution of the analysis, hence we avoided that. As described in \citet{2015AJ....150...44D}, we estimated the mean square due to groups ($MS_G$) and mean square due to nested observations in groups ($MS_{O(G)}$) with dof $\nu_{1} = a - 1$ and $\nu_{2}=a(b - 1)$, respectively. The statistic is given by 

\begin{equation} 
F_{} = \frac{MS_{G}}{MS_{O(G)}},
\end{equation}
 
If the $F-$value exceeds the critical value $F_{\nu_{1}, \nu_{2}}^{(\alpha)}$ at a significance level of 99\% ($\alpha= 0.01$), the null hypothesis will be rejected. \\

Comparing results from multiple tests can further increases the reliability of the variability analysis by avoiding the pseudo variability that might generated due to the odd behavior of a comparison star \citep{2011MNRAS.412.2717J}. Considering this way of multi-testing, we claim detection of micro-variability only if both the above mentioned tests could detect variability at a significance level of 99\%. Results of these two analysis are presented in Table \ref{tab5:long}.    

\begin{figure*}  
\centering
\mbox{\subfloat{\includegraphics[scale=0.47]{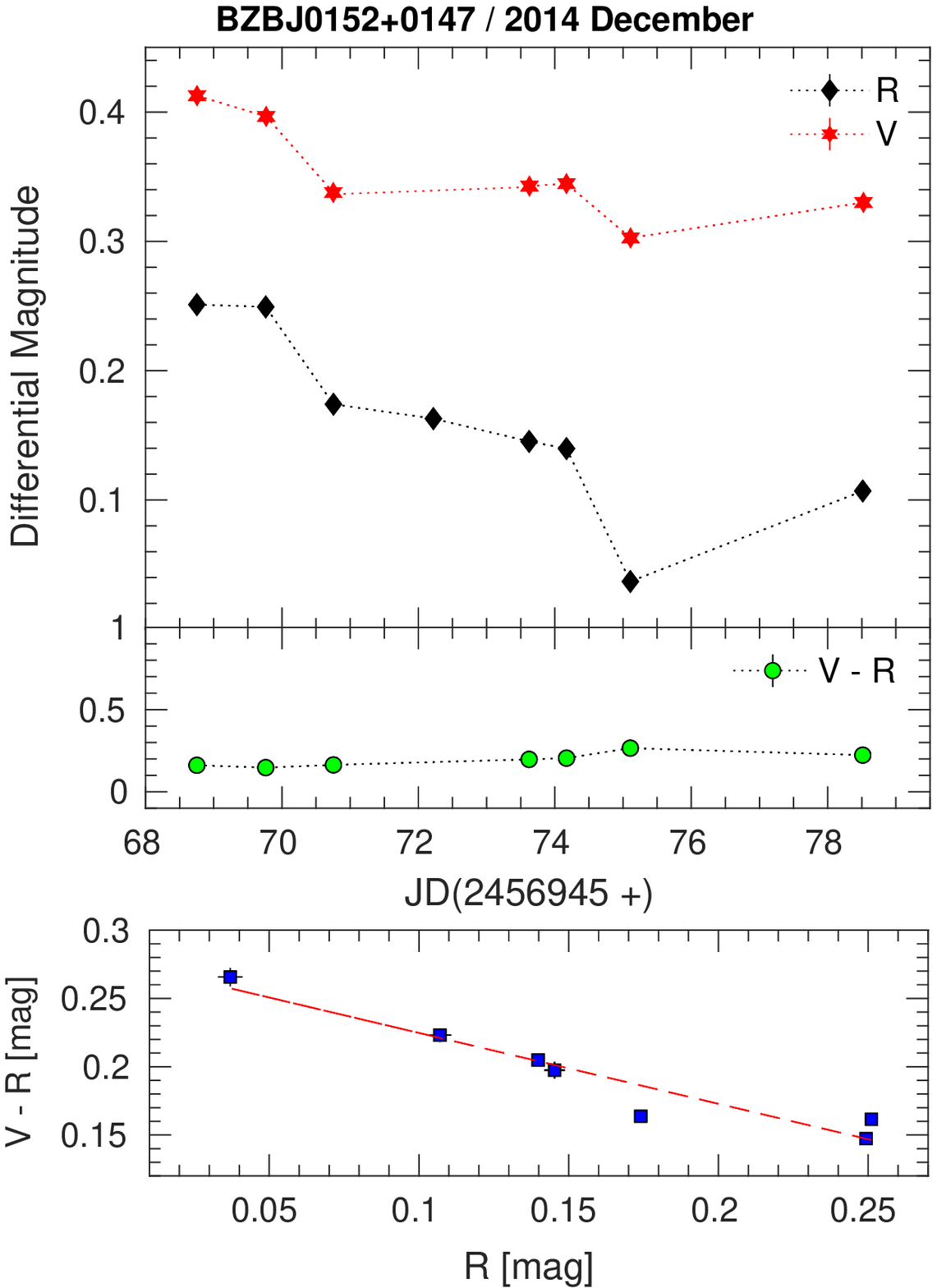} }\quad
\hspace*{0.0cm}
\subfloat{\includegraphics[scale=0.47]{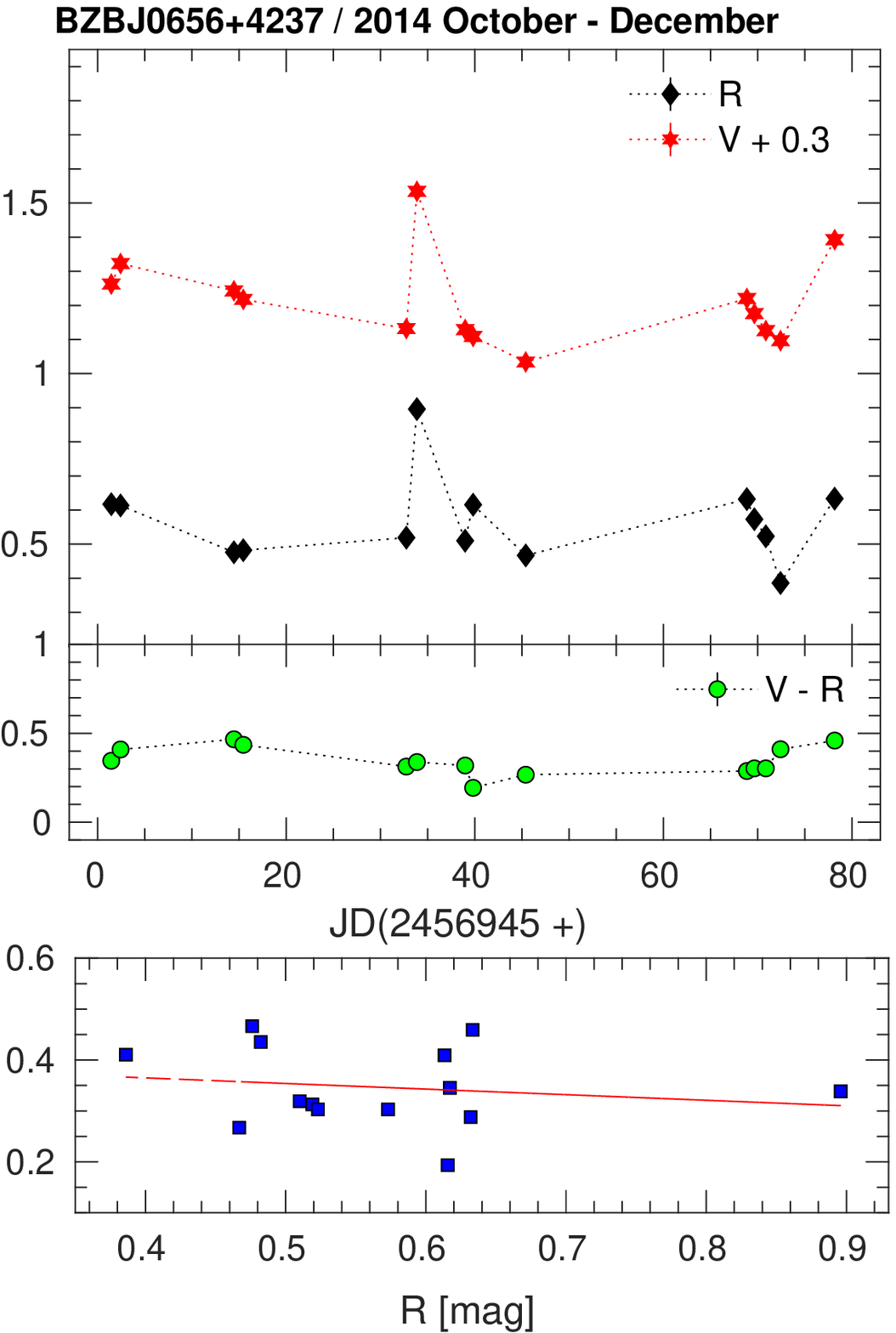}}\quad
\hspace*{0.0cm}
\subfloat{\includegraphics[scale=0.47]{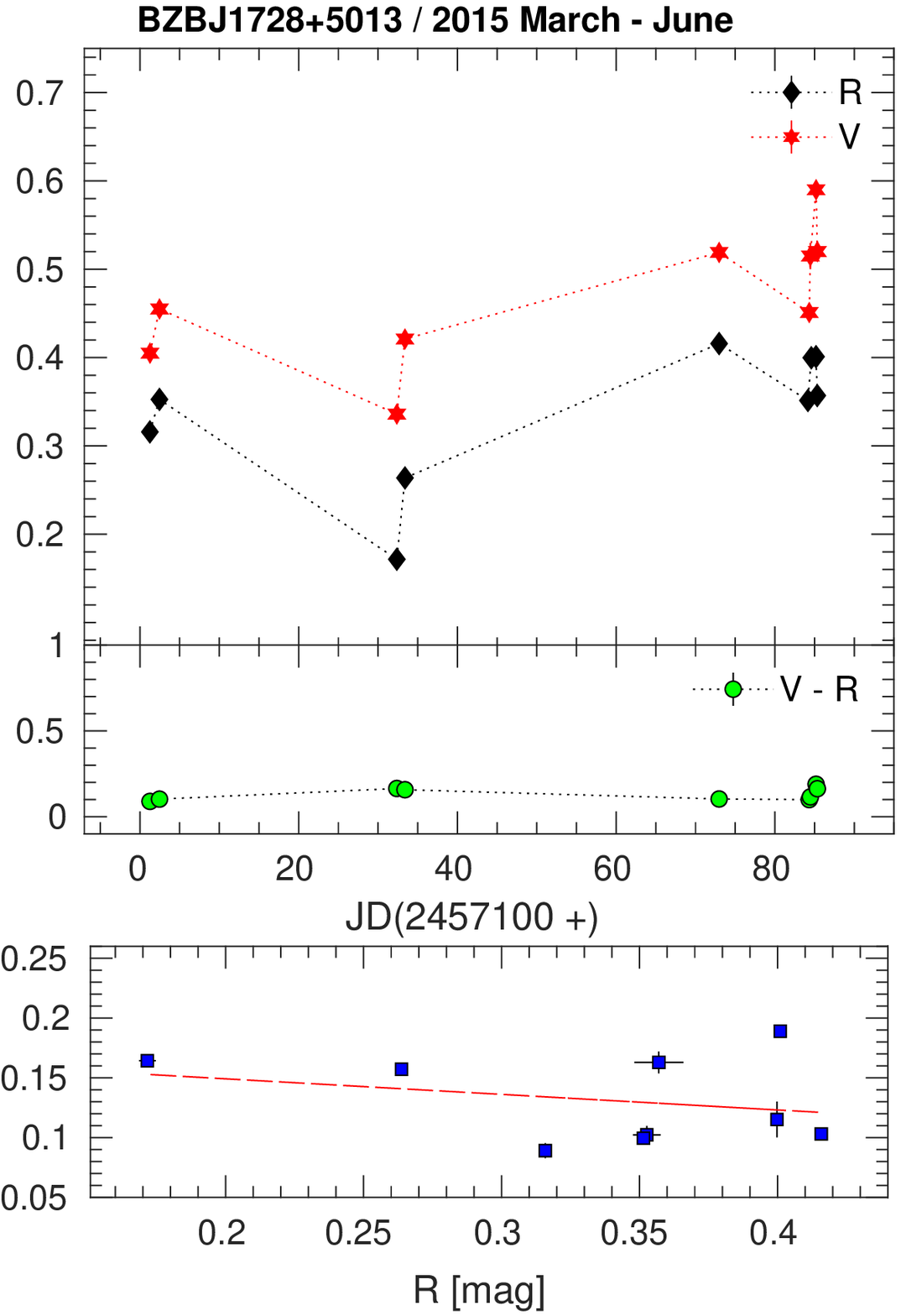}}}
\caption{Differential magnitude light curves on days to months timescales with observations taken during 2014 October to 2015 June for the blazars BZGJ0656+4237, BZGJ0152+0147 and BZBJ1728+5013 in R and V bands, respectively. Each data point represents average magnitude either taken from at least 3 CCD images or from multiple images belong to single night observation. The middle and bottom panels in each plot shows the differential (V--R) colour variation with time and differential R-magnitude, respectively.}
\end{figure*}

\begin{figure*}[hbt!]  
\centering
\begin{minipage}{0.4cm}
\rotatebox{90}{{\large {V--R color indices}}}
\end{minipage}
\begin{minipage}{\dimexpr\linewidth-1.10cm\relax}
\mbox{\subfloat{\includegraphics[scale=0.33]{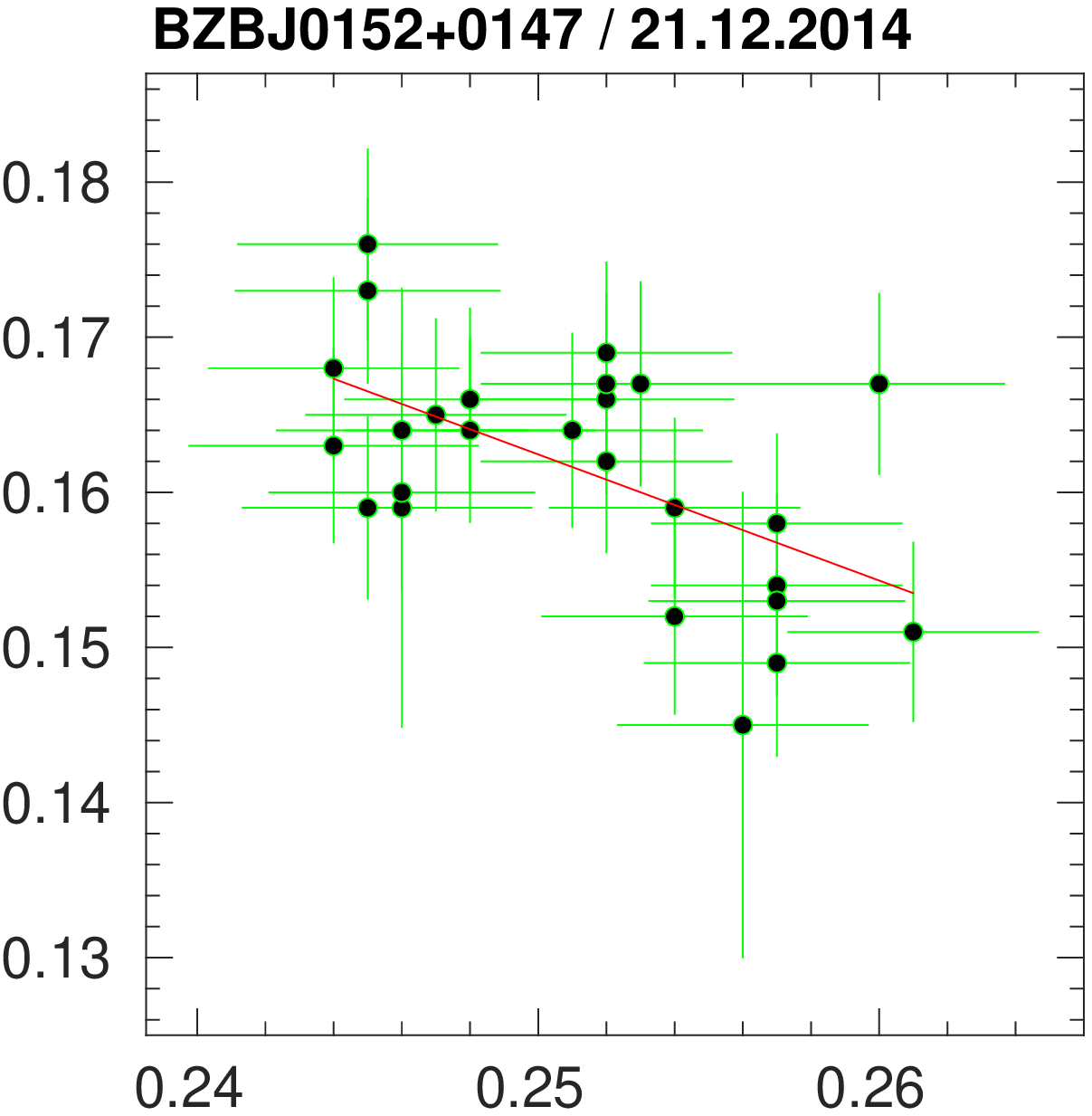}}\quad
\subfloat{\includegraphics[scale=0.33]{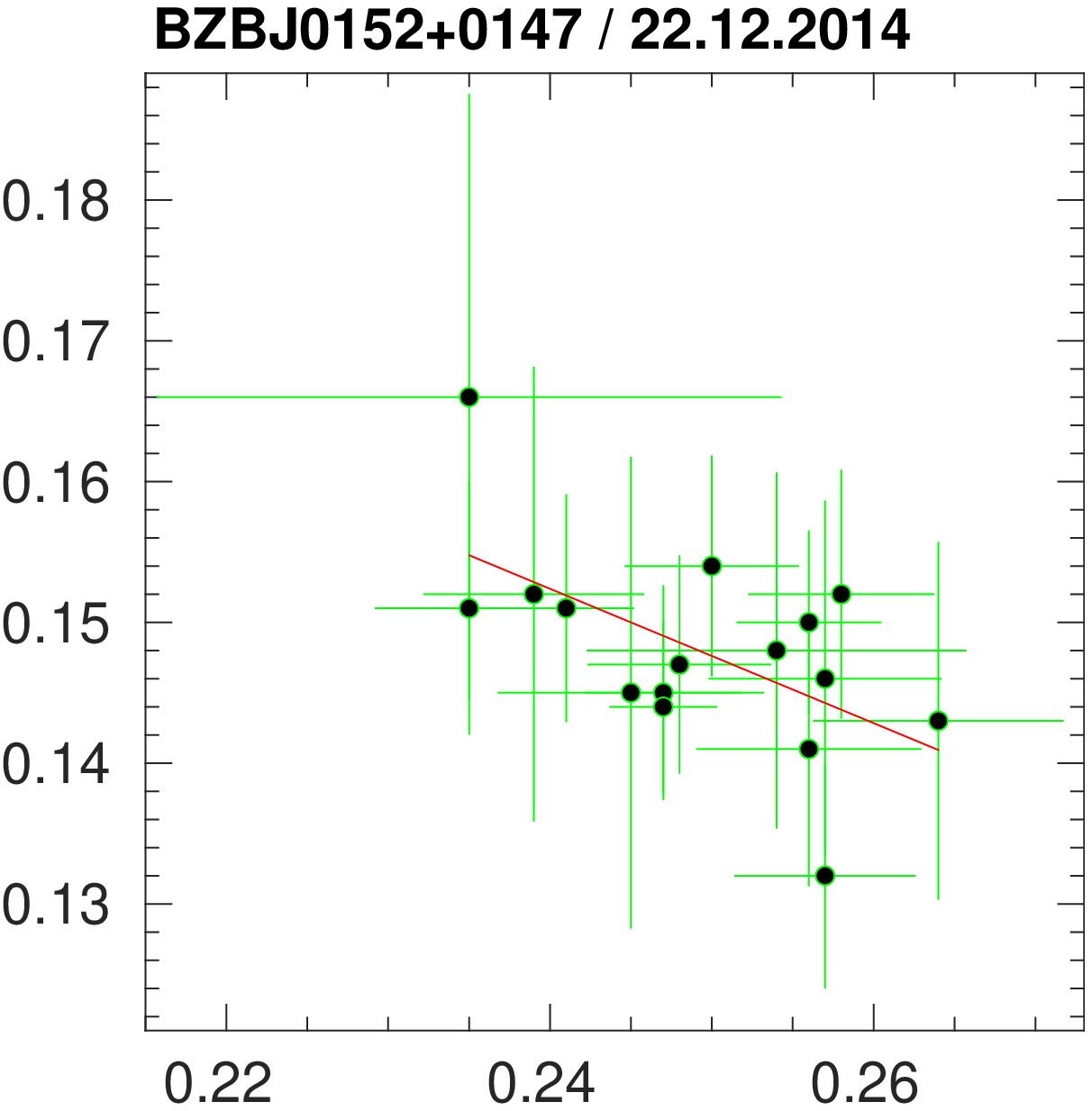}}\quad
\subfloat{\includegraphics[scale=0.33]{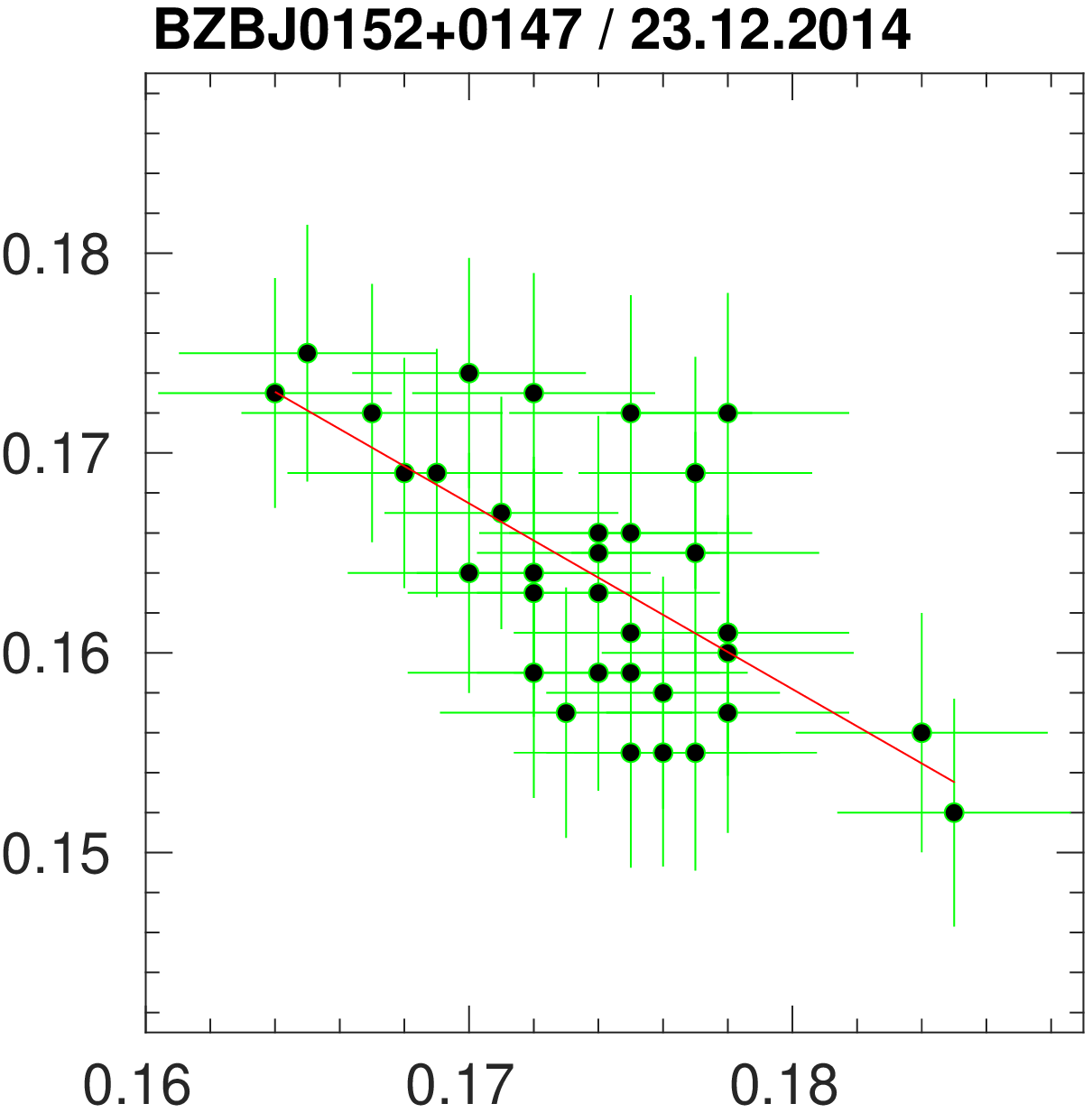}}\quad
\subfloat{\includegraphics[scale=0.33]{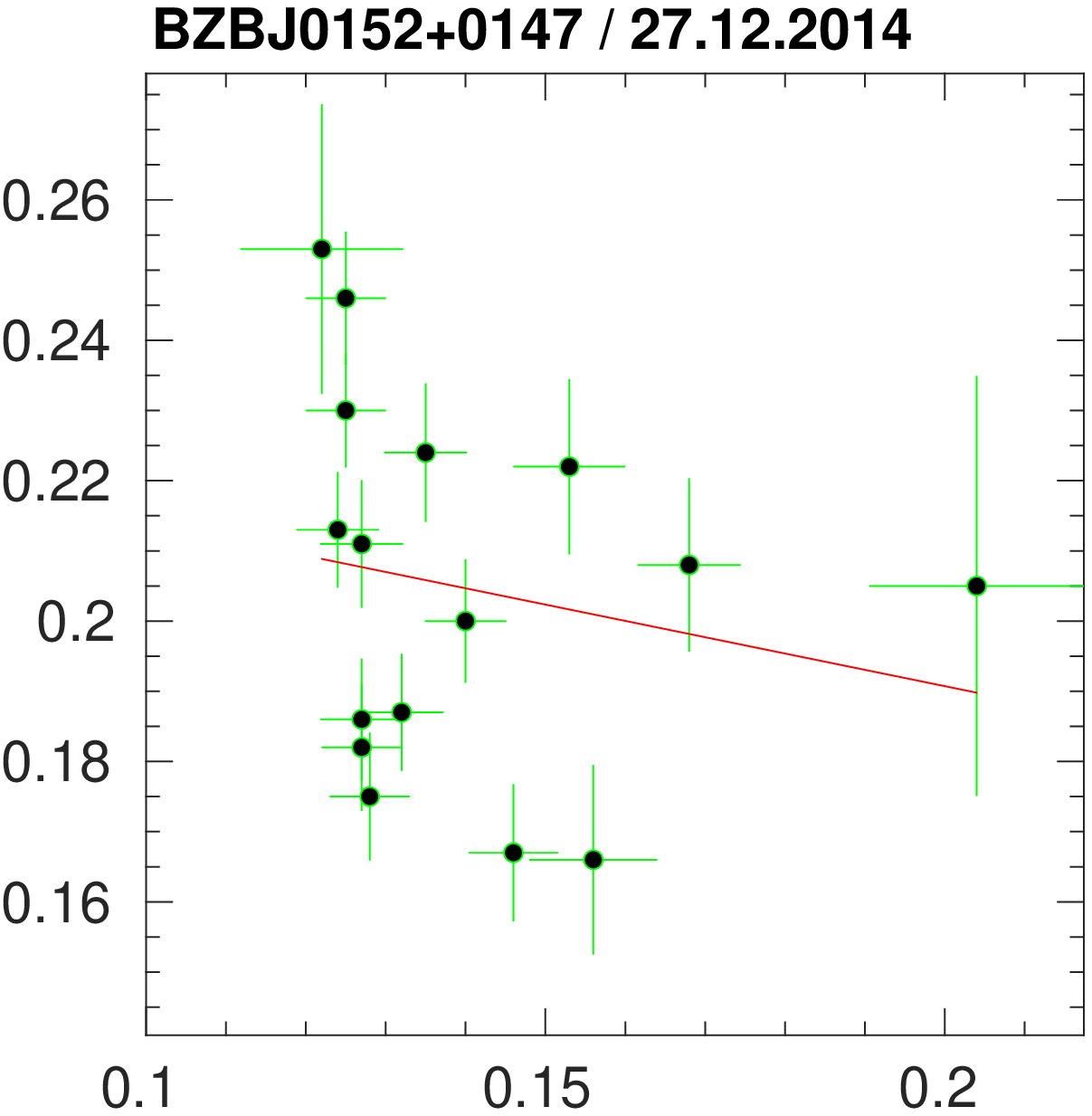}}}
\newline
\mbox{\subfloat{\includegraphics[scale=0.33]{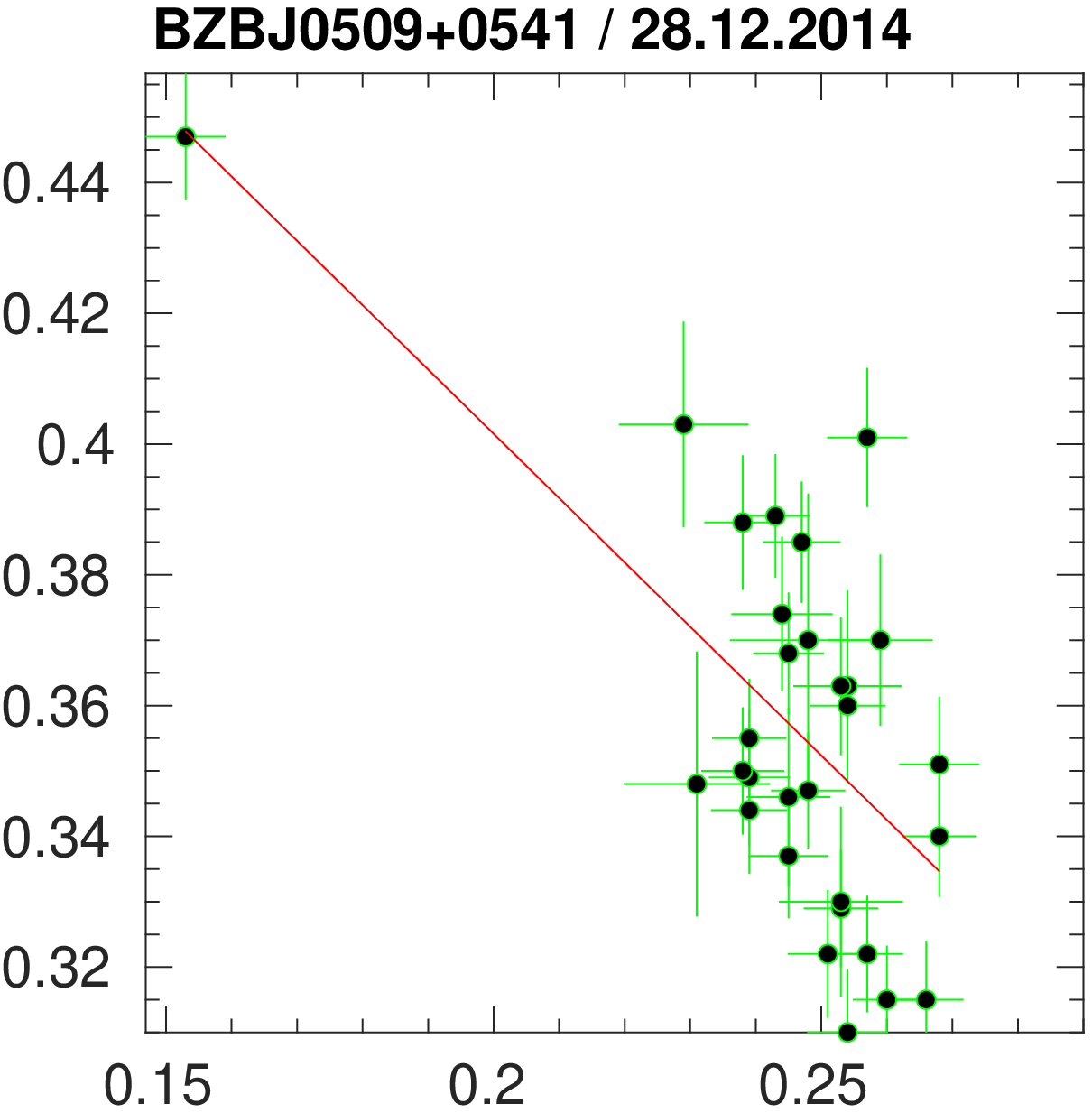}}\quad
\subfloat{\includegraphics[scale=0.33]{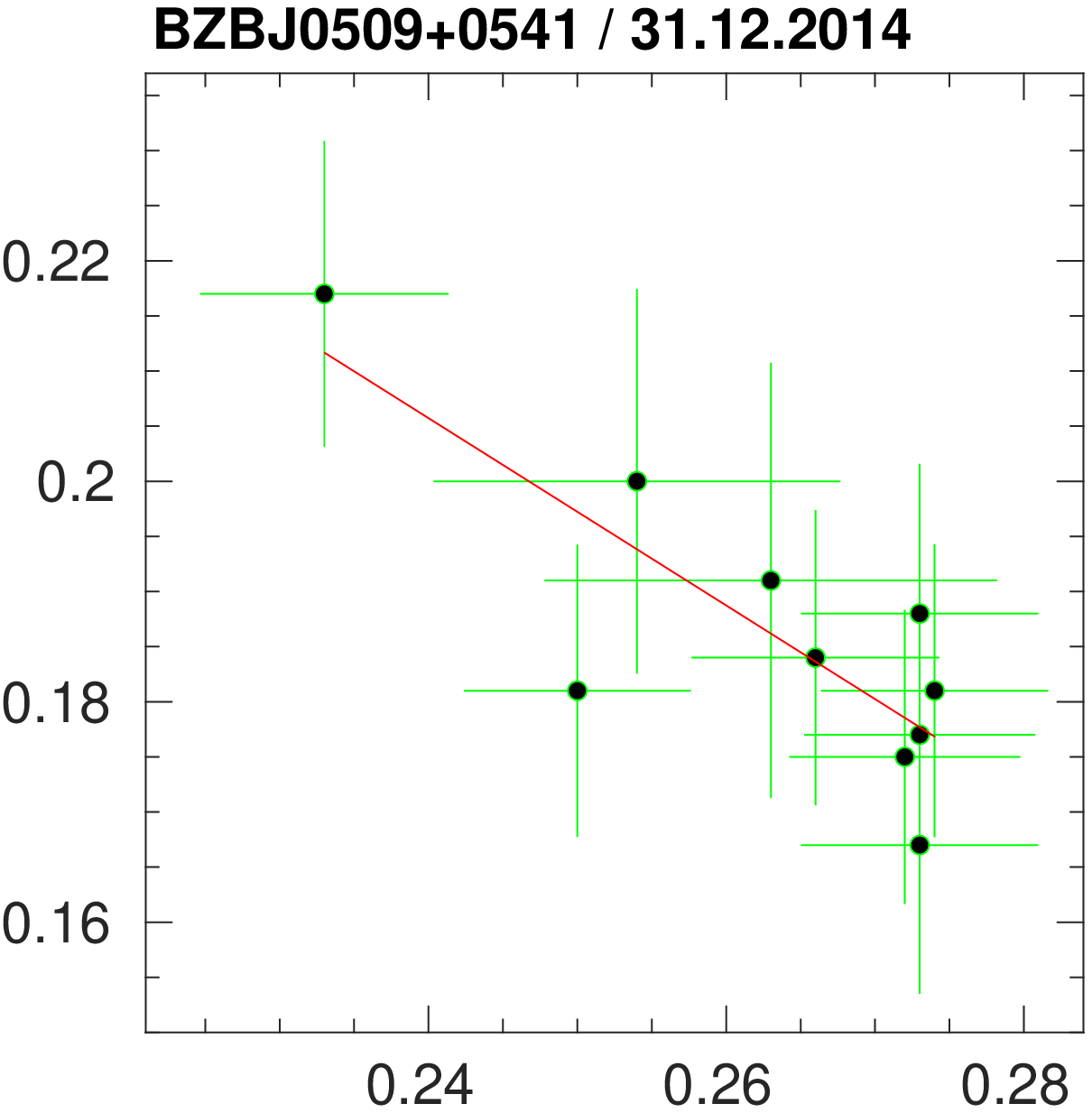}}\quad
\subfloat{\includegraphics[scale=0.33]{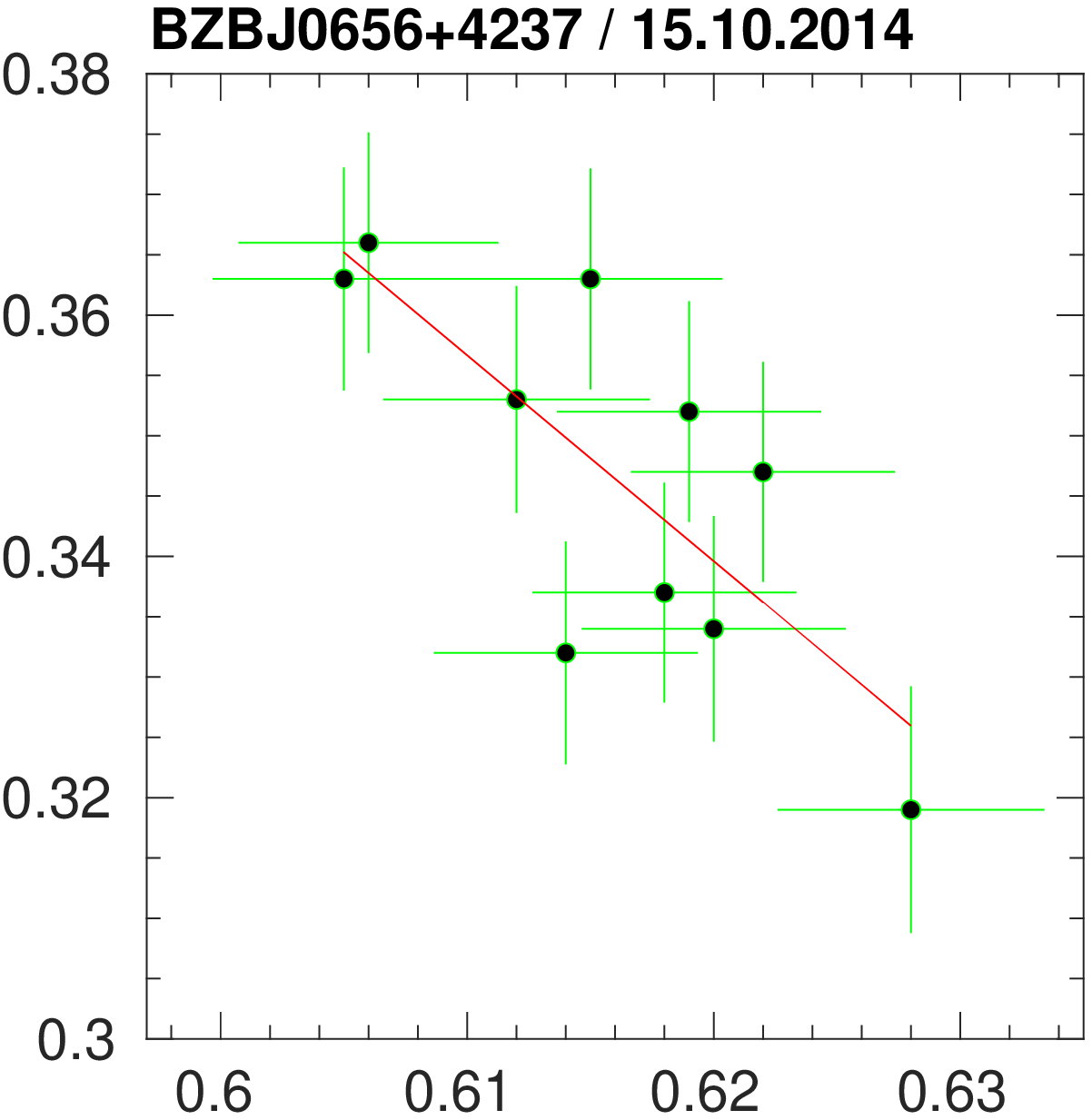}}\quad
\subfloat{\includegraphics[scale=0.33]{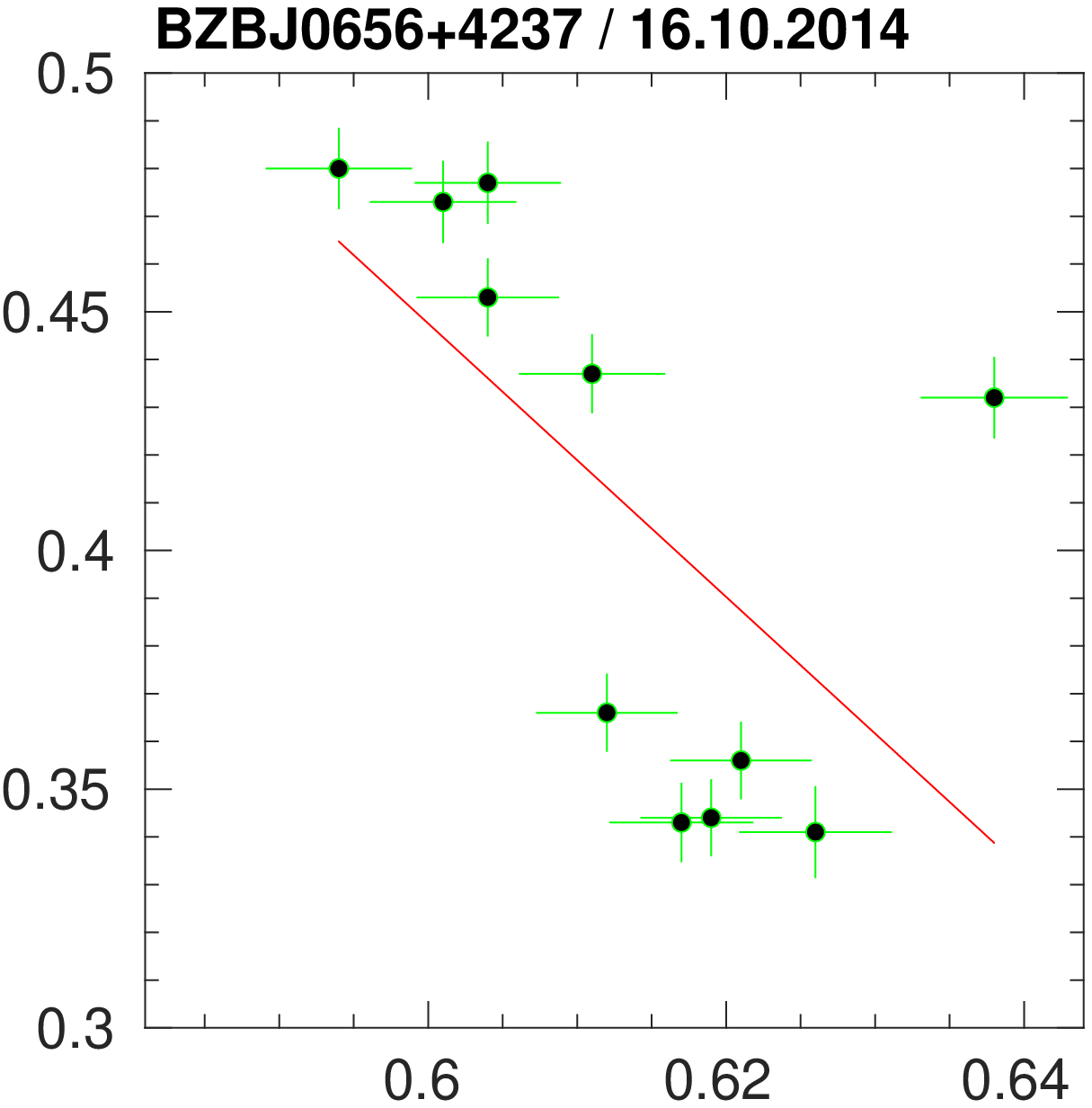}}}
\newline
\mbox{\subfloat{\includegraphics[scale=0.33]{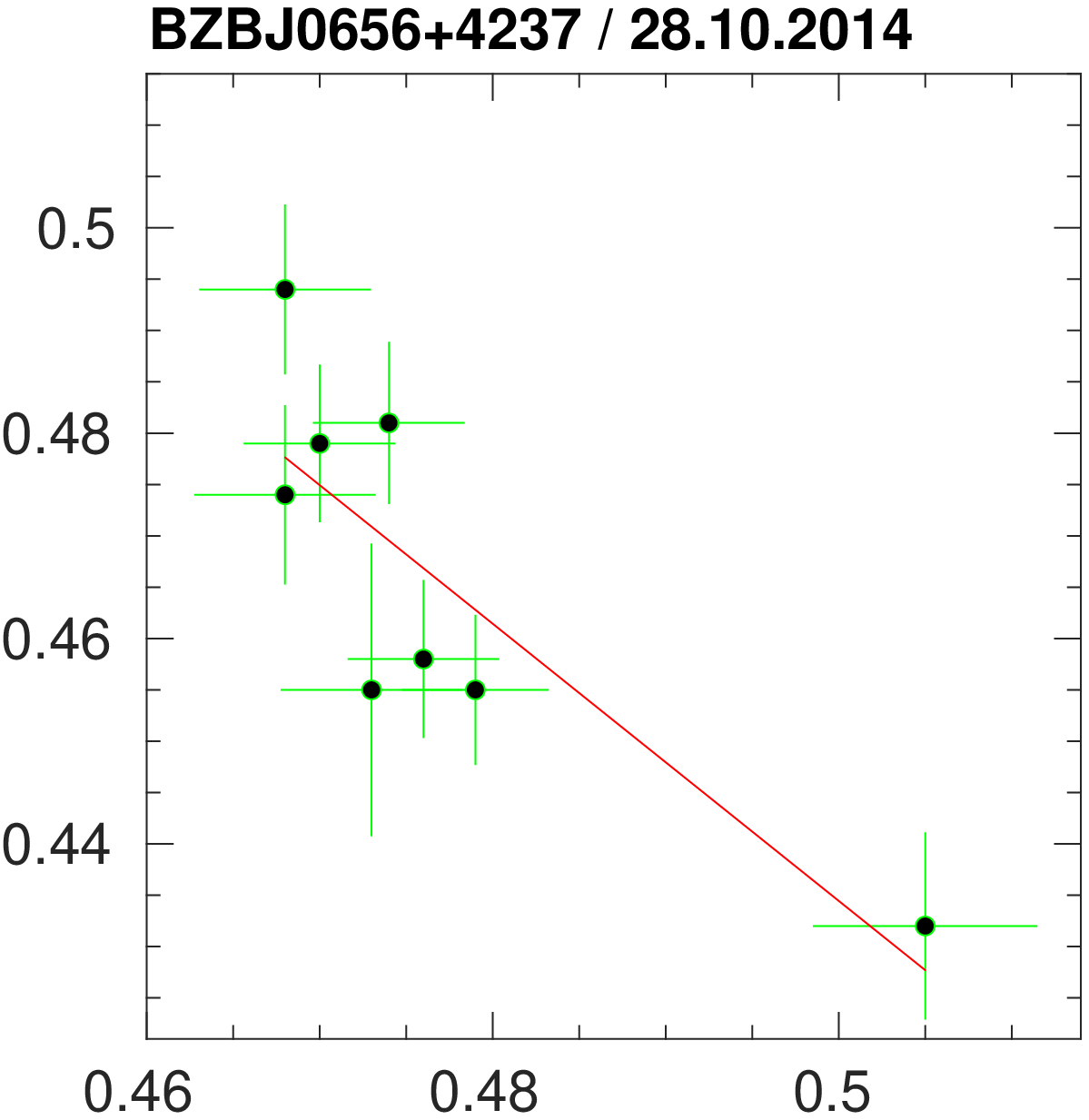}}\quad
\subfloat{\includegraphics[scale=0.33]{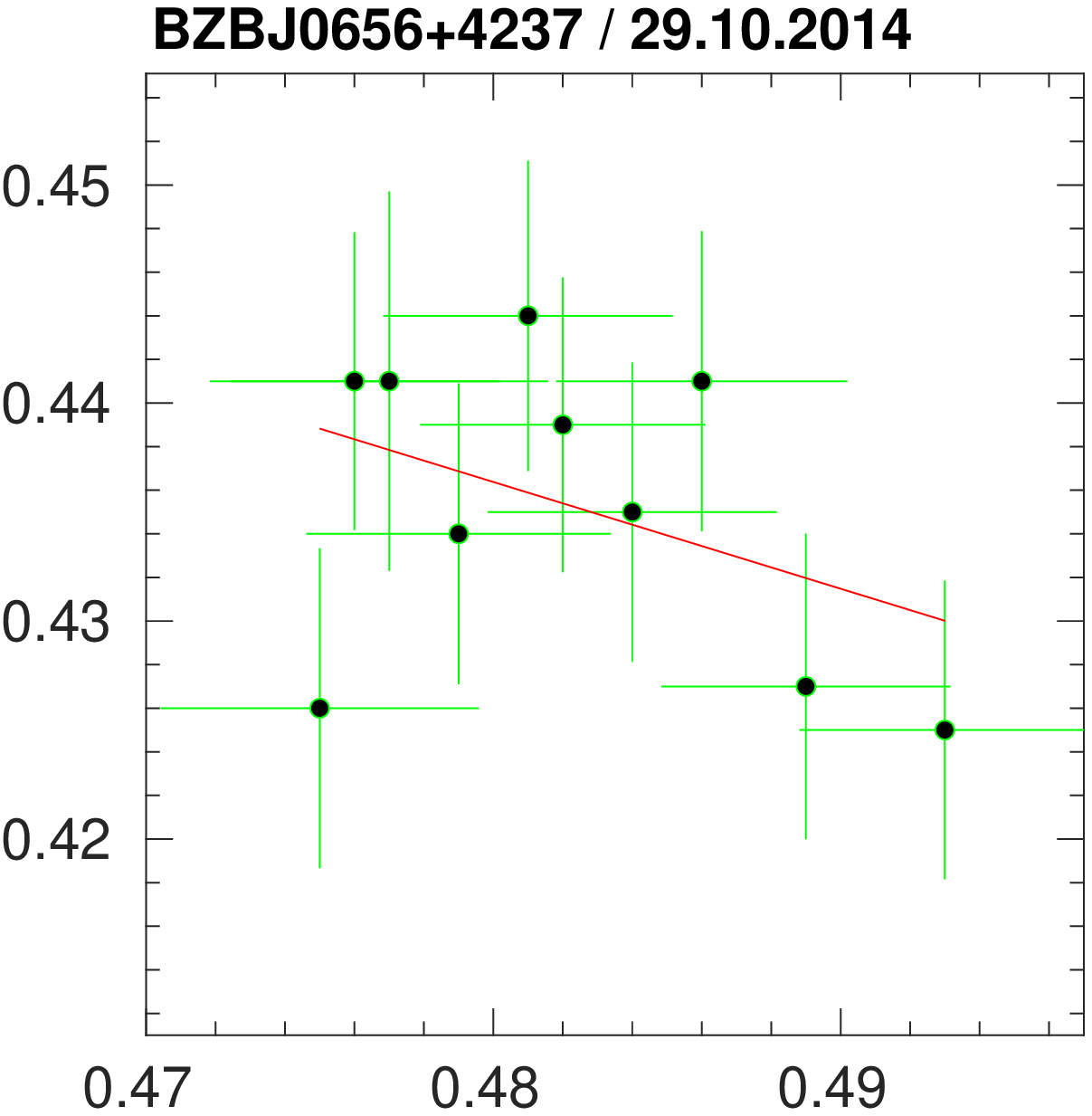}}\quad
\subfloat{\includegraphics[scale=0.33]{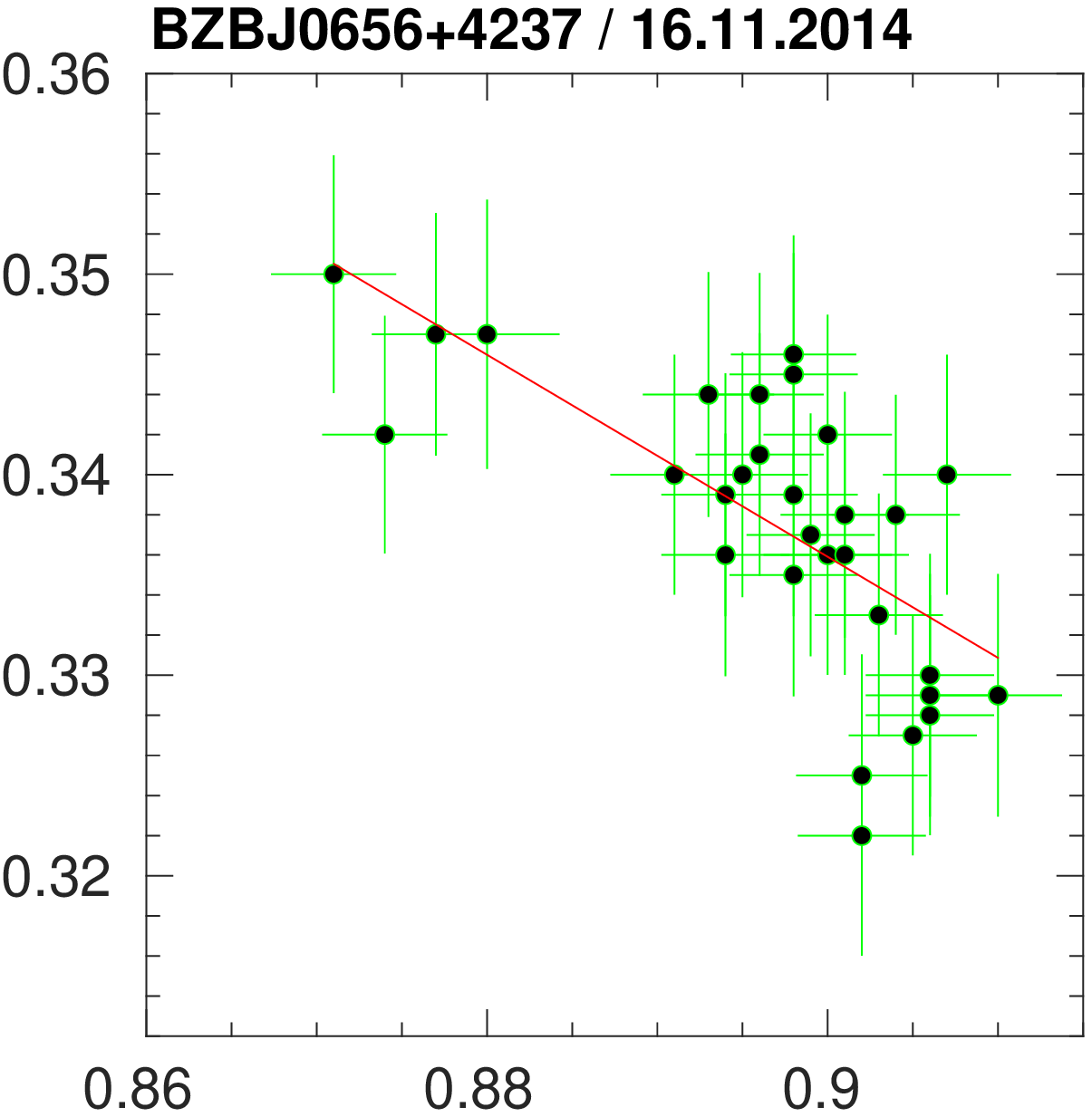}}\quad
\subfloat{\includegraphics[scale=0.33]{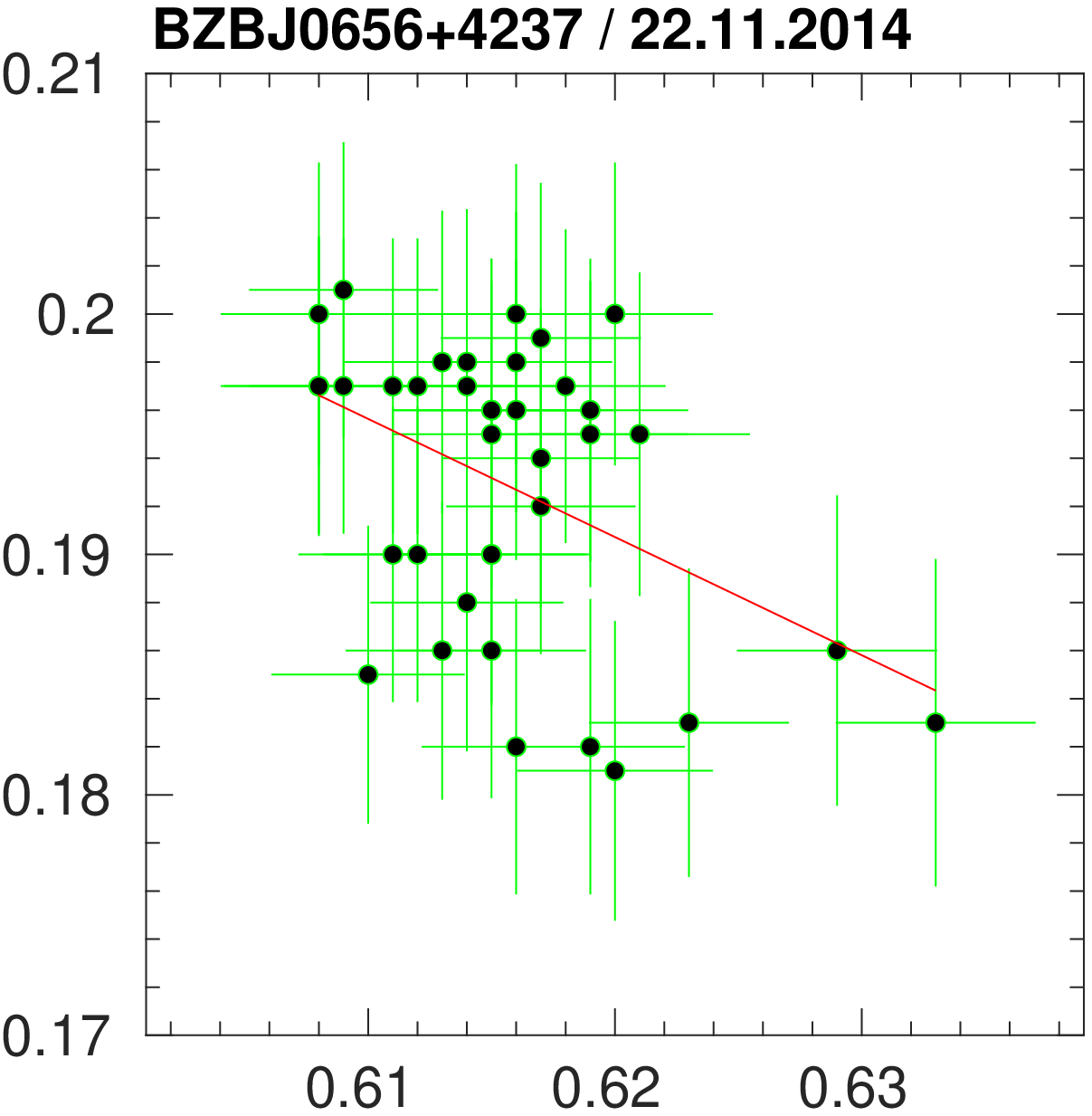}}}
\newline
\mbox{\subfloat{\includegraphics[scale=0.33]{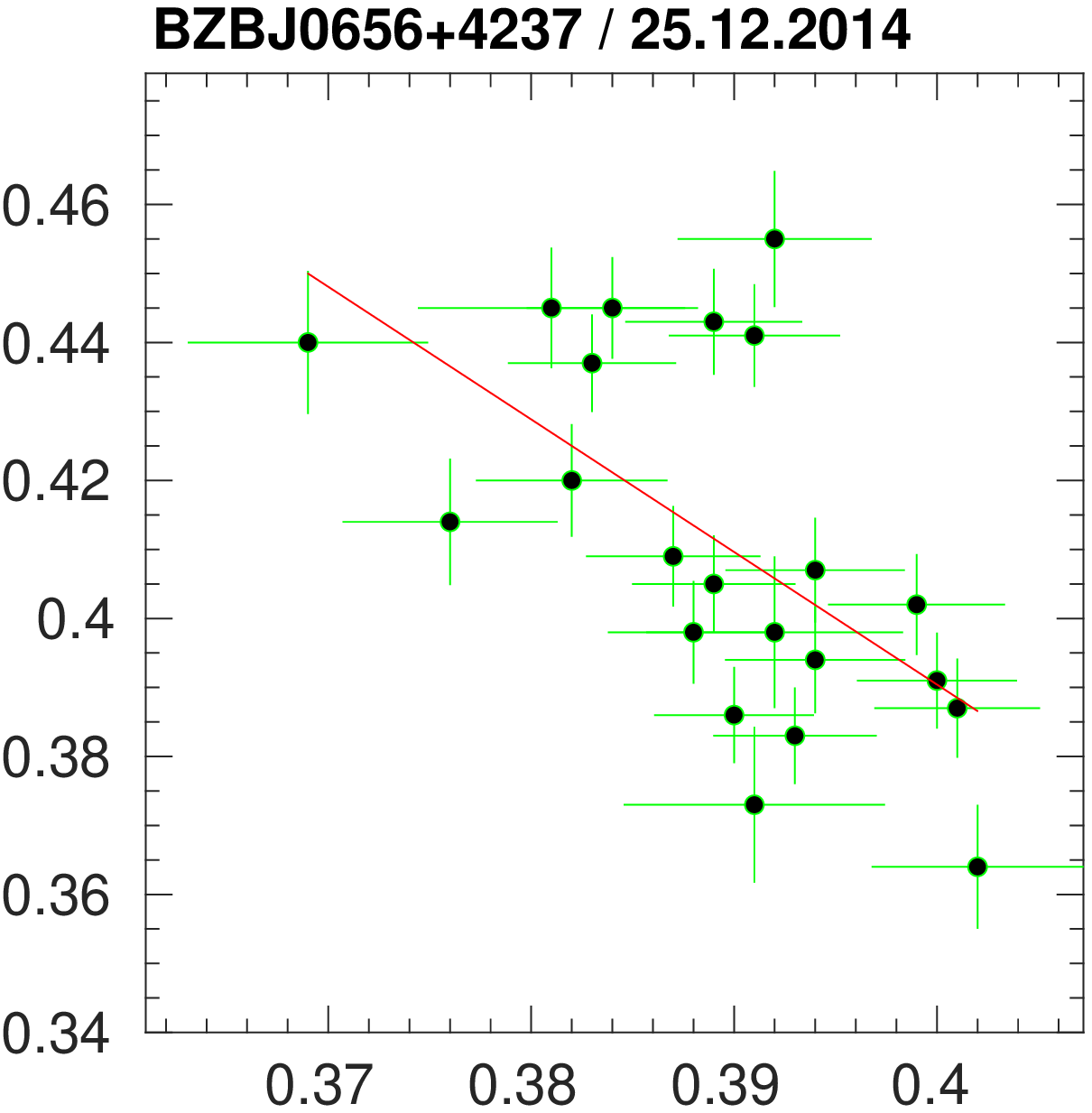}}\quad
\subfloat{\includegraphics[scale=0.33]{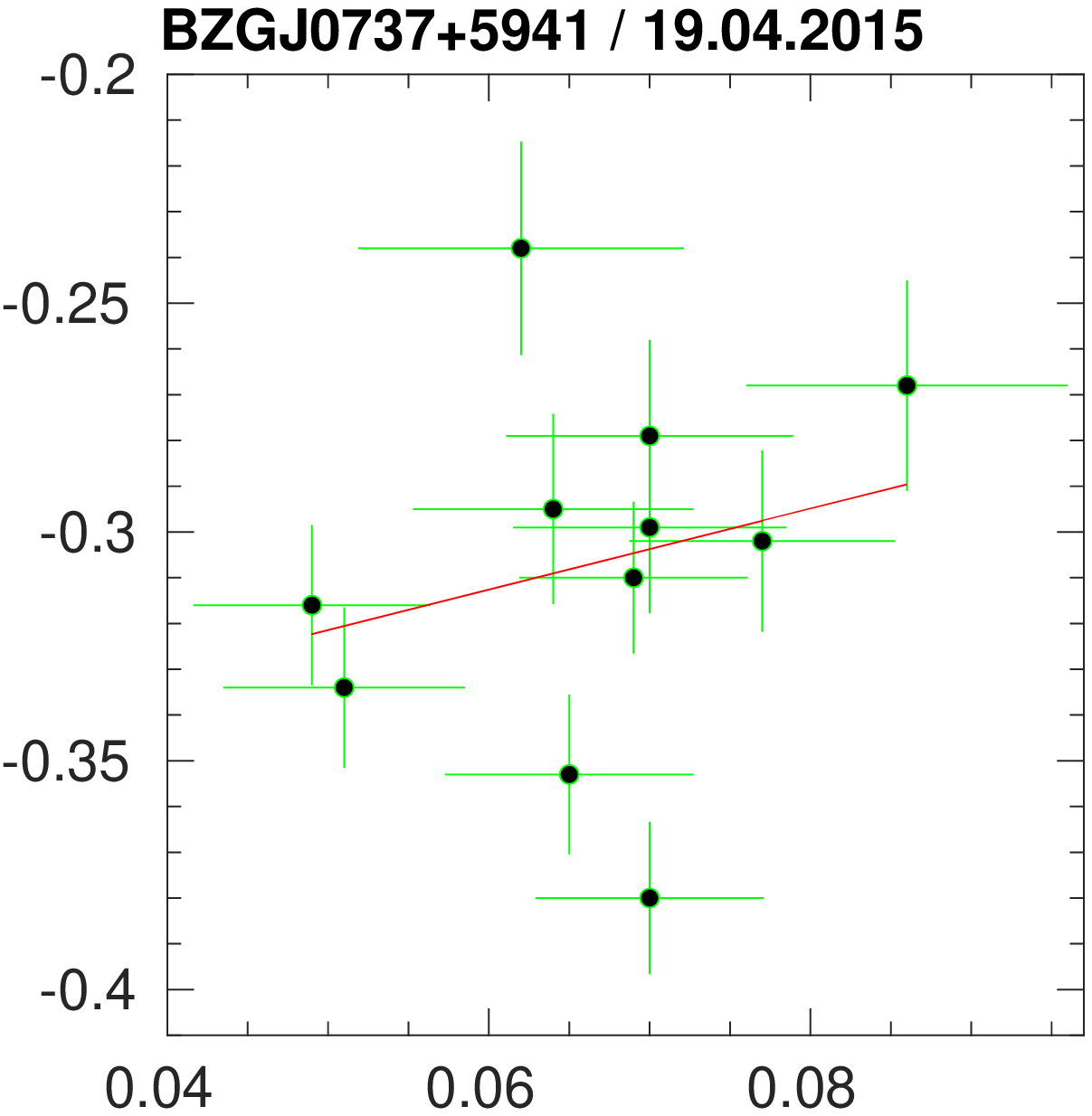}}\quad
\subfloat{\includegraphics[scale=0.33]{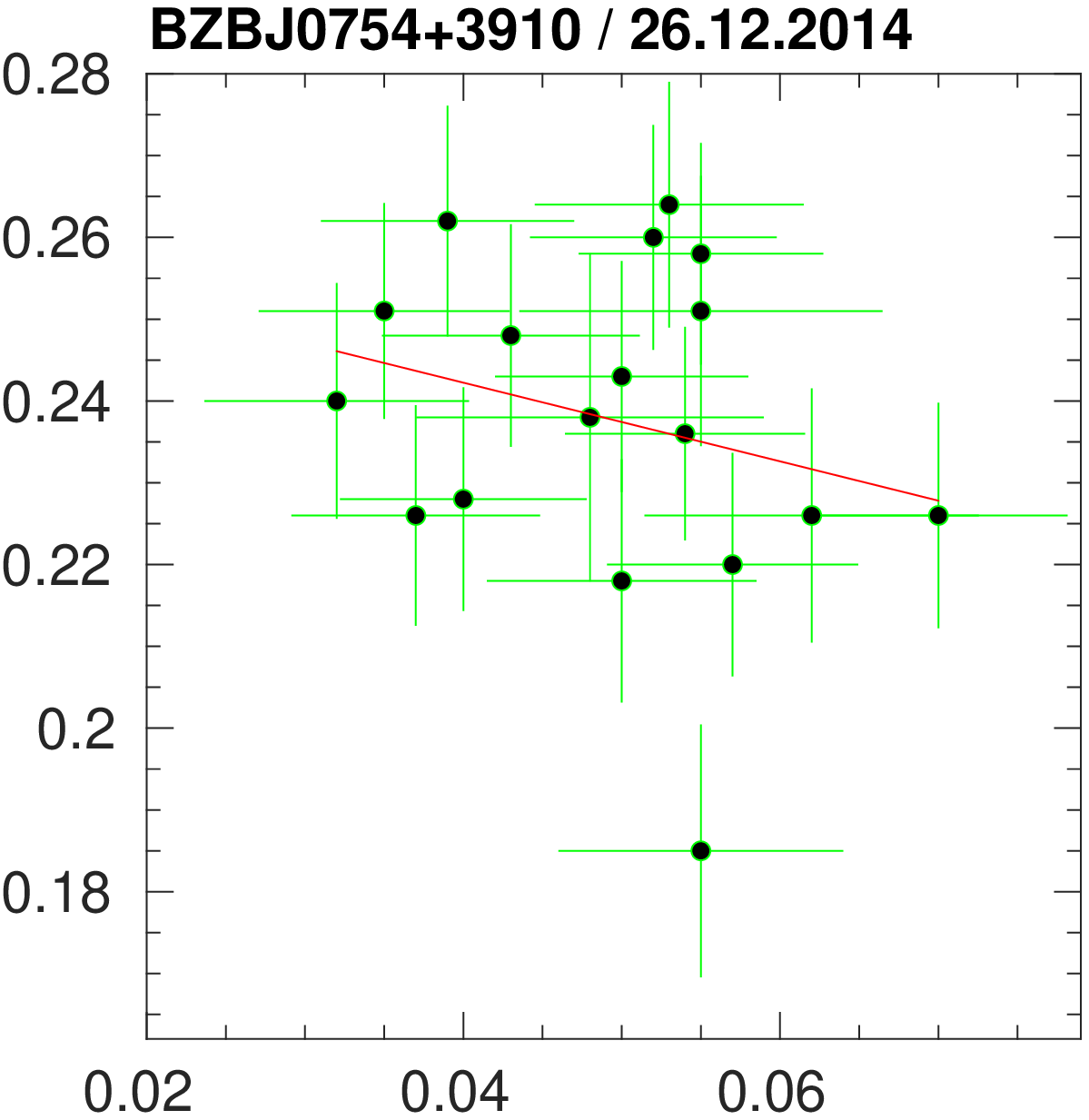}}\quad
\subfloat{\includegraphics[scale=0.33]{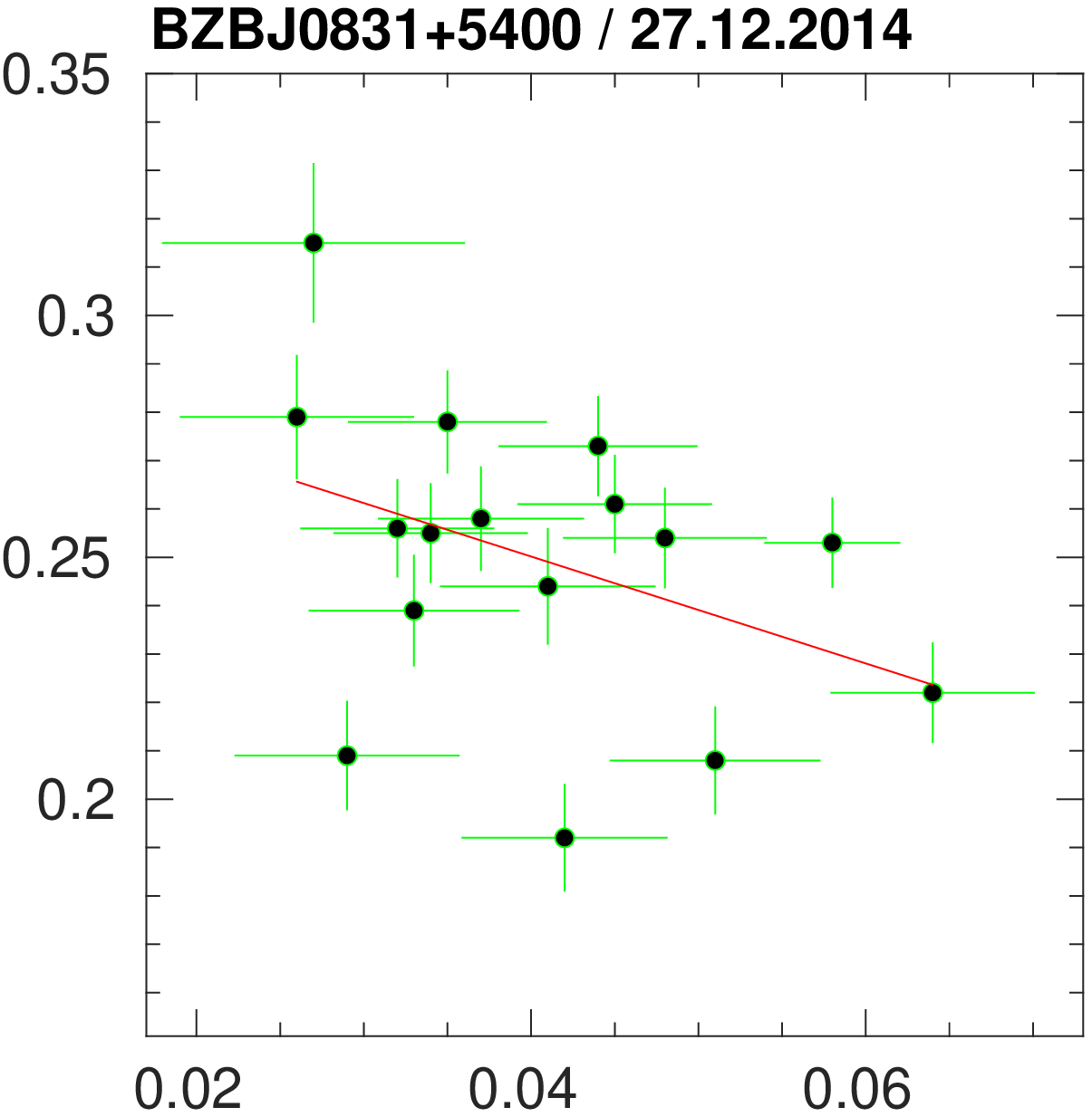}}}
\vspace*{0.1cm}\hspace*{7.0cm}{\large{R magnitude}}
\caption{V--R color indices vs. R-band magnitude diagrams for all the sources on intranight timescale. Source name and observing dates are given at the top of each plot. The red lines represent a linear regression least squared fitting to the data points.}
\end{minipage}
\end{figure*}

\begin{figure*}[hbt!]  
\centering
\begin{minipage}{0.4cm}
\rotatebox{90}{{\large {V--R color indices}}}
\end{minipage}
\begin{minipage}{\dimexpr\linewidth-1.1cm\relax}
\ContinuedFloat
\mbox{\subfloat{\includegraphics[scale=0.33]{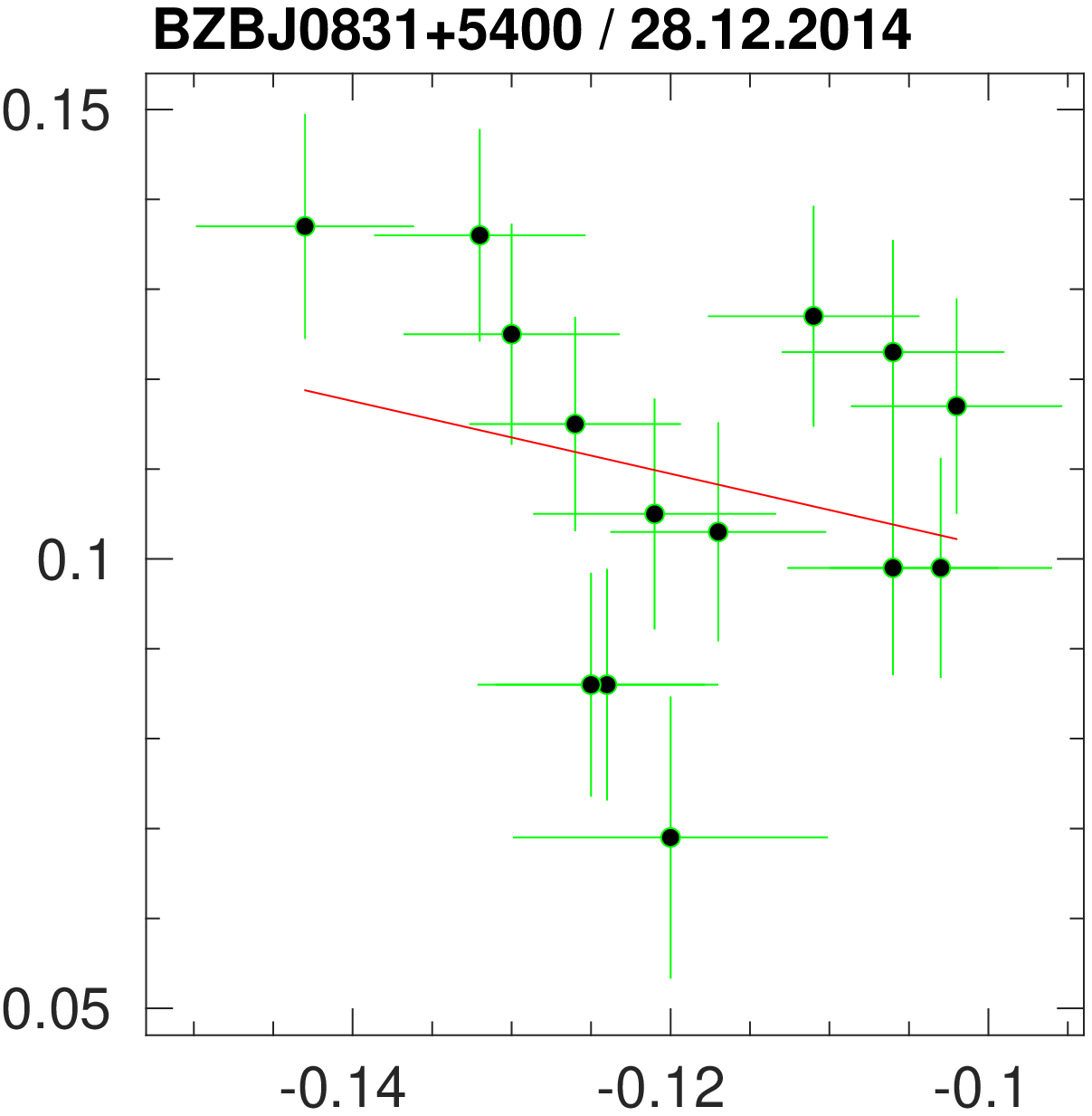}}\quad
\subfloat{\includegraphics[scale=0.33]{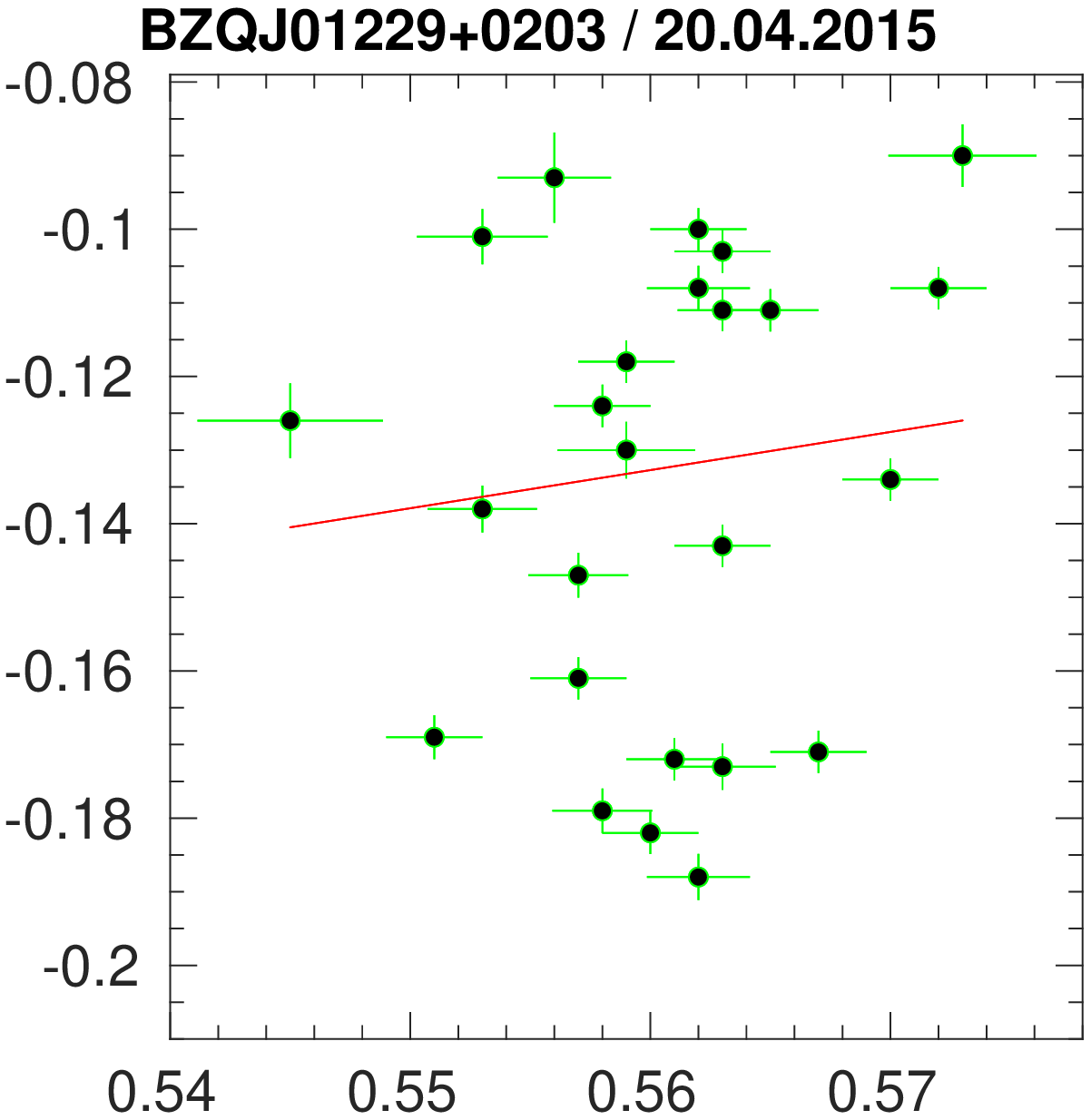}}\quad
\subfloat{\includegraphics[scale=0.33]{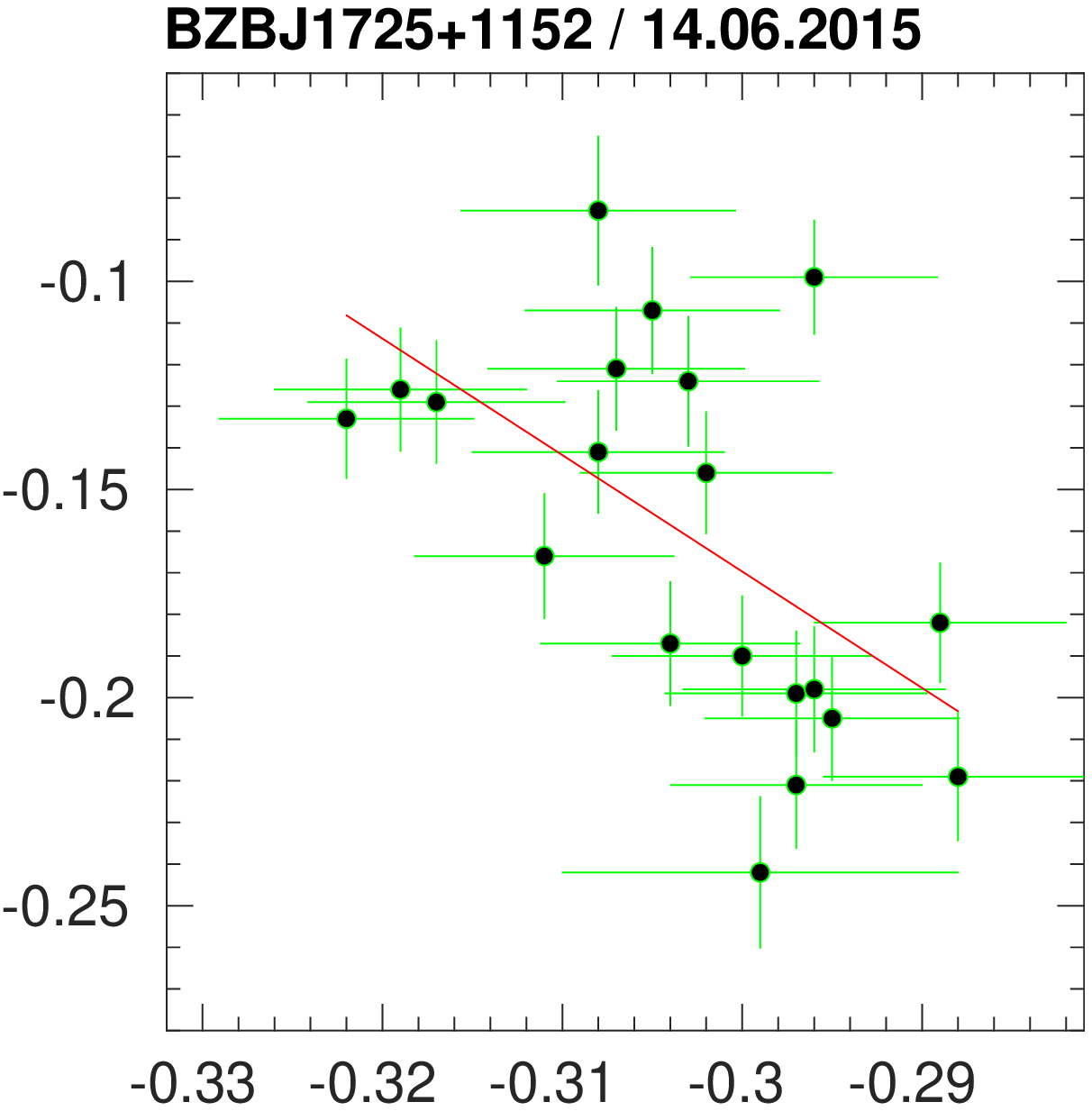}}\quad
\subfloat{\includegraphics[scale=0.33]{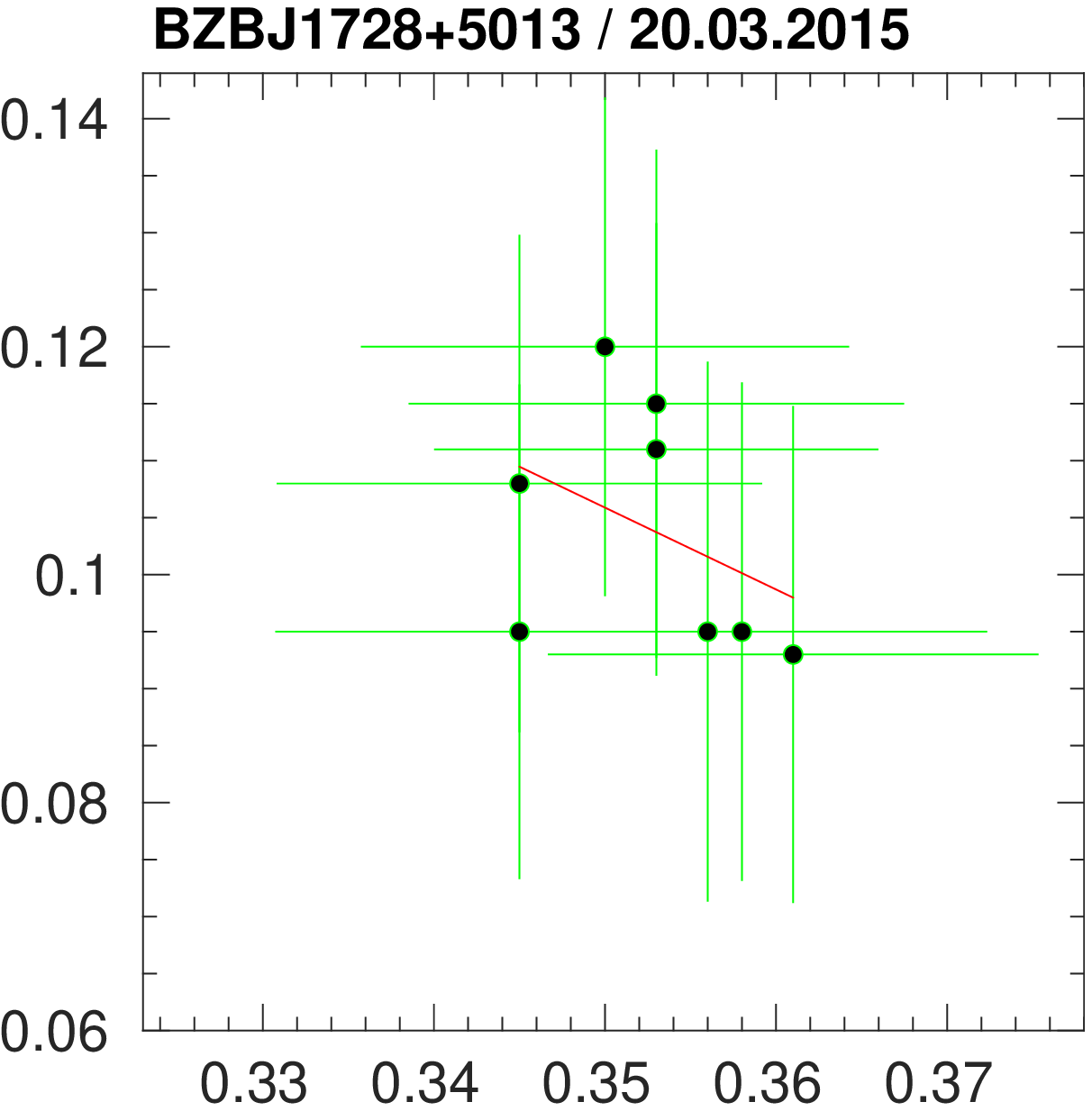}}}
\newline
\mbox{\subfloat{\includegraphics[scale=0.33]{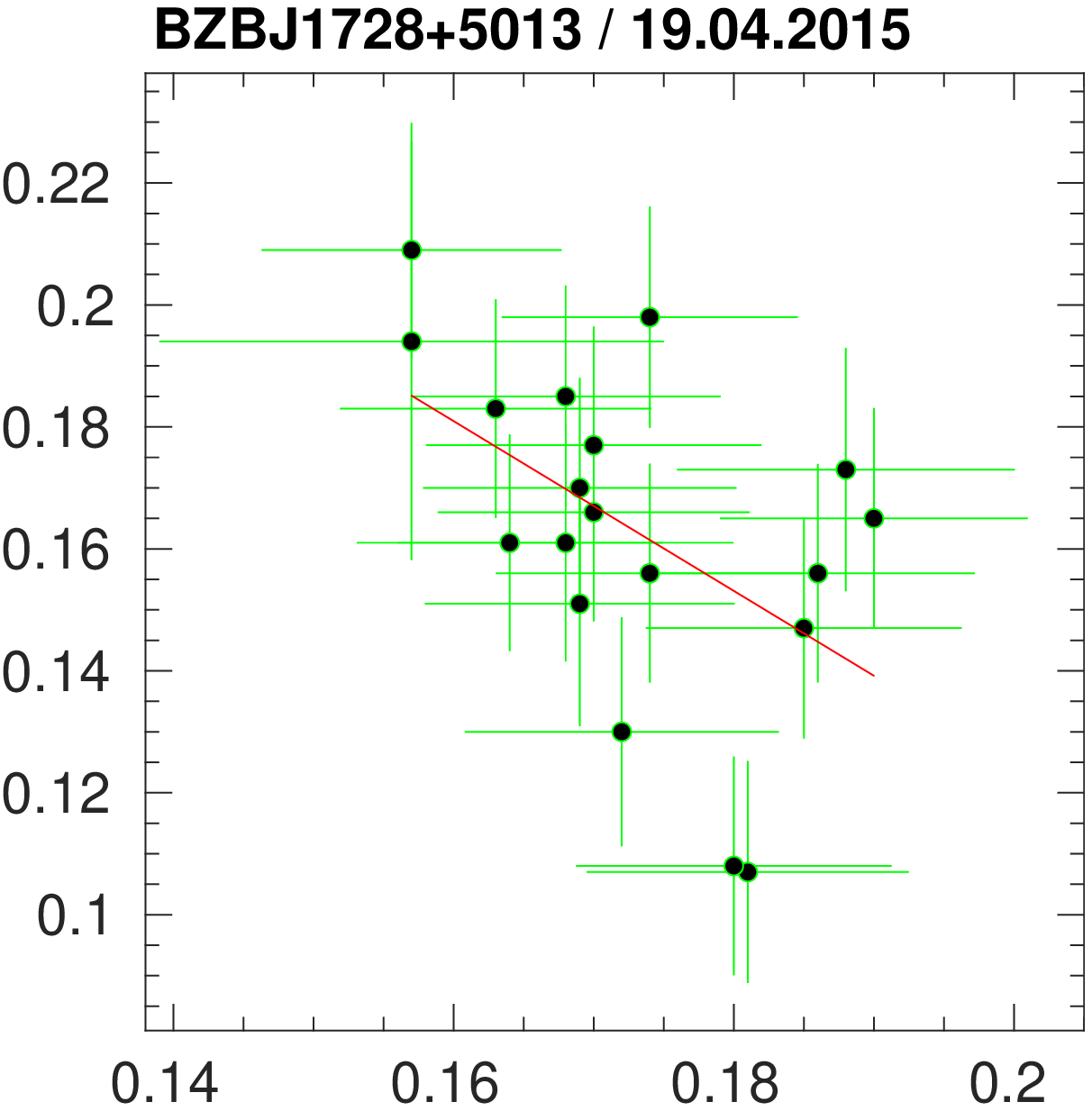}}\quad
\subfloat{\includegraphics[scale=0.33]{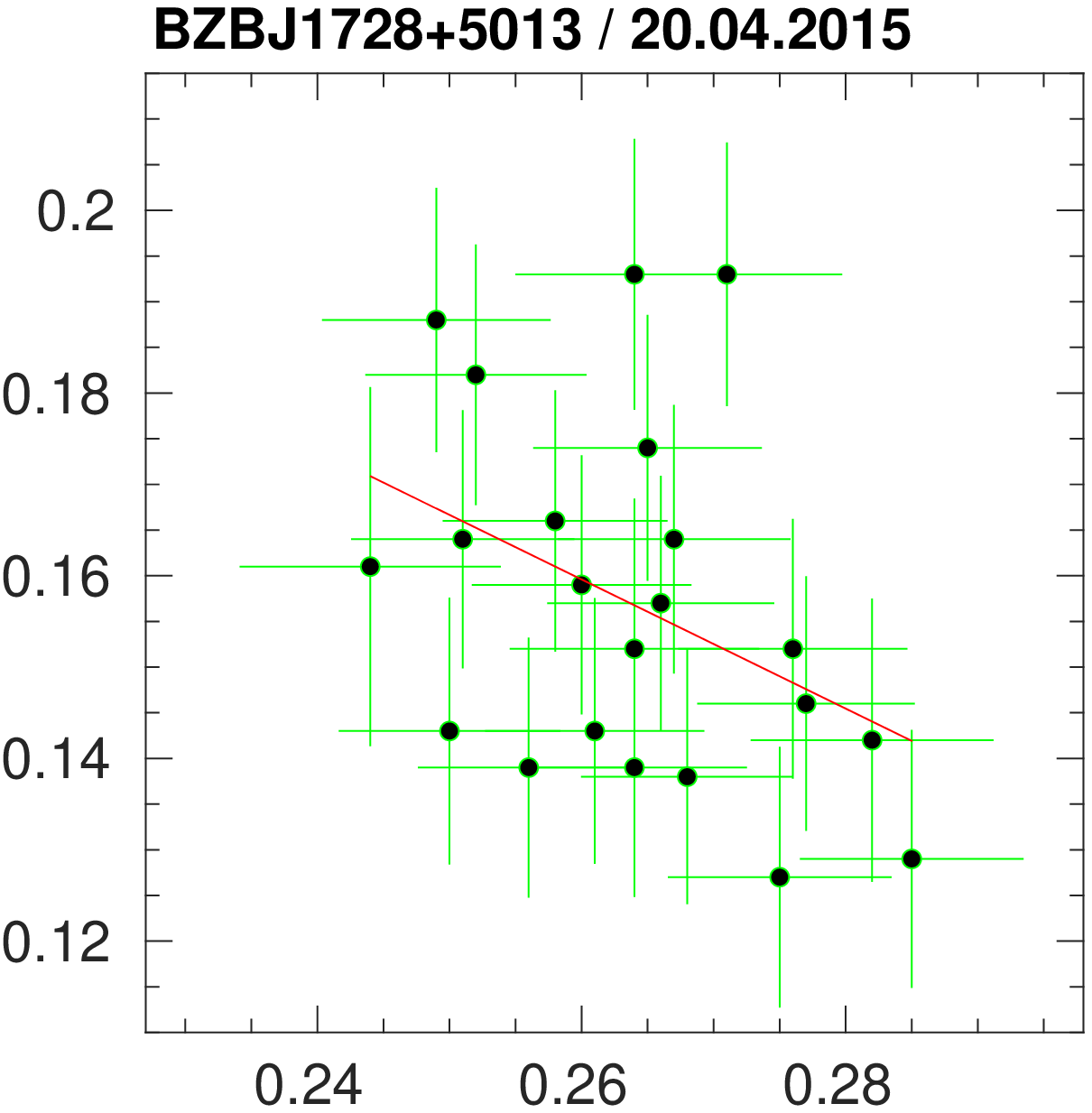}}\quad
\subfloat{\includegraphics[scale=0.33]{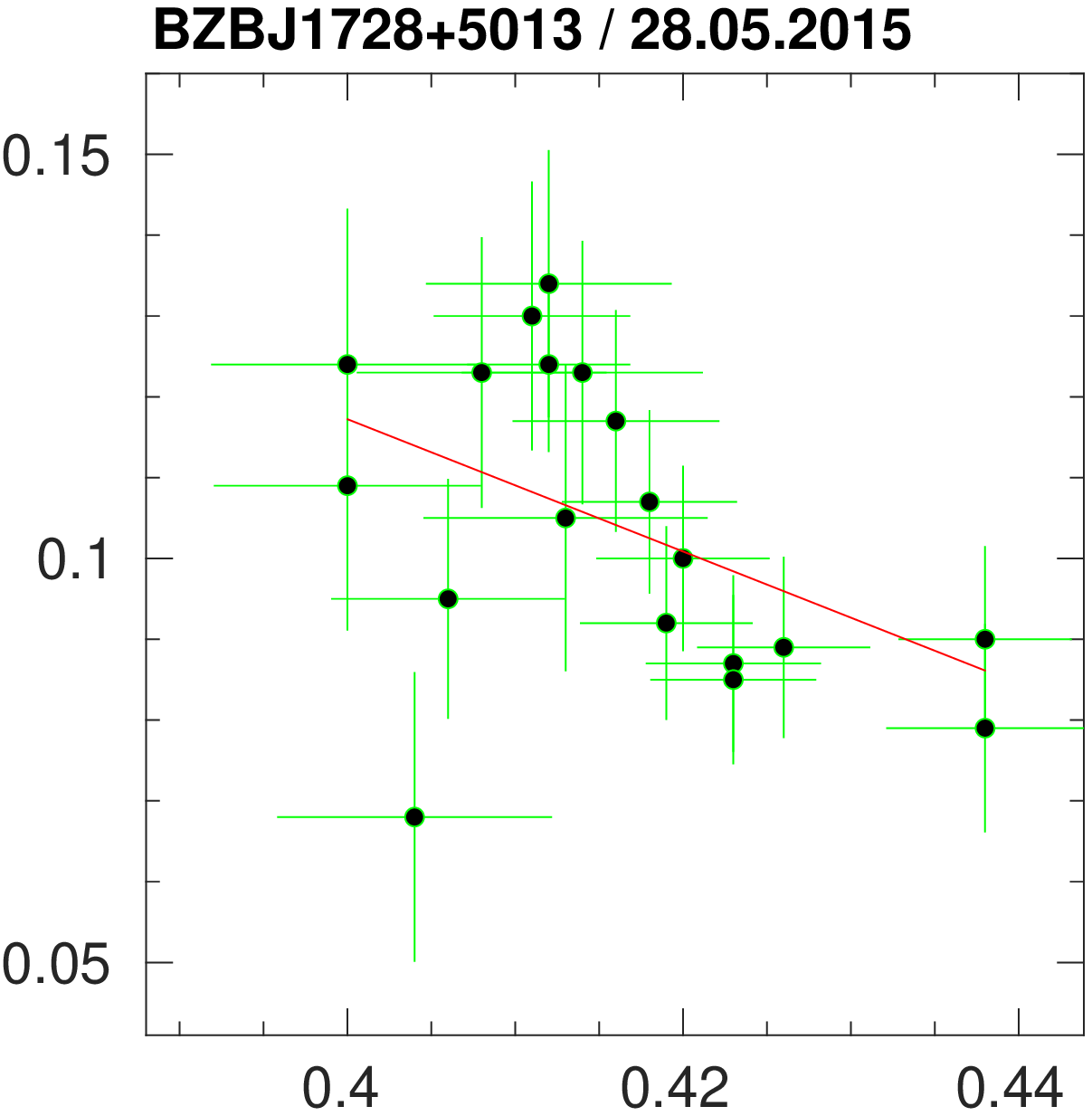}}}

\vspace*{0.1cm}\hspace*{7.0cm}{\large{R magnitude}}
\caption{Continued}
\end{minipage}
\end{figure*}


\subsection{\bf Variability Amplitude}
\noindent
To estimate the variability amplitude of the variable DLCs, we use the variability amplitude 
defined by \citet{1996A&A...305...42H}
\begin{eqnarray}
A = \sqrt {(A_{max} - A_{min})^{2} - 2\sigma^{2}} 
\end{eqnarray}

\noindent
where A$_{max}$ and A$_{min}$ are the maximum and minimum differential instrumental magnitudes in the blazar
LCs and $\sigma$ is the mean error. 


\vspace{1cm}
\section{{\bf Results}}
\noindent
\subsection{\bf Flux and colour variability}
In order to investigate the variability properties of the blazar sample, we study their magnitude and color variation with time on hourly and days to months timescales. We carried out the variability analysis with the PEF-test and nested-ANOVA test as mentioned in the previous section, in two filters with DLCs depicted in Figure 1. The analysis results are summarized in Table \ref{tab5:long}. The final remarks on the DLCs in the table was made based on F--statistic values estimated from both variability testing methods that we adopted for this study. The F--values measured from the tests are higher than the respective critical F--values ($F_{critical}$) at 99.9\% confidence level then we labeled the DLCs as variable (Var in the table). If any of the two F--statistic values are higher than the respective critical values at 99.9\% confidence level or both values are higher at 95\% confidence level, then it was termed as probable variable (PV). Finally the non-variable (NV) one where none of the F-statistic value satisfy the above two conditions. We also investigated the variation in V--R differential colour indices (DCIs) on intraday timescale using the above mentioned analysis methods and applied the same conditions to get their variability status. The V$-$R DCIs were generated by taking an average time for each set of alternate image frames in V and R bands and then subtracting the R band differential magnitude from that of corresponding V band. These results are also summarized in Table \ref{tab5:long}.

\subsection{\bf Spectral variation with colour--magnitude diagram}

To study the spectral behavior of the sources during our observing campaign, we investigated the colour--magnitude relationship via V--R DCIs vs. the R band magnitude diagrams, presented in Figure 3. As seen in the figure, two distinct trends stand out in most of the plots: a negative trend characterized by a steepening of the optical continuum with increasing flux, a trend popularly known as redder--when--brighter (RWB) and a opposite trend characterized by a flattening of the continuum with flux, also known as bluer--when--brighter (BWB). To quantify these correlations, we performed a linear regression fit on each plot using the least-squared method. The fitting is shown by a red straight line in these plots. Although, we see clear trend in all the plots; 20 anti- and 3 positive correlations, we considered them as significant only if the derived Pearson's correlation coefficient (P) lies at or above $P \geq 0.4$. In this way, we found significant correlations in 18 instances out of 23 and all of them follows a RWB trend i.e., brightest state of the sources have softest optical spectrum. On the other hand, a weak BWB trend is observed in only 2 cases with $P$ values 0.10 and 0.24. All the fitting results; slope of the fitting, $P$ values and the null-hypothesis of rejecting the model, are listed in Table \ref{tb}.

Using the average colours $\langle V-R \rangle$, the average spectral indices of the optical spectrum can be derived simply as 
\begin{equation}
\langle \alpha_{VR} \rangle = {0.4\, \langle V-R \rangle \over \log(\nu_V / \nu_R)} \, ,
\label{eq_1}
\end{equation}
where $\nu_V$ and $\nu_R$  are effective frequencies of the respective bands \citep{1998A&A...333..231B, 2015A&A...573A..69W}. The estimated values are also listed in Table \ref{tb}. We plotted the average spectral indices against the mean R-band magnitudes for all the sources in Figure 4, and modeled it with a linear equation of the form $y = 1.96$$\times$$x - 0.28$. We observed a global steepening of the optical spectrum with increasing flux on intranight timescale for the entire blazar sample during our observing run. \\

\subsection{\bf Notes on individual sources}
\vspace{0.4cm}

\hspace{2.8cm}{\it  BZGJ0152+0147}\vspace{0.3cm}

The Fermi-AGN catalog name of the source is PMN J0152+0146 and is also known as RGB J0152+017. It was detected in TeV energies by H.E.S.S. collaboration and belong to the extreme HBLs category. The SED of the source can be described by a two-component non-thermal synchrotron self-Compton (SSC) leptonic model, except in the optical band, which is dominated by a thermal host galaxy component \citep{2008A&A...481L.103A}. The authors also reported that the HBL show no variation in the contemporaneous radio, optical, X-ray, and VHE observations carried out in 2007. However, a separate study reported that the X-ray-UV flux varies on timescales of $\sim$ 10 days and the optical flux varies by $30\%$ on timescales of $\sim 1$ hour \citep{2008AIPC.1085..549K}. 

We observed the source for eight nights in December 2014 and collected data for 5 nights to study intranight flux variation. During the observing time period, the source was not significantly variable  however, we found it probably variable on one occasion. In all the nights, the source showed softening of the spectra with increasing flux in general, where there was a significant RWB trend in 3 nights.   

The short timescale LCs in V and R bands are shown in the second plot of Figure 2, along with color variation and CM diagram from top to bottom, respectively. As the previous source, it is also variable on short timescale which is clearly visible from the plot. Similar variation in the V and R bands results in non variable colour indices. The source's spectral indices show a strong declination with flux yielding a significant RWB trend on short timescales.\\

\hspace{2.8cm}{\it BZBJ0509+0541}\vspace{0.3cm}

This blazar has been recently got all the attention in 2017 due to the detection of neutrino emission from a region co-spatial to the source \citep{2018Sci...361..147I}. It is classified as an ISP blazar in the Fermi-AGN catalog having the commonly used name as TXS 0506+056. However, in a recent study, it has been classified as a FSRQ with hidden broad lines and a standard accretion disk based on the radio and O II luminosities, Eddington ratio and emission line ratios \citep{2019MNRAS.484L.104P}. The first cosmic neutron candidate source is also a TeV candidate. It was reported that after the event, the source's brightness changed by a factor of $\sim$ 1.0 magnitude on a timescale of several days, however no significant IDV was detected. A bluer-when-brighter trend was also observed in optical and near-infrared wavelengths \citep{2021PASJ...73...25M}. The BWB trend was detected in later period as well, during 2018--2020, along with rare, but prominent intra-night variability and significant STV (variation by $\sim$ 1 mag) on a month timescale \citep{2021BlgAJ..34...79B}. \\

The source was observed in two night on 28 and 31 December 2014, collecting intranight data with exposure length 5.3 and 1.5 hours. We did not detect intranight variability in any of these nights however, the V band magnitude varied by $\approx$ 0.15 mag between these two nights where the R-band magnitudes remain almost constant. Due to limited observation, we could not do a proper test for STV detection. Unlike the recent studies during and later the neutrino event, we detected significant RWB trends in both of our observing nights. \\

\hspace{2.8cm}{\it BZGJ0656+4237}\vspace{0.3cm}

The blazar is commonly also known as 4C 42.22 and is a possible TeV candidate \citep{2017A&A...608A..68F}. The source has not been studied in the multi-wavelength and neither it has a proper and detailed single band study. Thus, its properties and behavior are not known. The only study available for this source reported it as non-variable in optical R-band on intranight timescale \citep{2017ApJ...844...32P}.

In our optical monitoring program, this blazar has the highest number of nightly observations. We pointed the HSP for 14 nights between October and December 2014 and collected exposures on hourly timescales for 7 nights. The IDV LCs are shown in the 1st seven panels of Figure 1. From variability analysis, we found two of these DLCs showed probable variation, but none passes the variability tests at 99.9\% significance level. On intranight timescales, the optical spectrum of the source get softer with increasing brightness of the source i.e., a RWB trend. We investigated the temporal and spectral variation on short timescale as well, since the collected data for the blazar covers a time span of $\approx$ three months. The LCs for magnitude and and colour indices are shown in the first plot of Figure 2, from which temporal flux variation is clearly visible in both energy bands. The bottom panel of the plot shows the color-magnitude diagram (hereafter CM-diagram). The softening of the optical spectra with flux persist on the short timescale as well.  

Even though the  blazar did not show IDV, we see notable continuum variation during the observing period. The spectral slope $\alpha_{VR}$, varies between 1 to 2.42, from soft to extremely soft spectra. \\

\hspace{2.8cm}{\it BZGJ0737+5941}\vspace{0.3cm}

This is a BL Lac, also known as S40733+597. Like most of the sources in our sample, this blazar is also not explored yet and thus, we do not know its properties and SED class. In BZCAT, it is categorized as a BL Lac whose optical emission is contaminated by the hos galaxy emission. 

The blazar was observed for 2 and a half hour in April 19, 2015. The variability tests did not detect any temporal brightness variation or colour variation. This is one of the three sources which showed a BWB trend of the optical spectrum, usually observed in HSPs. However, the P--value of the correlation is small (= 0.24), which limit us to claim it a significant detection.\\

\hspace{2.8cm}{\it BZGJ0754+3910}\vspace{0.3cm}

This source is classified as BL Lac in the BZCAT, which shows a faint one-sided structure in NVSS radio images, and a core-jet morphology with a total flux density of $\sim$ 27.0 mJy in VLBA map \citep{2013A&A...560A..23L}. Though, it is classified as a Doppler dominated blazar, it has not been detected in $\gamma$-rays by Fermi-LAT up till now \citep{2018A&A...618A.175D}. Thus, its broadband SED properties are not known yet. No other study is available for this source so far. 

We observed this BL Lac for single night for more than 4 hours in December 2014. We did not detect intranight variability in the source, hence labeled it as a non-variable in Table \ref{tab5:long}.  We observed a weak negative correlation in the CM diagram, hinting presence of a RWB trend in the optical spectrum of the source (Figure 3). \\

\hspace{2.8cm}{\it BZGJ0831+5400}\vspace{0.3cm}

The blazar is classified as BL Lac in the Roma BZCAT catalog, at z = 0.062. The source has not been studied at any energy bands and therefore apart from above there is no other information in the literature regarding it's properties.

We observed this source in December 2014 for two consecutive nights with exposure times 5 hours 20 minutes, and 3 hours. The source was not variable on the observing nights, neither it showed probable variation. A RWB trend was observed in the CM diagram in both nights, however only one of them is significant with $P$ values at 0.40. \\    

\hspace{2.8cm}{\it BZGJ1154+1225}\vspace{0.3cm}

Another nearby BL Lac at a redshift of 0.081, whose optical flux has been labeled as contaminated by host galaxy emission in the BZCAT. The source was not studied before, hence there is no information regarding its behavior or features in the literature.

We pointed the source twice in March 20 and April 20, 2015 with the 40 inch telescope. Due to poor data quality, we had to discard the data taken in April and also the V--band data taken in March. We investigated the IDV of the source with the R--band data and found it to be non variable. Due to lack of multi-band optical data, we could not check the pattern of the optical spectrum. \\ 

\hspace{2.8cm}{\it BZQJ1229+0203}\vspace{0.3cm}

This was the first discovered quasar, commonly known as 3C 273, which is a very popular source and has been extensively studied across the complete EM spectrum. It has a well structured kiloparsec scale jet visible in both radio and optical bands. Due to comparatively high jet angle and also, depending on its different activity states, the source shows features resulted by both the accretion disk and the jet \citep{2008A&A...486..411S, 2017MNRAS.469.3824K}. It's optical flux changes with different amplitudes on diverse timescales revealing the non-linear variability characteristic of blazars \citep{2009AJ....138.1428F, 2019ApJ...880..155L, 2015MNRAS.451.1356K} and it often shows a “bluer-when-brighter” spectral evolution \citep{2017ApJS..229...21X, 2017A&A...605A..43Y}.

We observed the blazar for a single night with exposure time 4 hours 20 minutes with V and R filters. The source showed intraday variability in the V band and also showed temporal V--R colour variation with variability amplitude 11.7 \% and 12.19\%, respectively. Unlike most of the sources, we observed a bluer when brighter trend in the CM--diagram as seen in figure 3, second last panel. However, the pattern is not strong, having a Pearson's correlation coefficient value at 0.1. \\

\begin{figure}  
\centering
\includegraphics[scale=0.65]{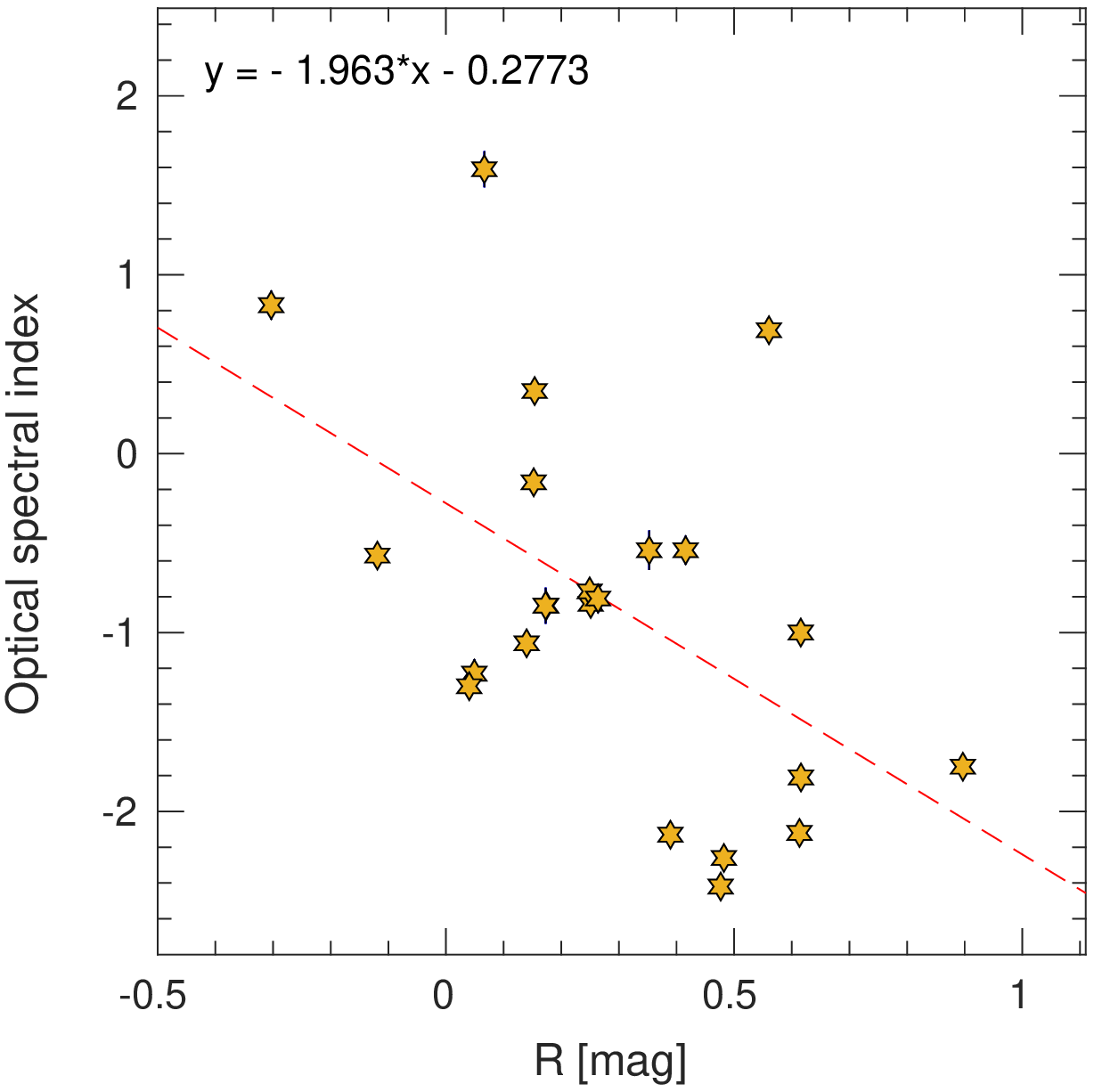}

\caption{Relationship between the optical spectral indices evaluated on nightly timescales and their corresponding averaged R-band magnitude for the entire blazar sample studied in this work. Vertical and horizontal error bars are not visible in most of the points due to their smaller sizes than that of the point symbols. The red dashed line denotes the linear regression fitting using the equation shown in the upper left side of the plot.}
\end{figure}

\hspace{2.8cm}{\it BZBJ1725+1152}\vspace{0.3cm}

This is an HBL blazar, also known as H 1722+119 and was detected in TeV with gamma-ray flux 0.02 of Crab Units \citep{2016MNRAS.459.3271A}. The source was described as a compact  object  with  a  short  jet with unidentified host galaxy morphology \citep{1990ApJ...350..578B, 2012ApJ...744..177L}. No intrinsic or intervening optical spectral features was detected in the source which could be resulted from extreme optical beaming of a closely aligned jet \citep{2014A&A...570A.126L}.  

In order to study the optical variability properties, observation of the blazar was carried out for more than 3 hours in June 2015. Although its an HBL, during our observing time the source did not show variability on IDV timescale with 99.9\% significance (Table \ref{tab5:long}). However, we found it as probably variable source. The source showed a strong RWB trend of the optical spectrum (Table \ref{tb}).  \\

\hspace{2.8cm}{\it BZBJ1728+5013}\vspace{0.3cm}
 
This HSP blazar is known by different names: IZw 187 or 1ES 1727+502 or OT 546. It was categorized as a strong X-ray emitter with a high X-ray to optical luminosity ratio \citep{1980Natur.288..323K} and was detected in TeV gamma-rays by MAGIC \citep{2014A&A...563A..90A}. The source was reported as variable where the variability in optical band was much higher than that in IR band \citep{2000ApJ...537..101F}. 

This source has the second highest number of observations in our sample, with six intra-nights and four nightly observations during 19 March to 16 June 2015. During our observing period, we did not detect any significant IDV in the blazar. We found probable variability only in one R--band LC, but none of the DLCs passed variability tests at the required significance level. We studied optical spectrum using only four CM--diagrams as in two nights, data were collected in only R bands. All of the spectrum showed a strong RWB trend with P--value higher than 0.4, yielding the spectra get softer with increasing flux. 

The temporal magnitude and colour variation on short timescale are shown in last plot of Figure 2 along with the CM-diagram. From this figure, we see that both energy bands are variable and follow the same pattern, and the V--R colour LC is non variable as expected, shown in the middle panel and the color and magnitude follows a weak RWB trend as seen in the bottom panel.\\

\section{\bf Discussion \& Conclusion}
\noindent

In this paper we investigated the optical variability properties of 10 blazars from the Roma BZCAT multi-frequency catalog on intraday and short timescales using multi-band optical observations carried out with 1.04m ST and 1.3m DOFT telescopes between 2014 and 2015. Most of the sources in our sample are newly discovered and has never been studied before. We present the differential magnitude variability study of the selected sources using the two most robust variability testing methods; Power-enhanced F-test and nested ANOVA test, along with their spectral behavior via color--magnitude diagrams. \\

Many of the blazar DLCs in the our sample passed either one of the variability test but none except the well known FSRQ 3C 273, could pass both the tests together at a 99.9\% significance level on a IDV timescale. It is important to note that some of them  showed probable variability as described in subsection 4.1. As mentioned before, the studied sample comprises new blazars that are not regularly monitored. Thus, we do not have information regarding the sources states during our observing run. Irrespective of the sources state, we observed a overall redder-when-brighter trends in the gathered data sets for all the objects, where 78\% of the sample showed significance steepening of the optical spectra with increasing flux with a Pearson’s correlation  coefficients $\geq$ 0.4. \\

The blazar jet is directly pointing towards us, due to that the broad-band emission of these extragalactic sources is predominantly coming from the jet. It is widely believed that most of the flux variation originates from enhanced emission in a tiny region within the jets which could be triggered by dissipation of particle energy inside a shells of relativistic plasma ejected at the jet base. Intermittent ejection of these shells at varying speeds can set up shock fronts which travel down the jet. Multiple shocks propagation can lead to shock-shock interactions, which can increase the magnetic field strength and the particle acceleration. This phenomena is highly efficient in enhancing the emission which sometimes leads to emergence of multiwavelength flares. More complex scenario of variable flux emission might involved magneto-hydrodynamic instabilities. Consequently, the synchrotron emission from the jet will be highly amplified where the frequency of the emitted photons depends on the energy of the accelerated particles. In addition to that the relativistic beaming changes the apparent brightness and the relative prominence of the approaching and receding components (shell of particles) in the observer's rest frame. Due to the extreme Doppler boosting, even a tiny variation in these parameters may cause a huge difference in the observed emission and the variability timescale.\\

Consequently, blazars show unpredictable and rapidly variable emission at all frequencies on different timescales. Their intensity and amplitude of variability are highly dependent on the source states, basically the jet activity. During a flaring phase, extreme flux variation may occur on a time period as short as few minutes, while in a quiescent state it is possible to not even observe any variability. In general BL Lac objects show high degree of variability at higher frequencies, while the opposite is observed in case of FSRQs. \\
 
The sample we studied here mainly comprises of BL Lacs and the non detection of significant IDV in them is a clear indication that most of the sources were in quiescent state during our observing time where the jet emission was not dominant. At optical frequencies, in addition to the intrinsic non-thermal polarized jet emission, unpolarized thermal radiation from the accretion disk also make a good contribution. Now, prominence of the later component varies based on the jet power, which eventually can influence the observed variability. It has been observed that BL Lacs show increasing polarization towards the blue, likely related to intrinsic jet emission, while FSRQs show opposite behavior; dilution effect towards the blue resulted by the thermal emission \citep{2012A&A...545A..48R,2017Natur.552..374R}. Thus, a possible explanation for non variability in the BL Lacs could be resulted by distinct contribution from the disk as well as other components such as the broad line region (BLR) region. The collective output might have canceled out the flux fluctuation of individual components and stabilized the overall light curve. Lack of IDV in optical band has been reported in some well known BL Lacs particularly HSPs, such as Mrk 421, 1ES 1553+113, 1ES 1959+650, and 1ES 2344+514 from time to time in different studies \citep{2012MNRAS.425.3002G, 2019ApJ...871..192P}. A spectroscopic investigation could be useful in such cases to disentangle different emission components contributed by different regions which could provide a more complete picture. \\

An alternative explanation could be given in terms of the synchrotron peak frequency and its location in the broadband SED. In HSP blazars, the optical bands lie below the synchrotron peak, hence, we should see changes in the efficiency of acceleration of the highest energetic particles available for synchrotron emission, and/or in their rate of energy radiation, which would have a retarded effect on optical variability \citep{2012MNRAS.420.3147G}.\\  

On the days to months long timescales (STV), unlike IDV, the optical light curves of the three HSP blazars showed magnitude variation. This could be an indication that the short term variabilities are due to the changes in the accretion disk which could drive the evolution of the optical spectra to a softer side while the source brightness increases as we observed for these HSPs in Figure 2.\\

\begin{longtable*}{lccccccccccccccc}
\caption{\bf Results of variability analysis} \label{tab:results} 
\label{tab5:long}\\

\hline\hline \multicolumn{1}{l}{\textbf{Source name}}  &\multicolumn{1}{c}{\textbf{Date of Obs.}}    &\multicolumn{1}{c}{\textbf{Band}}   &\multicolumn{4}{c}{\textbf{Power-enhanced {\it F}-test}}   &&& \multicolumn{3}{c}{\textbf{Nested {\it ANOVA} test}}   &&&\multicolumn{1}{c}{\textbf{FS}}  &\multicolumn{1}{c}{\textbf{A}} \\ 

                               \cline{4-7} \cline{10-13}

\multicolumn{1}{c}{\textbf{}}   &\multicolumn{1}{c}{\textbf{(dd.mm.yyyy)}} & &\multicolumn{1}{c}{\textbf{dof($\nu_{1}$, $\nu_{2}$)}} &\multicolumn{1}{c}{\textbf{$F_{enh}$}}    &\multicolumn{1}{c}{\textbf{ $F_{c1}$}}    &\multicolumn{1}{c}{\textbf{ $F_{c2}$}}   &&&\multicolumn{1}{c}{\textbf{dof($\nu_{1}$, $\nu_{2}$)}}    &\multicolumn{1}{c}{\textbf{$F$}}   &\multicolumn{1}{c}{\textbf{$F_{c1}$}}    &\multicolumn{1}{c}{\textbf{$F_{c2}$}}    &&\multicolumn{1}{c}{\textbf{}}    &\multicolumn{1}{c}{\textbf{$\%$}}\\

&\multicolumn{1}{c}{} & &(1) &(2) &(3) &(4) &&&(5) &(6) &(7)&(8)&&(9)&(10)\\\hline
\endfirsthead
\multicolumn{3}{c}%
{{\bfseries \tablename\ \thetable{} -- continued from previous page}} \\
\hline\hline \multicolumn{1}{l}{\textbf{Source name}}  &\multicolumn{1}{c}{\textbf{Date of Obs.}}    &\multicolumn{1}{c}{\textbf{Band}}   &\multicolumn{4}{c}{\textbf{Power-enhanced {\it F}-test}}   &&& \multicolumn{3}{c}{\textbf{Nested {\it ANOVA} test}}   &&&\multicolumn{1}{c}{\textbf{FS}}  &\multicolumn{1}{c}{\textbf{A}} \\ 

                               \cline{4-7} \cline{10-13}

\multicolumn{1}{c}{\textbf{}}   &\multicolumn{1}{c}{\textbf{(dd.mm.yyyy)}} & &\multicolumn{1}{c}{\textbf{dof($\nu_{1}$, $\nu_{2}$)}} &\multicolumn{1}{c}{\textbf{$F_{enh}$}}    &\multicolumn{1}{c}{\textbf{ $F_{c1}$}}    &\multicolumn{1}{c}{\textbf{ $F_{c2}$}}   &&&\multicolumn{1}{c}{\textbf{dof($\nu_{1}$, $\nu_{2}$)}}    &\multicolumn{1}{c}{\textbf{$F$}}   &\multicolumn{1}{c}{\textbf{$F_{c1}$}}    &\multicolumn{1}{c}{\textbf{$F_{c2}$}}    &&\multicolumn{1}{c}{\textbf{}}    &\multicolumn{1}{c}{\textbf{$\%$}}\\

&\multicolumn{1}{c}{} & &(1) &(2) &(3) &(4) &&&(5) &(6) &(7)&(8)&&(9)&(10)\\\hline
\endhead

\hline \multicolumn{15}{r}{{Continued on next page}} \\\hline
\endfoot

\hline \multicolumn{16}{l}{{   {\bf Column:}(1) degrees of freedom (dof) in the numerator and the denominator in F-statistic; (2) F value for Power-enhanced F-test; (3), (4), (7)  }} \\ 
\multicolumn{16}{l}{{          \& (8) Critical values at 99$\%$ ($F_{c1}$) and 95$\%$ ($F_{c2}$); (5) degrees of freedom (dof) in the numerator and the denominator in ANOVA-statistics; }} \\ 
\multicolumn{16}{l}{{ (6) F value for nested-ANOVA test; (9) final variability status (V = variable; NV = non--variable; PV = probable variable); (10) Percentage }}\\
\multicolumn{16}{l}{{ variability amplitude.}} \\

\endlastfoot

BZGJ0152+0147& 21.12.2014  &V    &  25, 50   & 0.58   &2.12   &1.79  &&&   5, 18   &2.38  &4.25    &2.77  && NV  &...   \\
             &             &R    &  25, 50   & 0.80   &2.12   &1.79  &&&   5, 18   &3.21  &4.25    &2.77  && NV  &...   \\
             &             &V--R &  25, 50   & 0.80   &2.12   &1.79  &&&   5, 18   &3.54  &4.25    &2.77  && NV  &...   \\
             & 22.12.2014  &V    &  17, 34   & 0.54   &2.70   &2.02  &&&   3, 12   &2.06  &5.95    &3.49  && NV  &...   \\
             &             &R    &  15, 30   & 0.47   &2.70   &2.02  &&&   3, 12   &1.82  &5.95    &3.49  && NV  &...   \\
             &             &V--R &  15, 30   & 0.24   &2.70   &2.02  &&&   3, 12   &0.13  &5.95    &3.49  && NV  &...   \\
             & 23.12.2014  &V    &  33, 66   & 0.59   &2.03   &1.65  &&&   7, 24   &1.10  & 3.50   &2.42  && NV  &...   \\    
             &             &R    &  33, 66   & 0.65   &2.03   &1.65  &&&   7, 24   &1.16  & 3.50   &2.42  && NV  &...   \\
             &             &V--R &  33, 66   & 1.00   &2.03   &1.65  &&&   7, 24   &0.96  & 3.50   &2.42  && NV  &...   \\
             &25.12.2014   &V    &   7  14   & 5.22   &4.28   &2.76  &&&   1, 6    &2.23  &13.75   &5.99  && PV  &...   \\
             &             &R    &  14, 28   & 5.41   &2.76   &2.04  &&&   2, 9    &0.63  & 8.02   &4.26  && PV  &...   \\
                                                                                                                      
             &27.12.2014   &V    &  15, 30   & 0.36   &2.70   &2.02  &&&   3, 12   &2.40  & 5.95   &3.49  && NV  &...   \\
             &             &R    &  16, 32   & 2.06   &2.70   &2.02  &&&   3, 12   &0.55  & 5.95   &3.49  && NV  &...   \\
             &             &V--R &  15, 30   & 0.26   &2.70   &2.02  &&&   3, 12   &2.97  & 5.95   &3.49  && NV  &...   \\
                                                                                                                      
BZBJ0509+0541&28.12.2014   &V     &29, 58   &0.49   & 2.03    &1.65  &&& 6, 21  &  2.29  & 3.81  &2.57  && NV  &...   \\
             &             &R     &30, 60   &0.01   & 2.03    &1.65  &&& 6, 21  &  1.79  & 3.81  &2.57  && NV  &...   \\
             &             &V--R  &29, 58   &0.01   & 2.03    &1.65  &&& 6, 21  &  3.46  & 3.81  &2.57  && NV  &...   \\

             &31.12.2014   &V     & 9, 18   &0.57   & 3.60    &2.46  &&& 1, 6   &  0.09  &13.75  &5.99  && NV  &...   \\
             &             &R     &10, 20   &0.13   & 3.37    &2.35  &&& 1, 6   &  6.25  &13.75  &5.99  && NV  &...   \\
             &             &V--R  & 9, 18   &0.10   & 3.37    &2.35  &&& 1, 6   &  3.43  &13.75  &5.99  && NV  &...   \\

BZGJ0656+4237& 15.10.2014  &V    &   9, 18   & 0.19   &3.60   &2.46  &&&   1, 6  & 1.16   &13.75   &5.99  && NV  &...   \\
             &             &R    &  11, 22   & 0.29   &3.26   &2.30  &&&   2, 9  & 0.30   &13.75   &4.26  && NV  &...   \\
             &             &V--R &   9, 18   & 0.25   &3.60   &2.46  &&&   1, 6  & 1.34   &13.75   &5.99  && NV  &...   \\
             & 16.10.2014  &V    &  10, 20   & 0.78   &3.37   &2.35  &&&   1, 6  & 7.12   &13.75   &5.99  && NV  &...   \\
             &             &R    &  10, 20   & 0.19   &3.37   &2.35  &&&   1, 6  & 0.38   &13.75   &5.99  && NV  &...   \\
             &             &V--R &  10, 20   & 0.61   &3.37   &2.35  &&&   1, 6  & 5.98   &13.75   &5.99  && NV  &...   \\
             & 28.10.2014  &V    &   7, 14   & 0.62   &4.28   &2.76  &&&   1, 6  & 0.06   &13.75   &5.99  && NV  &...   \\
             &             &R    &   8, 16   & 0.05   &3.89   &2.59  &&&   1, 6  & 0.16   &13.75   &5.99  && NV  &...   \\
             &             &V--R &   7, 14   & 0.20   &3.89   &2.59  &&&   1, 6  & 0.36   &13.75   &5.99  && NV  &...   \\
             & 29.10.2014  &V    &   9, 18   & 0.04   &3.60   &2.46  &&&   1, 6  & 2.01   &13.75   &5.99  && NV  &...   \\   
             &             &R    &   9, 18   & 0.01   &3.60   &2.46  &&&   1, 6  & 0.03   &13.75   &5.99  && NV  &...   \\
             &             &V--R &   9, 18   & 0.02   &3.60   &2.46  &&&   1, 6  & 1.71   &13.75   &5.99  && NV  &...   \\
             & 16.11.2014  &V    &  29, 58   & 0.48   &2.03   &1.65  &&&   6, 21 & 2.59   &  3.81  &2.57  && NV  &...   \\
             &             &R    &  31, 62   & 0.52   &2.03   &1.65  &&&   7, 24 & 3.14   &  3.50  &2.42  && NV  &...   \\
             &             &V--R &  29, 58   & 0.64   &2.03   &1.65  &&&   6, 21 & 0.57   &  3.81  &2.57  && NV  &...   \\
             & 22.11.2014  &V    &  37, 74   & 0.82   &1.94   &1.59  &&&   8, 27 & 4.57   &  3.26  &2.31  && PV  &...   \\
             &             &R    &  37, 74   & 1.26   &1.94   &1.59  &&&   8, 27 & 4.82   &  3.26  &2.31  && PV  &...   \\
             &             &V--R &  37, 74   & 0.54   &1.94   &1.59  &&&   8, 27 & 1.59   &  3.26  &2.31  && NV  &...   \\
             & 25.12.2014  &V    &  22, 44,  & 0.26   &2.37   &1.84  &&&   4, 15 & 3.15   &  4.89  &3.06  && NV  &...   \\
             &             &R    &  23, 46   & 0.68   &2.29   &1.79  &&&   5, 18 & 6.45   &  4.25  &2.77  && PV  &...   \\
             &             &V--R &  21, 42   & 0.54   &2.29   &1.79  &&&   4, 15 & 7.19   &  4.89  &3.06  && PV  &...   \\
                                                                                                                   

BZGJ0737+5941&19.04.2015   &V     &10, 20   &1.50   & 3.37    &2.35  &&& 1, 6   &  1.04  &13.75  &5.99  && NV  &...   \\ 
             &             &R     &10, 20   &2.08   & 3.37    &2.35  &&& 1, 6   &  1.63  &13.75  &5.99  && NV  &...   \\
             &             &V--R  &10, 20   &1.25   & 3.37    &2.35  &&& 1, 6   &  0.15  &13.75  &5.99  && NV  &...   \\     

BZGJ0754+3910&26.12.2014   &V     &17, 34   &1.04   & 2.70    &2.02  &&& 3, 12  &  1.38  & 5.95  &3.49  && NV  &...   \\
             &             &R     &19, 38   &2.28   & 2.37    &1.84  &&& 4, 15  &  0.19  & 4.89  &3.06  && NV  &...   \\
             &             &V--R  &17, 34   &1.16   & 2.70    &2.02  &&& 3, 12  &  1.30  & 5.95  &3.49  && NV  &...   \\

BZGJ0831+5400&27.12.2014   &V     &19, 38   &0.05   & 2.37    &1.84  &&& 4, 15  &  1.10  & 4.89  &3.06  && NV  &...   \\
             &             &R     &16, 32   &2.56   & 2.70    &2.02  &&& 3, 12  &  1.01  & 5.95  &3.49  && NV  &...   \\
             &             &V--R  &15, 30   &0.04   & 2.70    &2.02  &&& 3, 12  &  1.33  & 5.95  &3.49  && NV  &...   \\ 
             &28.12.2014   &V     &16, 32   &0.9    & 2.70    &2.02  &&& 3, 12  &  5.07  & 5.95  &3.49  && NV  &...   \\ 
             &             &R     &14, 28   &2.19   & 2.75    &2.04  &&& 2, 9   &  2.13  & 8.02  &4.26  && NV  &...   \\
             &             &V--R  &13, 26   &0.75   & 2.75    &2.04  &&& 2, 9   &  0.25  & 8.02  &4.26  && NV  &...   \\      
BZGJ1154+1225&20.03.2015   &R     &15, 30   &1.32   & 2.70    &2.02  &&& 3, 12  &  2.81  & 5.95  &3.49  && NV  &...   \\

BZQJ1229+0203&20.04.2015   &V     &25, 50   &4.96   & 2.29    &1.79  &&& 5, 18  &  8.55  & 4.25  &2.77  && Var &11.70 \\
             &             &R     &25, 50   &1.97   & 2.29    &1.79  &&& 5, 18  &  3.54  & 4.25  &2.77  && PV  &...   \\
             &             &V--R  &25, 50   &4.27   & 2.29    &1.79  &&& 5, 18  &  5.32  & 4.25  &2.77  && Var &12.19 \\  
BZBJ1725+1152&14.06.2015   &V     &19, 38   &0.74   & 2.37    &1.84  &&& 4, 15  &  4.22  & 4.89  &3.06  && NV  &...   \\  
             &             &R     &19, 38   &0.69   & 2.37    &1.84  &&& 4, 15  & 11.73  & 4.89  &3.06  && PV  &...   \\
             &             &V--R  &19, 38   &0.99   & 2.37    &1.84  &&& 4, 15  &  9.91  & 4.89  &3.06  && PV  &...   \\
BZBJ1728+5013&20.03.2015   &V     & 8, 16   &1.55   & 3.89    &2.59  &&& 1, 6   & 11.28  &13.75  &5.99  && NV  &...   \\
             &             &R     & 7, 14   &0.56   & 4.28    &2.76  &&& 1, 6   &  0.88  &13.75  &5.99  && NV  &...   \\
             &             &V--R  & 7, 14   &0.73   & 4.28    &2.76  &&& 1, 6   & 11.08  &13.75  &5.99  && NV  &...   \\    
             &19.04.2015   &V     &19, 38   &0.68   & 2.37    &1.84  &&& 3, 12  &  3.24  & 5.95  &3.49  && NV  &...   \\
             &             &R     &18, 36   &1.22   & 2.37    &1.84  &&& 4, 15  &  1.05  & 4.89  &3.06  && NV  &...   \\
             &             &V--R  &18, 36   &1.06   & 2.37    &1.84  &&& 3, 12  &  2.35  & 5.95  &3.49  && NV  &...   \\
             &20.04.2015   &V     &21, 42   &1.78   & 2.37    &1.84  &&& 4, 15  &  5.22  & 4.89  &3.06  && NV  &...   \\
             &             &R     &22, 44   &1.20   & 2.37    &1.84  &&& 4, 15  &  1.36  & 4.89  &3.06  && NV  &...   \\ 
             &             &V--R  &21, 42   &1.23   & 2.37    &1.84  &&& 4, 15  &  2.90  & 4.89  &3.06  && NV  &...   \\
             &28.05.2015   &V     &20, 40   &0.05   & 2.37    &1.84  &&& 4, 15  &  0.82  & 4.89  &3.06  && NV  &...   \\ 
             &             &R     &18, 36   &1.51   & 2.37    &1.84  &&& 3, 12  &  1.69  & 5.95  &3.49  && NV  &...   \\
             &             &V--R  &18, 36   &0.06   & 2.37    &1.84  &&& 3, 12  &  1.25  & 5.95  &3.49  && NV  &...   \\      
             &11.06.2015   &R     &60, 120  &1.33   & 1.66    &1.43  &&&14, 45  &  3.58  & 2.52  &1.92  && PV  &...   \\
             &12.06.2015   &R     &28,  56  &0.58   & 2.03    &1.65  &&& 6, 21  &  2.00  & 3.81  &2.57  && NV  &...   \\

\end{longtable*}

\begin{table*}
\label{tb}
\centering
\caption{\bf Linear regression fit to the color--magnitude diagrams.}\label{tb}
\begin{tabular}{lccccc} \hline \hline
Source name   &Date of obs.& $a$       & P$_{(R, V-R)}$         & N   &$\alpha_{\small{VR}}$  \\
              &dd.mm.yy    &  (1)      &    (2)    & (3)        & (4)   \\\hline

BZGJ0152+0147 &21.12.2014  & -0.81  & -0.57   & 2.27e-03  & -0.84 $\pm$ 0.04 \\
              &22.12.2014  & -0.48  & -0.58   & 1.98e-02  & -0.77 $\pm$ 0.06 \\ 
              &23.12.2014  & -0.93  & -0.66   & 2.21e-05  & -0.85 $\pm$ 0.03 \\
              &27.12.2014  & -0.23  & -0.19   & 4.78e-01  & -1.06 $\pm$ 0.07 \\ 
                                                                       
BZBJ0509+0541 &28.12.2014  &  0.98  & -0.65   & 8.99e-05  & -1.85 $\pm$ 0.06 \\
              &31.12.2014  & -0.85  & -0.82   & 4.03e-03  &  0.97 $\pm$ 0.08 \\

BZGJ0656+4237 &15.10.2014  & -1.71  & -0.77   & 9.00e-03  & -1.81 $\pm$ 0.05 \\
              &16.10.2014  & -2.86  & -0.61   & 4.60e-02  & -2.12 $\pm$ 0.04 \\
              &28.10.2014  & -1.35  & -0.83   & 1.06e-02  & -2.42 $\pm$ 0.05 \\
              &29.10.2014  & -0.49  & -0.41   & 2.44e-01  & -2.26 $\pm$ 0.04 \\
              &16.11.2014  & -0.50  & -0.69   & 2.41e-05  & -1.75 $\pm$ 0.03 \\
              &22.11.2014  & -0.49  & -0.43   & 6.80e-03  & -1.00 $\pm$ 0.03 \\
              &25.12.2014  & -1.92  & -0.59   & 3.90e-03  & -2.13 $\pm$ 0.04 \\

BZGJ0737+5941 &19.04.2015  &  0.88  &  0.24   & 4.84e-01  &  0.69 $\pm$ 0.02 \\

BZGJ0754+3910 &26.12.2014  & -0.48  & -0.24   & 3.34e-01  & -1.23 $\pm$ 0.08 \\

BZGJ0831+5400 &27.12.2014  & -1.11  & -0.40   & 1.34e-01  & -1.30 $\pm$ 0.06 \\
              &28.12.2014  & -0.40  & -0.24   & 4.01e-01  & -0.57 $\pm$ 0.07 \\
BZQJ1229+0203 &20.04.2015  &  0.52  &  0.10   & 6.36e-01  &  1.59 $\pm$ 0.10 \\

BZBJ1725+1152 &14.06.2015  & -2.80  & -0.57   & 9.20e-03  &  0.83 $\pm$ 0.08 \\ 
                                                                       
BZBJ1728+5013 &20.03.2015  & -0.72  & -0.41   & 3.43e-01  & -0.54 $\pm$ 0.11 \\
              &19.04.2015  & -1.39  & -0.50   & 2.82e-02  & -0.85 $\pm$ 0.10 \\
              &20.04.2015  & -0.71  & -0.40   & 6.92e-02  & -0.81 $\pm$ 0.08 \\
              &28.05.2015  & -0.82  & -0.46   & 4.62e-02  & -0.54 $\pm$ 0.08 \\\hline
\\
\end{tabular} \\   
{\bf Columns:} (1) Slope of least squared fitting; (2) correlation co-efficient; (3) null-hypothesis\\ of rejecting the model; (4) average spectral indices of the optical spectrum. \\ 
\end{table*}

\acknowledgments 

\noindent
We thank the anonymous referee for constructive comments and suggestions. N.K. acknowledges funding from the Chinese Academy of Sciences President’s International Fellowship
Initiative Grant No. 2020PM0029. ACG is partially supported by Chinese Academy of Sciences (CAS) President’s International Fellowship Initiative (PIFI) grant no. 2016VMB073. MFG acknowledges support from the National Science Foundation of China (grant 11873073).


\end{document}